\documentclass[12pt, a4paper]{article}
 \usepackage[font=small,format=plain,labelfont=bf,up,textfont=normal,up,justification=justified,singlelinecheck=false]{caption}
\usepackage{subcaption}
\usepackage{mathrsfs}
\usepackage[a4paper, left=2cm,right=2cm]{geometry}
\usepackage[colorlinks=true,linkcolor=black,citecolor=teal,urlcolor=MidnightBlue,filecolor=black]{hyperref}
\usepackage{amsfonts}
\usepackage{amsmath,amssymb}
\usepackage{setspace}
\usepackage{slashed}
\usepackage{braket}

\usepackage[dvipsnames]{xcolor}
\definecolor{SchoolColor}{rgb}{0.6471, 0.1098, 0.1882} 
\usepackage{subfloat}

\usepackage[utf8,applemac]{inputenc}
\usepackage{tensor}
\usepackage{cite}
\usepackage{tikz}
\usetikzlibrary{calc}
\usetikzlibrary{patterns}
\usetikzlibrary{arrows.meta}
\usetikzlibrary{decorations.text}

\usepackage{graphicx}
\usepackage{bm} 
\allowdisplaybreaks[4]
\usepackage[utf8,applemac]{inputenc}
\usepackage{tensor}
\usepackage{cite}
\usepackage{tikz}
\usepackage{graphicx}
\usepackage{graphics}
\graphicspath{{figure/}}
\usepackage{array}
\usepackage{booktabs}

\usepackage{multirow}

\usepackage{dcolumn}
\usepackage{bm}

\usepackage{verbatim}

\usepackage{textcomp} 
\usepackage{graphicx} 

\numberwithin{equation}{section}
\newcommand{\bea}{\begin{eqnarray}}
\newcommand{\eea}{\end{eqnarray}}
\newcommand{\be}{\begin{equation}}
\newcommand{\ee}{\end{equation}}
\newcommand{\bs}{\begin{subequations}}
\newcommand{\es}{\end{subequations}}
\def\nn{\nonumber}

\newcommand{\beqs}{\begin{eqnarray}}
\newcommand{\eeqs}{\end{eqnarray}}
\numberwithin{equation}{section}

\newcommand{\Rmnum}[1]{\uppercase\expandafter{\romannumeral #1\relax}}

\def\c.c.{\mathrm{c.c.}}

\begin{document}
\begin{titlepage}

\begin{flushright}\vspace{-3cm}
{\small
\today }\end{flushright}
\vspace{0.5cm}
\begin{center}
	{{ \LARGE{\bf{Carrollian propagator and amplitude \\\vspace{8pt}in Rindler spacetime   }}}}\vspace{5mm}

	\centerline{Ang Li\footnote{liang121@hust.edu.cn}, Jiang Long\footnote{longjiang@hust.edu.cn} \& Jing-Long Yang\footnote{yangjinglong@hust.edu.cn}}
	\vspace{2mm}
	\normalsize
	\bigskip\medskip

	\textit{School of Physics, Huazhong University of Science and Technology, \\ Luoyu Road 1037, Wuhan, Hubei 430074, China
	}
	
	\vspace{25mm}
	
	\begin{abstract}
		\noindent
		{We study the three-dimensional Carrollian field theory  on the Rindler horizon which is dual to a bulk massless scalar field theory in the four-dimensional Rindler wedge. The Carrollian field theory could be mapped to a two-dimensional Euclidean field theory in the transverse plane by a Fourier transform.  After defining the incoming and outgoing states at the future and past Rindler horizon respectively, we construct the boundary-to-boundary and  bulk-to-boundary propagators that are consistent with the bulk Green's function in the literature. We investigate the tree-level Carrollian amplitudes up to four points. The tree-level four-point Carrollian amplitude in $\Phi^4$ theory has the same structure as the one-loop triangle Feynman integral in the Lee-Pomeransky representation with complex powers in the propagators and spacetime dimension. Moreover, the four-point Carrollian amplitude with a zero energy state inserted at infinity in $\Phi^4$ theory is proportional to the three-point Carrollian amplitude in $\Phi^3$ theory.} \end{abstract} 
	

\end{center}

\end{titlepage}
\tableofcontents

\section{Introduction}
Recently, there are numerous works on holographic principle \cite{1993gr.qc....10026T,Susskind:1994vu,Polchinski:1999ry,Susskind:1998vk,Giddings:1999jq,deBoer:2003vf,Arcioni:2003xx,Arcioni:2003td,Mann:2005yr}  in flat spacetime in the framework of celestial holography \cite{Pasterski:2016qvg,Pasterski:2017kqt,Pasterski:2017ylz}  and Carrollian holography \cite{Donnay:2022aba,Bagchi:2022emh}. The former claims that the boundary two-dimensional field theory lives on the celestial sphere whose correlators are defined through celestial amplitudes. The latter conjectures that the boundary field theory lives on a three-dimensional Carrollian manifold \cite{Une, Gupta1966OnAA,Duval:2014uva,Duval:2014lpa} whose correlators are mapped to the Carrollian amplitudes \cite{Liu:2022mne, Donnay:2022wvx,Salzer:2023jqv,Nguyen:2023miw,Mason:2023mti,Liu:2024nfc,Stieberger:2024shv,Adamo:2024mqn,Alday:2024yyj,Ruzziconi:2024zkr,Liu:2024llk} in the bulk. Both of  the celestial amplitude and the Carrollian amplitude are equivalent to the standard S-matrix by integral transforms. 
The  celestial holography can be adapted to  two-dimensional conformal field theory. On the other hand, the Carrollian holography is based on geometric properties of the Carrollian manifold and matches perfectly with asymptotic symmetries \cite{Barnich:2010eb,Sachs:1962wk,Campiglia:2014yka, Campiglia:2015yka,Duval:2014uva,Duval:2014lpa} and field quantization \cite{Liu:2022mne,Liu:2023qtr,Liu:2023gwa,Li:2023xrr,Liu:2023jnc,Liu:2024nkc,Liu:2024rvz}.

Most of the works  focus on the Carrollian field theories at future/past null infinity ($\mathscr{I}^{\pm}$) in Carrollian holography. In this case, one can utilize the well-established S-matrix, transform it to the Carrollian amplitude and define the correlation functions of the putative Carrollian field theory. On the other hand, there are various Carrollian manifolds in physics, including the Rindler horizon of an accelerating observer and the event horizon of a black hole. Both of them are extremely important for us to explore the fundamental properties of spacetime. From the intrinsic perspective of Carrollian physics, one should be able to study the correlators of the putative field theory on these manifolds. Unfortunately, besides the equivalence between the Carrollian correlator and amplitude, there are no known work on the Carrollian amplitude in these curious spacetimes and it is not clear how to extract the information of the putative field theory from the bulk side. 

However, we have shown in a previous paper that one can quantize the field theory on an arbitrary null hypersurface \cite{Li:2023xrr} in the framework of bulk reduction \cite{Liu:2022mne,Liu:2024nkc}.
This work builds a connection between the boundary and  bulk field which is the key to define Carrollian amplitude in these non-Minkowski spacetimes. For a globally hyperbolic spacetime, there are two null boundaries, one is in the past and the other is in the future, which can be used to define the incoming and outgoing states. The Carrollian amplitude is still the S-matrix between these incoming and outgoing states in the Carrollian space, and could be connected to the bulk S-matrix in the momentum space by an integral transform which  is not necessary the Fourier transform in the general cases. For the Rindler spacetime, the Poincar\'e group $\text{ISO}(1,3)$ is broken to $\text{SO}(1,1)\times \text{ISO}(2)$ due to  the presence of the Rindler horizons. We investigate the Ward identities associated with these residual global symmetries and the boundary-to-boundary, bulk-to-boundary and bulk-to-bulk propagators in the right Rindler wedge (RRW). After switching to the  Fourier space, the boundary-to-boundary propagator in RRW is the same as the two-point correlation function of a primary operator with a complex conformal dimension in a two-dimensional conformal field theory, which is quite different from the propagator from $\mathscr{I}^-$ to $\mathscr{I}^+$ of Minkowski spacetime. Using the split representation of the bulk-to-bulk propagator \cite{Liu:2024nfc}, we could reproduce the Feynman propagator \cite{Dowker:1978aza} in the Rindler vacuum. We have also computed 
the tree-level three-point Carrollian amplitude in $\Phi^3$ and four-point Carrollian amplitude in $\Phi^4$ massless scalar theory in the RRW. Interestingly, the three-point zero-energy Carrollian amplitude (ZECA) in $\Phi^3$ theory has the same form of the four-point ZECA in $\Phi^4$ theory up to some kinematic factors. The tree-level four-point Carrollian amplitude in $\Phi^4$ theory shows a two-dimensional conformal invariance in the transverse plane. On the other hand, the three-point Carrollian amplitude in $\Phi^3$ theory breaks the conformal symmetry due to the dimensional parameter $\lambda_3$ in the action.

The layout of this  paper is  as follows. In section \ref{rindlerspacetime}, we review various aspects of Rindler spacetime, including the coordinate systems and  the residual global symmetries  used in this work.  In section \ref{Scalarfield}, we will explore the definition of the incoming and outgoing Rindler states and study the definition of the Carrollian amplitude in RRW. We calculate the boundary-to-boundary, bulk-to-boundary and bulk-to-bulk propagators in the following section. We use the propagators and Feynman rules in the Carrollian space to investigate various Carrollian amplitudes in section \ref{app}. We will conclude in section \ref{conc}. Technical details are  relegated to several appendices.

\section{Rindler spacetime}\label{rindlerspacetime}
\subsection{Coordinate systems}
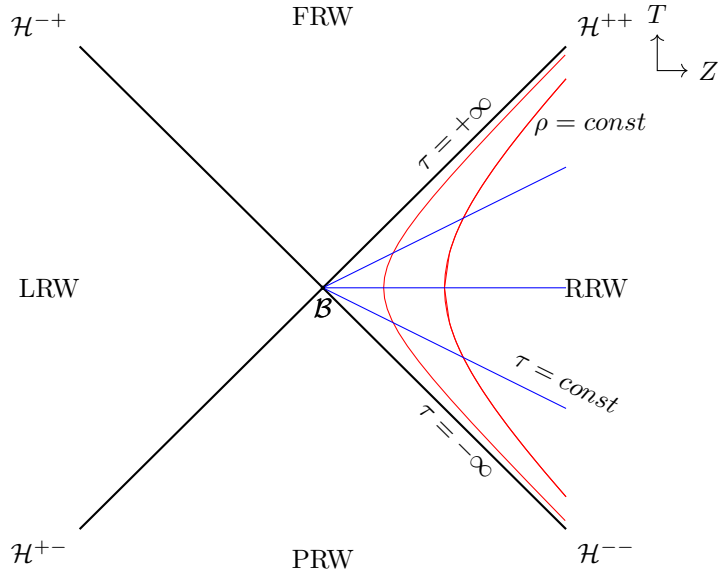
\begin{figure}
  \centering
  \usetikzlibrary{decorations.text}
  \begin{tikzpicture}[scale=0.8]
    \draw[draw,thick] (4,4) node[above right]{\footnotesize $\mathcal{H}^{++}$} -- (0,0) node[below]{\footnotesize $\mathcal{B}$} -- (4,-4) node[below right]{\footnotesize $\mathcal{H}^{--}$};
    \draw[draw,thick] (-4,4) node[above left]{\footnotesize $\mathcal{H}^{-+}$} -- (0,0) node[below]{\footnotesize $\mathcal{B}$} -- (-4,-4) node[below left]{\footnotesize $\mathcal{H}^{+-}$};

\draw[color=red,domain=-2.06:2.06]    plot ({cosh(\x)},{sinh(\x)})  ;
  
  \draw[color=red,domain=-1.3169:1.3169]    plot ({2*cosh(\x)},{2*sinh(\x)})  ;
 \draw[color=red,domain=2:4]    plot (\x,{((\x)^2-4)^0.5})  ;
\draw[color=red,domain=2:4]    plot (\x,{-((\x)^2-4)^0.5})  ;

  \draw[color=blue,domain=0:4]    plot (\x,{\x*(0.5)})  ;
   \draw[color=blue,domain=0:4]    plot (\x,{\x*tanh(0)})  ;  
    \draw[color=blue,domain=0:4]    plot (\x,{\x*(-0.5)})  ;

    \fill [fill,use as bounding box](0,0) circle (1pt);

    \node at (4.4,2.7) {\footnotesize $\rho=const$};
 
    \node at (4.5,0) {\footnotesize RRW};
    \node at (-4.5,0) {\footnotesize LRW};
       \node at (0,4.5) {\footnotesize FRW};
    \node at (0,-4.5) {\footnotesize PRW};
    
\path decorate [decoration={text along path,raise=4pt,
text={{\footnotesize$\tau=+\infty$}{}}}]{ (1.8,1.8) -- (4.4,4.4) };

 \path decorate [decoration={text along path,raise=6pt,
text={{\footnotesize$\tau=const$}{} }}]{ (3,-1.5) -- (5,-2.5) };
       
 \path decorate [decoration={text along path,raise=-8pt,
text={{\footnotesize$\tau=-\infty$}{} }}]{ (1.8,-1.8) -- (4,-4) };

    \draw[<->] (5.5,4.2) node[above] {\footnotesize $T$} -- (5.5,3.6) -- (6,3.6) node[right] {\footnotesize $Z$};
  \end{tikzpicture}
  \caption{Rindler spacetime. The Minkowski spacetime is divided into four patches by the two null hypersurfaces $\mathcal{H}^+$ and $\mathcal{H}^-$. The four patches are named as left Rindler wedge (LRW), right Rindler wedge (RRW), future Rindler wedge (FRW) and past Rindler wedge (PRW), respectively.   The RRW is parameterized by the Rindler coordinates $(\tau,\rho,\bm x)$. The blue straight lines are the constant $\tau$ slices whose value increases in an anticlockwise manner. The null hypersurface $\mathcal{H}^{--}$ corresponds to  $\tau=-\infty$ and $\mathcal{H}^{++}$ corresponds to $\tau=+\infty$. Therefore, the Rindler incoming states may be defined at $\mathcal{H}^{--}$ and the Rindler outgoing states are defined at $\mathcal{H}^{++}$. The red curves (hyperbolic curves in the figure) are constant $\rho$ surfaces. }\label{RRW}
\end{figure}

In this paper, the metric of the Minkowski spacetime $\mathbb{R}^{1,3}$ in Cartesian coordinates $X^\mu=(T,X,Y,Z)$,
$\mu=0,1,2,3$ is 
\be 
ds^2_{\text{Mink}}=-dT^2+dX^2+dY^2+dZ^2.
\ee 
Rindler spacetime is a patch of the Minkowski spacetime which may be obtained  from the coordinate transformation 
\be 
T=\rho \sinh\tau,\quad Z=\rho\cosh\tau,\quad X=x,\quad Y=y.\label{transrindler}
\ee The spatial coordinates $(X,Y,Z)$ are collected as $X^i,\ i=1,2,3$ and the transverse coordinates $(X,Y)$ are denoted as $x_{}^A,\ A=1,2$. We may also use $\bm x$ to denote the transverse coordinates to simplify notation. The RRW is the patch that satisfies the inequality
\be 
Z>|T|.
\ee The Rindler coordinates $(\tau,\rho,\bm x_{})$ are in the domain
\be 
\rho>0,\quad -\infty<\tau<\infty,\quad -\infty<x_{}^A<\infty,
\ee and the metric of the RRW is 
\be 
ds^2=-\rho^2d\tau^2+d\rho^2+\delta_{AB}dx_{}^A dx_{}^B,
\ee which can be transformed to the form 
\be 
ds^2=-\rho^2 du^2-2\rho du d\rho+\delta_{AB}dx_{}^A dx_{}^B
\ee in advanced coordinates $(u,\rho,\bm x_{})$ with \footnote{The notation of advanced/retarded time $u$/$v$  in this article is opposite to that in asymptotically flat spacetime. }
\be 
u=\tau-\log\rho
\ee and 
\be 
ds^2=-\rho^2 dv^2+2\rho dv d\rho+\delta_{AB}dx_{}^A dx_{}^B
\ee
in retarded coordinates $(v,\rho,\bm x_{})$ with 
\be 
v=\tau+\log\rho.
\ee 
The transformation from Cartesian coordinates to advanced/retarded Rindler coordinates are 
\bs\begin{align}
    T&=\frac{1}{2}(-e^{-u}+\rho^2 e^u),\quad Z=\frac{1}{2}(e^{-u}+\rho^2 e^u),\quad X=x,\quad Y=y,\quad \text{advanced},\\
    T&=\frac{1}{2}(e^v-\rho^2 e^{-v}),\quad Z=\frac{1}{2}(e^v+\rho^2 e^{-v}),\quad X=x,\quad Y=y,\quad \text{retarded}.
\end{align}\es 
The null boundary of the RRW is 
\be 
Z=|T|,\quad Z>0
\ee which is split into two parts according to the sign of the Cartesian time. The null hypersurface $\mathcal{H}^{++}$ is the boundary with positive Cartesian time 
\be 
\mathcal{H}^{++}=\{ Z=T>0\}
\ee while the null hypersurface $\mathcal{H}^{--}$ is the boundary with negative Cartesian time 
\be 
\mathcal{H}^{--}=\{Z=-T>0\}.
\ee Both of them  are Killing horizons associated with the Lorentz boost generator along $Z$ direction.
Since 
\be 
Z^2-T^2=\rho^2, 
\ee the null boundaries correspond to the limit $\rho\to0$. Practically, we may choose a cutoff $\rho=\epsilon>0$ and consider the hypersurface $\mathcal{H}_{\epsilon}$
\bea 
\mathcal{H}_\epsilon=\{Z^2-T^2=\epsilon^2\}.
\eea The Killing horizon $\mathcal{H}^{--}$ is the $\epsilon\to 0$ limit of a series hypersurfaces $\mathcal{H}_{\epsilon}$ while keeping the advanced time $u$ finite
\be 
\mathcal{H}^{--}=\lim_{\epsilon\to 0,\ u\ \text{finite}} \mathcal{H}_{\epsilon}.
\ee 
 Therefore, the Killing horizon $\mathcal{H}^{--}$ could be parameterized by three coordinates $(u,\bm x_{})$  and its metric is degenerate 
\be ds^2_{\mathcal{H}^{--}}=\delta_{AB}dx^A dx^B.\label{metrichmm}
\ee This is exactly a Carrollian manifold. Note that to keep the advanced time $u$ finite, the Rindler time $\tau$ should be sent to $-\infty$. Similarly, the Killing horizon $\mathcal{H}^{++}$ is the $\epsilon\to0$ limit of a series of hypersurfaces $\mathcal{H}_\epsilon$ while keeping the retarded time $v$ finite
\be 
\mathcal{H}^{++}  =\lim_{\epsilon\to 0,\ v\ \text{finite}} \mathcal{H}_{\epsilon}.
\ee We can still use three coordinates $(v,\bm x_{})$ to describe $\mathcal{H}^{++}$ whose metric is the same as \eqref{metrichmm}.
To keep the retarded time $v$  finite, the Rindler time $\tau$ should be  $\tau=+\infty$. As has been shown in \cite{Li:2023xrr}, one can also define the null boundaries of left Rindler wedge (LRW) 
\be 
\mathcal{H}^{+-}=\{T=Z<0\},\quad
\mathcal{H}^{-+}=\{T=-Z>0\}
\ee
and a bifurcation surface 
\be 
\mathcal{B}=\{T=Z=0\}.
\ee For latter convenience, we define two other null hypersurfaces $\mathcal{H}^{\pm}$ 
\bs\begin{align}
    \mathcal{H}^+&=\{T=Z\}=\mathcal{H}^{++}\cup\mathcal{H}^{+-}\cup\mathcal{B},\\
    \mathcal{H}^-&=\{T=-Z\}=\mathcal{H}^{--}\cup \mathcal{H}^{-+}\cup\mathcal{B}
\end{align}\es which could be described by lightcone coordinates
\be 
V=T+Z,\quad U=T-Z.
\ee The null hypersurface $\mathcal{H}^+$ is parameterized by $U=0$ and one can use coordinates $(V,X,Y)$ to describe $\mathcal{H}^+$. Similarly, $V=0$ corresponds to the null hypersurface $\mathcal{H}^-$ and one can use  $(U,X,Y)$ to describe it.
In Figure \ref{RRW}, we have separated the Minkowski spacetime to four parts according to the two null hypersurfaces $\mathcal{H}^{\pm}$.

 For completeness, we should also consider the future/past null infinity of RRW. They are portions of the future/past null infinity of Minkowski spacetime. We define the retarded and advanced coordinates in Minkowski spacetime as usual
\bs\begin{align}
\bar U&=T-R=\rho\sinh\tau-\sqrt{\rho^2\cosh^2\tau+X^2+Y^2},\label{barU}\\ \bar V&=T+R=\rho\sinh\tau+\sqrt{\rho^2\cosh^2\tau+X^2+Y^2},
\end{align}\es where $R=\sqrt{X^2+Y^2+Z^2}$ is the radial coordinate.
The future null infinity of Minkowski spacetime $\mathscr{I}^+$ corresponds to $R\to+\infty$ with $\bar U$ finite. In terms of Rindler coordinates, this may be realized by setting $\rho\to+\infty,\ \tau\to+\infty$ and keep $
u=\tau-\log\rho
$ finite\footnote{We have assumed $X^2+Y^2$ to be finite here. To keep $\bar U$ finite,  $X^2+Y^2$ and $\rho$ should obey the condition 
\be 
\lim_{\rho\to\infty}\frac{\rho^2+X^2+Y^2}{\rho\sinh\tau}=\text{finite}.
\ee Since $X^2+Y^2\ge 0$, it follows that $\sinh\tau$ is order $\mathcal{O}(\rho)$. Combining with the fact $\tau>0$ at the future null infinity of Rindler spacetime , we find the conclusion in the context. It is possible that $X^2+Y^2\to\infty$, one can find more details in Appendix \ref{rindlerconformal}.}.  Similarly, the past null infinity of RRW corresponds to $\rho\to+\infty,\ \tau\to-\infty$ with $v=\tau+\log\rho$ finite. In Figure \ref{conf}, we have shown  RRW as a portion of the Minkowski spacetime in conformal diagram \cite{Birrell:1982ix}. The $\mathscr{I}_R^{\pm}$ are the future/past null infinity and $i^{\pm}_R$ are the future/past timelike infinity of RRW. $i^{\pm}_R$ can be approached by taking the limit $\tau\to\pm\infty$ with $\rho$ finite, respectively. 

\begin{figure}\centering
\begin{tikzpicture}[scale=0.6]

 \node (I)    at (4,0)  {\Large R};
 \node (II)   at (-4,0) {\Large L};
 \node (III)  at (0, 4) {\Large F};
 \node (IV)   at (0,-4) {\Large P};
\node (IIItop) at(0,8)  {};
\node (IVbot)  at(0,-8) {};
 \node (FH)   at (1.6,2.6) {{$\mathcal{H}^{++}$}};
 \node (PH)   at (1.6,-2.6) { {$\mathcal{H}^{--}$}};
  \node (FH)   at (-1.6,2.6) {{$\mathcal{H}^{-+}$}};
 \node (PH)   at (-1.6,-2.6) { {$\mathcal{H}^{+-}$}};

 \path
 (II) +(90:4)  coordinate[label=90:]  (IItop)
      +(-90:4) coordinate[label=-90:] (IIbot)
      +(0:4)   coordinate                  (IIright)
      +(180:4) coordinate[label=180:$i^{0}$] (IIleft);

 \draw[fill opacity=0.2,text opacity=1]
 (IIleft) -- node[above left]   {$\mathscr{I}_L^{+}$}
 (IItop) --  node[below, sloped] {}node[at start,above left] {$i_L^+$}
 (IIright)-- node[below, sloped] {}node[at end,below left] {$i_L^-$}
 (IIbot) -- 
             node[below left]    {$\mathscr{I}_L^{-}$}  
 (IIleft)  -- cycle;

 \path
 (I) +(90:4)  coordinate[label=90:]   (Itop)
     +(-90:4) coordinate  (Ibot)
     +(180:4) coordinate                   (Ileft)
     +(0:4)   coordinate[label=0:$i^0$]  (Iright);

 \draw[fill opacity=0.2,text opacity=1]
 (Ileft)node[below=2pt]{$\mathcal{B}$} -- node[above, sloped] {}
 (Itop) --  node[above right]   {$\mathscr{I}_R^+$}node[at start,above right] {$i_R^+$}
 (Iright)--
            node[below right]   {$\mathscr{I}_R^-$}node[at end,below right] {$i_R^-$}
 (Ibot) --  node[below, sloped] {}
 (Ileft) -- cycle;

\draw[fill opacity=0.2,text opacity=1]
(Itop) --node[above right]   {$\mathscr{I}_F^+$}node[at end,above]   {$i^+$} (0,8)--node[above left]   {$\mathscr{I}_F^+$}(IItop);
\draw[fill opacity=0.2,text opacity=1]
(Ibot) --node[below right,]   {$\mathscr{I}_P^-$}node[at end,below]   {$i^-$}(0,-8)--node[below left]   {$\mathscr{I}_P^-$}(IIbot);

 \draw[thick] (Ibot) .. controls (1.5,0) .. (Itop);

\end{tikzpicture}
      \caption{Rindler spacetime as a portion of Minkowski spacetime in conformal diagram. Besides the Rindler horizons, we have also presented the future/past null infinity for each Rindler wedge. For example, the notation $\mathscr{I}_L^+$ is the future null infinity of left Rindler wedge.}\label{conf}
\end{figure}
\subsection{Global symmetries}
In Minkowski spacetime, the ten global transformations that preserve the metric form the Poincar\'e group. These include four spacetime translations 
\be 
\bm\xi_T=\partial_T,\quad \bm\xi_X=\partial_X,\quad\bm\xi_Y=\partial_Y,\quad \bm\xi_Z=\partial_Z,
\ee 
three spatial rotations 
\be 
\bm\xi_{XY}=X\partial_Y-Y\partial_X,\quad \bm\xi_{YZ}=Y\partial_Z-Z\partial_Y,\quad \bm\xi_{XZ}=X\partial_Z-Z\partial_X,
\ee and three Lorentz boosts 
\be 
\bm\xi_{TX}=T\partial_X+X\partial_T,\quad \bm\xi_{TY}=T\partial_Y+Y\partial_T,\quad \bm\xi_{TZ}=T\partial_Z+Z\partial_T.
\ee The ten Killing vectors in Rindler coordinates are collected in Appendix \ref{kv}.
\paragraph{In the presence of $\mathcal{H}^+$.}
To preserve the position of the  hypersurface $\mathcal{H}$, the global Poincar\'e symmetries are broken to a subgroup
\be 
\delta_{\bm\xi}f(T,X,Y,Z)=0,
\ee where $f(T,X,Y,Z)$ is the function that characterizes the hypersurface $\mathcal{H}$ and the Killing vector $\bm\xi$ is a superposition of the ten Killing vectors 
\be 
\bm\xi=a_0\bm\xi_T+a_1\bm\xi_X+a_2\bm\xi_Y+a_3\bm\xi_Z+a_{12}\bm\xi_{XY}+a_{13}\bm\xi_{XZ}+a_{23}\bm\xi_{YZ}+a_{01}\bm\xi_{TX}+a_{02}\bm\xi_{TY}+a_{03}\bm\xi_{TZ}.
\ee  In the presence of $\mathcal{H}^+$, this condition becomes 
\be 
\delta_{\bm\xi}(T-Z)=0\label{hyperHp}
\ee that is solved by 
\bea 
a_0=a_3,\quad a_{01}=a_{13},\quad a_{02}=a_{23}.
\eea It follows that $\bm\xi$ is
\bea 
\bm\xi=a_0(\bm\xi_T+\bm\xi_Z)+a_{01}(\bm\xi_{TX}+\bm\xi_{XZ})+a_{02}(\bm\xi_{TY}+\bm\xi_{YZ})+a_1\bm\xi_X+a_2\bm\xi_Y+a_{12}\bm\xi_{XY}+a_{03}\bm\xi_{TZ}.
\eea Therefore, the subgroup that preserves the condition \eqref{hyperHp} is generated by the following seven Killing vectors
\begin{align}
&\bm\xi_+=\bm\xi_T+\bm\xi_Z,\quad \bm\xi_{+X}=\bm\xi_{TX}+\bm\xi_{XZ},\quad \bm\xi_{+Y}=\bm\xi_{TY}+\bm\xi_{YZ},\quad
\bm\xi_X,\quad \bm\xi_Y,\quad \bm\xi_{XY},\quad \bm\xi_{TZ}.
\end{align}They form a Lie algebra \footnote{The corresponding group is a seven-dimentional connected subgroup of $\text{ISO}(1,3)$ which is denoted as $T^0(3)\Box S(2)$ in \cite{Bacry:1974af}. This is the semidirect product of $S(2)$ and $T^0(3)$. The former is generated by $\bm \xi_{X},\bm \xi_Y,\bm \xi_{XY},\bm \xi_{TZ}$ and the latter is generated by $\bm \xi_-,\bm \xi_{-X},\bm \xi_{-Y}$. }
\bs\label{Liehp}\begin{align} 
&[\bm\xi_+,\bm\xi_{TZ}]=\bm\xi_+,\quad  [\bm\xi_{+A},\bm\xi_B]=-\delta_{AB}\bm\xi_+,\quad [\bm\xi_{+A},\bm\xi_{BC}]=\delta_{AB}\bm\xi_{+C}-\delta_{AC}\bm\xi_{+B},\\
&[\bm\xi_{+A},\bm\xi_{TZ}]=\bm\xi_{+A},\quad  [\bm\xi_A,\bm\xi_{BC}]=\delta_{AB}\bm\xi_C-\delta_{AC}\bm\xi_B,\\
& [\bm\xi_{AB},\bm\xi_{CD}]=\delta_{BC}\bm\xi_{AD}-\delta_{BD}\bm\xi_{AC}-\delta_{AC}\bm\xi_{BD}+\delta_{AD}\bm\xi_{BC},\label{abcd}\\
& [\bm\xi_+,\bm\xi_{A}]=[\bm\xi_+,\bm\xi_{AB}]= [\bm\xi_{+A},\bm\xi_{+B}]=[\bm\xi_{A},\bm\xi_{B}]=[\bm\xi_A,\bm\xi_{TZ}]=[\bm\xi_+,\bm\xi_{+A}]=0.
\end{align}\es 
In above commutators, we have used $\bm\xi_{+A}$ to denote $\bm\xi_{+X}$ or $\bm\xi_{+Y}$, $\bm\xi_{AB}$ to denote $\bm\xi_{XY}$.
The finite transformations in the bulk and the boundary are listed in Table \ref{finHp} and \ref{finHp2}, respectively. Note that the right hand side \eqref{abcd} is zero since there is only one independent rotation generator in the transverse plane. However, in general dimensions, the right hand side is non-zero.
\begin{table}
\begin{center}
\renewcommand\arraystretch{1.5}
    \begin{tabular}{|c||c|}\hline
Killing vectors $\bm\xi$&Finite transformations in the bulk\\\hline\hline
$\bm\xi_+$&$T'=T+\alpha,\quad Z'=Z+\alpha$\\\hline
$\bm\xi_{+X}$&$T'=T+\beta X+\frac{\beta^2}{2}(T-Z),\quad X'=X+\beta(T-Z),\quad Z'=Z+\beta X+\frac{\beta^2}{2}(T-Z)$\\\hline
$\bm\xi_{+Y}$&$T'=T+\delta Y+\frac{\delta^2}{2}(T-Z),\quad Y'=Y+\delta(T-Z),\quad Z'=Z+\delta Y+\frac{\delta^2}{2}(T-Z)$\\\hline 
$\bm\xi_X$&$X'=X+X_0$\\\hline 
$\bm\xi_Y$&$Y'=Y+Y_0$\\\hline
$\bm\xi_{XY}$&$X'=X\cos\varphi+Y\sin\varphi,\quad Y'=-X\sin\varphi+Y\cos\varphi$\\\hline
$\bm\xi_{TZ}$&$T'=T\cosh\gamma+Z\sinh\gamma,\quad Z'=Z\cosh\gamma+T\sinh\gamma$\\\hline
\end{tabular}
\caption{Correspondence between Killing vectors and finite transformations in the bulk. The constants $\alpha,\beta,\gamma,\delta, X_0,Y_0,\varphi$ are seven parameters to represent the corresponding finite transformations. Coordinates that are invariant have been omitted.}\label{finHp}
\end{center}
\end{table}
\begin{table}
\begin{center}
\renewcommand\arraystretch{1.5}
    \begin{tabular}{|c||c|}\hline
Killing vectors $\bm\xi$&Finite transformations on $\mathcal{H}^+$\\\hline\hline
$\bm\xi_+$&$V'=V+2\alpha$\\\hline
$\bm\xi_{+X}$&$V'=V+2\beta X$\\\hline
$\bm\xi_{+Y}$&$V'=V+2\delta Y$\\\hline 
$\bm\xi_X$&$X'=X+X_0$\\\hline 
$\bm\xi_Y$&$Y'=Y+Y_0$\\\hline
$\bm\xi_{XY}$&$X'=X\cos\varphi+Y\sin\varphi,\quad Y'=-X\sin\varphi+Y\cos\varphi$\\\hline
$\bm\xi_{TZ}$&$V'=V e^{\gamma}$\\\hline
\end{tabular}
\caption{Correspondence between Killing vectors and finite transformations on $\mathcal{H}^+$. The seven constants $\alpha,\beta,\gamma,\delta, X_0,Y_0,\varphi$ are exactly the same ones in Table \ref{finHp}. Coordinates that are invariant have been omitted. }\label{finHp2}
\end{center}
\end{table}
\paragraph{In the presence of $\mathcal{H}^-$.}
In this case, the transformations should preserve the function $T+Z=0$
and the general solution of $\bm\xi$ is
\bea 
\bm\xi=a_0(\bm\xi_T-\bm\xi_Z)+a_{01}(\bm\xi_{TX}-\bm\xi_{XZ})+a_{02}(\bm\xi_{TY}-\bm\xi_{YZ})+a_1\bm\xi_X+a_2\bm\xi_Y+a_{12}\bm\xi_{XY}+a_{03}\bm\xi_{TZ}.
\eea Therefore, the subgroup that preserves the position of $\mathcal{H}^-$ is generated by the following seven Killing vectors
\begin{align}
&\bm\xi_-=\bm\xi_T-\bm\xi_Z,\quad \bm\xi_{-X}=\bm\xi_{TX}-\bm\xi_{XZ},\quad \bm\xi_{-Y}=\bm\xi_{TY}-\bm\xi_{YZ},\quad
\bm\xi_X,\quad \bm\xi_Y,\quad \bm\xi_{XY},\quad \bm\xi_{TZ}.
\end{align} They form a Lie algebra that is isomorphic to \eqref{Liehp}
\bs\begin{align}
    &[\bm\xi_-,\bm\xi_{TZ}]=-\bm\xi_-,\quad  [\bm\xi_{-A},\bm\xi_B]=-\delta_{AB}\bm\xi_-,\quad [\bm\xi_{-A},\bm\xi_{BC}]=\delta_{AB}\bm\xi_{-C}-\delta_{AC}\bm\xi_{-B},\\
&[\bm\xi_{-A},\bm\xi_{TZ}]=-\bm\xi_{-A},\quad  [\bm\xi_A,\bm\xi_{BC}]=\delta_{AB}\bm\xi_C-\delta_{AC}\bm\xi_B,\\
& [\bm\xi_{AB},\bm\xi_{CD}]=\delta_{BC}\bm\xi_{AD}-\delta_{BD}\bm\xi_{AC}-\delta_{AC}\bm\xi_{BD}+\delta_{AD}\bm\xi_{BC},\\
& [\bm\xi_-,\bm\xi_{A}]=[\bm\xi_-,\bm\xi_{AB}]= [\bm\xi_{-A},\bm\xi_{-B}]=[\bm\xi_{A},\bm\xi_{B}]=[\bm\xi_A,\bm\xi_{TZ}]=[\bm\xi_-,\bm\xi_{-A}]=0.
\end{align}\es The finite transformations in the bulk and the boundary are given in Table \ref{finHm} and \ref{finHm2}, respectively.
\begin{table}
\begin{center}
\renewcommand\arraystretch{1.5}
    \begin{tabular}{|c||c|}\hline
Killing vectors $\bm\xi$&Finite transformations in the bulk\\\hline\hline
$\bm\xi_-$&$T'=T+\overline{\alpha},\quad Z'=Z-\overline{\alpha}$\\\hline
$\bm\xi_{-X}$&$T'=T+\overline{\beta} X+\frac{\overline{\beta}^2}{2}(T+Z),\quad X'=X+\overline{\beta}(T+Z),\quad Z'=Z-\overline{\beta} X-\frac{\overline{\beta}^2}{2}(T+Z)$\\\hline
$\bm\xi_{-Y}$&$T'=T+\overline{\delta} Y+\frac{\overline{\delta}^2}{2}(T+Z),\quad Y'=Y+\overline{\delta}(T+Z),\quad Z'=Z-\overline{\delta} Y-\frac{\overline{\delta}^2}{2}(T+Z)$\\\hline 
$\bm\xi_X$&$X'=X+X_0$\\\hline 
$\bm\xi_Y$&$Y'=Y+Y_0$\\\hline
$\bm\xi_{XY}$&$X'=X\cos\varphi+Y\sin\varphi,\quad Y'=-X\sin\varphi+Y\cos\varphi$\\\hline
$\bm\xi_{TZ}$&$T'=T\cosh\gamma+Z\sinh\gamma,\quad Z'=Z\cosh\gamma+T\sinh\gamma$\\\hline
\end{tabular}
\caption{Killing vectors that preserve the null hypersurface $\mathcal{H}^-$ and the corresponding finite transformations in the bulk. The constants $\gamma, X_0,Y_0,\varphi$ are the same parameters for $\mathcal{H}^+$ while $\overline{\alpha},\overline{\beta},\overline{\delta}$ are three new parameters since they correspond to three different isometric transformations.  Coordinates that are invariant are not written out. The finite transformations generated by $\bm\xi_X,\bm\xi_Y,\bm\xi_{XY}$ and $\bm\xi_{TZ}$ match with those for $\mathcal{H}^+$.}\label{finHm}
\end{center}
\end{table}
\begin{table}
\begin{center}
\renewcommand\arraystretch{1.5}
    \begin{tabular}{|c||c|}\hline
Killing vectors $\bm\xi$&Finite transformations on $\mathcal{H}^-$\\\hline\hline
$\bm\xi_-$&$U'=U+2\overline{\alpha}$\\\hline
$\bm\xi_{-X}$&$U'=U+2\overline{\beta} X$\\\hline
$\bm\xi_{-Y}$&$U'=U+2\overline{\delta} Y$\\\hline 
$\bm\xi_X$&$X'=X+X_0$\\\hline 
$\bm\xi_Y$&$Y'=Y+Y_0$\\\hline
$\bm\xi_{XY}$&$X'=X\cos\varphi+Y\sin\varphi,\quad Y'=-X\sin\varphi+Y\cos\varphi$\\\hline
$\bm\xi_{TZ}$&$U'=U e^{-\gamma}$\\\hline
\end{tabular}
\caption{Killing vectors that preserve the null hypersurface $\mathcal{H}^-$ and the  corresponding finite transformations on $\mathcal{H}^-$. Constants are exactly the same as the ones in Table \ref{finHm}. Coordinates that are invariant have been omitted. }\label{finHm2}
\end{center}
\end{table}
 
\paragraph{Rindler wedge}
To study the global symmetries that preserve the Rindler wedge, we should impose the condition that leave both of $\mathcal{H}^+$ and $\mathcal{H}^-$ invariant
\be 
\delta_{\bm\xi}(T\pm Z)=0.
\ee The solution $\bm\xi$ is a superposition of the four Killing vectors 
\be 
\bm\xi_X,\quad \bm\xi_Y,\quad\bm\xi_{XY},\quad \bm\xi_{TZ}
\ee that generate the group $\text{SO}(1,1)\times \text{ISO}(2)$ 
\bs\begin{align}
 &   [\bm\xi_A,\bm\xi_{BC}]=\delta_{AB}\bm\xi_C-\delta_{AC}\bm\xi_B,\quad [\bm\xi_A,\bm\xi_B]=[\bm\xi_{TZ},\bm\xi_A]=[\bm\xi_{TZ},\bm\xi_{AB}]=0,\\
& [\bm\xi_{AB},\bm\xi_{CD}]=\delta_{BC}\bm\xi_{AD}-\delta_{BD}\bm\xi_{AC}-\delta_{AC}\bm\xi_{BD}+\delta_{AD}\bm\xi_{BC}.
\end{align}\es 
The finite transformations could be reduced either to $\mathcal{H}^{++}$ or $\mathcal{H}^{--}$, which are shown in Table \ref{iso}. Note that the subgroup $\text{SO}(1,1)$ is the time translation along the Rindler time, which is also the Lorentz boost along $Z$ direction in Minkowski spacetime. The subgroup $\text{ISO}(2)$ is the Euclidean group of the transverse plane. In Minkowski spacetime, the Poincar\'e transformation could preserve the locations of the future and past null infinity. Therefore, there is no symmetry breaking in Minkowski vacuum. However, to preserve the positions of the Rindler horizons, the Poincar\'e group is broken and the corresponding Rindler vacuum is only invariant under the residual subgroup.
\begin{table}
\begin{center}
\renewcommand\arraystretch{1.5}
    \begin{tabular}{|c||c|c|}\hline
Killing vectors $\bm\xi$&Finite transformations on $\mathcal{H}^{++}$&Finite transformations on $\mathcal{H}^{--}$\\\hline\hline 
$\bm\xi_X$&$x'=x+X_0$&$x'=x+X_0$\\\hline 
$\bm\xi_Y$&$y'=y+Y_0$&$y'=y+Y_0$\\\hline
$\bm\xi_{XY}$&$\bm x'=R\bm x$&$\bm x'=R\bm x$\\\hline
$\bm\xi_{TZ}$&$v'=v+\gamma$&$u'=u+\gamma$\\\hline
\end{tabular}
\caption{Killing vectors that preserve the Rindler wedge and the corresponding finite transformations on $\mathcal{H}^{\pm\pm}$. Constants are exactly the same ones in previous tables. Coordinates that are invariant have been omitted. The orthogonal matrix $R$ is the rotation matrix in $x$-$y$ plane
\be 
R=\left(\begin{array}{cc}\cos\varphi&\sin\varphi\\-\sin\varphi&\cos\varphi\end{array}\right).
\ee  }\label{iso}
\end{center}
\end{table}
\section{Scalar field}\label{Scalarfield}
In this section, we will discuss various properties of the scalar field on the null  boundary of the RRW. To  extract the fundamental field, we may impose the fall-off condition near $\mathcal{H}^{++}$ and $\mathcal{H}^{--}$
 \be 
\Phi(x)=\left\{\begin{array}{cc}\Sigma(u,\bm x_{})+\mathcal{O}(\rho),&\text{near}\ \mathcal{H}^{--},\\ \Xi(v,\bm x_{})+\mathcal{O}(\rho),&\text{near}\ \mathcal{H}^{++}.\end{array}\right.
\ee  We introduce a symbol $\sigma$ to distinguish the field on $\mathcal{H}^{++}$ and $\mathcal{H}^{--}$
 \bea 
\sigma=\left\{\begin{array}{cc}-,&\text{on}\ \mathcal{H}^{--},\\ +,&\text{on}\ \mathcal{H}^{++}.\end{array}\right.
\eea The field on $\mathcal{H}^{--}$ is denoted as 
\be 
\Sigma(u,\bm x,-)=\Sigma(u,\bm x)
\ee and the field on $\mathcal{H}^{++}$ is denoted as
\bea 
\Sigma(u,\bm x_{},+)=\Xi(v\to u,\bm x).
\eea Note that we have introduced an  ``antipodal map'' in Rindler wedge, generalizing the one in Minkowski spacetime. Under this map, we may treat the field at $\mathcal{H}^{++}$ and $\mathcal{H}^{--}$ equally. The incoming and outgoing states are distinguished by the symbol $\sigma$. Then the fall-off condition becomes
\be 
\Phi(x)=\left\{\begin{array}{cc}\Sigma(u,\bm x_{},-)+\mathcal{O}(\rho)&\text{near}\ \mathcal{H}^{--},\\ \Sigma(u,\bm x_{},+)+\mathcal{O}(\rho)&\text{near}\ \mathcal{H}^{++}.\end{array}\right.
\ee
A scalar field in RRW obeys the transformation law 
\be 
\Phi'(x')=\Phi(x),\quad x\to x'.
\ee For any transformation in $\text{SO}(1,1)\times \text{ISO}(2)$, the field $\Sigma(u,\bm x_{},\sigma)$ transforms as follows
\be 
\Sigma'(u',\bm x_{}',\sigma)=\Sigma(u,\bm x_{},\sigma).
\ee 
\subsection{Bulk and boundary fields}
The mode expansion of the bulk field in RRW is
\bea 
\Phi(x)=\int_{0}^{\infty}d\omega \int_{-\infty}^{+\infty}d\bm k\ \chi_{\omega,\bm k}(\rho)(c_{\omega,\bm k} e^{-i\omega\tau+i\bm k\cdot\bm x_{}}+c^{\dagger}_{\omega,\bm k} e^{i\omega\tau-i\bm k\cdot\bm x_{}}),\label{modePhi}
\eea 
where the function $\chi_{\omega,\bm k}$ is obtained by solving the Klein-Gordon equation in Rindler coordinates with sufficient fall-off condition at $\rho\to\infty$ 
\bea 
\chi_{\omega,\bm k}(\rho)=\sqrt{\frac{4\sinh\pi \omega }{(2\pi)^4}}K_{i\omega}(\bar{k}\rho),\quad \bar{k}=\sqrt{\bm k^2+m^2},\label{solutionK}
\eea where the function $K_{i\omega}(\bar{k}\rho)$ is the Modified Bessel function of the second kind and $m$ is the mass of the scalar field. The annihilation and creation operators $c_{\omega,\bm k},c^\dagger_{\omega,\bm k}$ satisfy the commutation relations 
\bea 
[c_{\omega,\bm k},c^\dagger_{\omega',\bm k'}]=\delta(\omega-\omega')\delta^{(2)}(\bm k-\bm k'),\quad [c_{\omega,\bm k},c_{\omega',\bm k'}]=[c^\dagger_{\omega,\bm k},c^\dagger_{\omega',\bm k'}]=0.\label{com}
\eea Rindler vacuum $|0\rangle_{\text{R}}$ is annihilated by the operators $c_{\omega,\bm k}$
\be 
c_{\omega,\bm k}|0\rangle_{\text{R}}=0.
\ee It is well known that the Rindler vacuum is not equivalent to the Minkowski vacuum $|0\rangle_{\text{M}}$. To simplify notation, we will omit the subscript $\text{R}$ and the Rindler vacuum is written as $|0\rangle$. The Modified Bessel function $K_{i\omega}(\bar k \rho)$ has the asymptotic behaviour near $\rho\to 0$ 
\be 
K_{i\omega}(\bar k\rho)\sim 2^{-1-i\omega}\Gamma(-i\omega)(\bar k\rho)^{i\omega}+2^{-1+i\omega}\Gamma(i\omega)(\bar k\rho)^{-i\omega},
\ee from which we can read out the boundary fields 
\bs\label{modeSigma}\begin{align}
    \Sigma(u,\bm x_{},-)&=\int_0^\infty \frac{d\omega}{\sqrt{4\pi\omega}}{ \int_{-\infty}^{\infty}}\frac{d\bm k}{\sqrt{(2\pi)^{2}}}[a_{\omega,\bm k,-} e^{-i\omega u+i\bm k\cdot\bm x_{}}+a_{\omega,\bm k,-}^{\dagger} e^{i\omega u-i\bm k\cdot\bm x_{}}\label{expsig1}],\\
    \Sigma(u,\bm x_{},+)&=\int_0^\infty \frac{d\omega}{\sqrt{4\pi\omega}}{ \int_{-\infty}^{\infty}}\frac{d\bm k_{}}{\sqrt{(2\pi)^{2}}}[{a}_{\omega,\bm k,+} e^{-i\omega u+i\bm k\cdot\bm x_{}}+{a}_{\omega,\bm k,+}^{\dagger} e^{i\omega u-i\bm k\cdot\bm x_{}}],
\end{align}\es where 
\bs\label{amap}\begin{align}
   a_{\omega,\bm k,-}&=\sqrt{\frac{\omega\sinh\pi\omega}{\pi}}(\frac{\bar{k}_{}}{2})^{i\omega}\Gamma(-i\omega)c_{\omega,\bm k},\\
   a_{\omega,\bm k,-}^{\dagger}&=\sqrt{\frac{\omega\sinh\pi\omega}{\pi}}(\frac{\bar{k}}{2})^{-i\omega}\Gamma(i\omega)c^{\dagger}_{\omega,\bm k},\\{a}_{\omega,\bm k,+}&=\sqrt{\frac{\omega\sinh\pi\omega}{\pi}}(\frac{\bar{k}}{2})^{-i\omega}\Gamma(i\omega)c_{\omega,\bm k},\\
 {a}_{\omega,\bm k,+}^{\dagger}&=\sqrt{\frac{\omega\sinh\pi\omega}{\pi}}(\frac{\bar{k}}{2})^{i\omega}\Gamma(-i\omega)c^{\dagger}_{\omega,\bm k}.
\end{align}
\es
The creation and annihilation operators satisfy the identities 
\be 
a_{\omega,\bm k,-}=\left(\frac{\bar{k}}{2}\right)^{2i\omega}\Gamma(-i\omega)\Gamma(i\omega)^{-1}{a}_{\omega,\bm k,+},\quad a_{\omega,\bm k,-}^\dagger=\left(\frac{\bar k}{2}\right)^{-2i\omega}\Gamma(i\omega)\Gamma(-i\omega)^{-1}{a}_{\omega,\bm k,+}^\dagger.
\ee It is clear that the Rindler vacuum is  also annihilated by the operators $a_{\omega,\bm k,\sigma}$
\be 
a_{\omega,\bm k,-}|0\rangle={a}_{\omega,\bm k,+}|0\rangle=0.
\ee We can reverse \eqref{modeSigma} to obtain 
\bs\begin{align}
    a_{\omega,\bm k,\sigma}&=\sqrt{\frac{2\omega}{(2\pi)^3}}\int_{-\infty}^\infty du \int_{-\infty}^\infty d\bm x \Sigma(u,\bm x,\sigma)e^{i\omega u-i\bm k\cdot\bm x_{}},\\
    a_{\omega,\bm k,\sigma}^\dagger&=\sqrt{\frac{2\omega}{(2\pi)^3}}\int_{-\infty}^\infty du \int_{-\infty}^\infty d\bm x \Sigma(u,\bm x,\sigma)e^{-i\omega u+i\bm k\cdot\bm x}.
\end{align}\es Under the Lorentz boost along $Z$ direction, we have 
\be 
\Sigma'(u+\gamma,\bm x,\sigma)=\Sigma(u,\bm x,\sigma).
\ee Therefore, the annihilation and creation operators transform as 
\be 
a'_{\omega,\bm k,\sigma}=e^{i\omega\gamma}a_{\omega,\bm k,\sigma},\quad
a'^\dagger_{\omega,\bm k,\sigma}=e^{-i\omega\gamma}a^\dagger_{\omega,\bm k,\sigma}.
\ee Similarly, one can obtain the transformation law of the annihilation and creation operators associated with other residual global symmetries. In Table \ref{ann}, we summarize these transformation laws.  It is clear that the Rindler vacuum is invariant under the symmetry group $\text{SO}(1,1)\times \text{ISO}(2)$ since it doesn't  mix the positive and negative frequency modes.
\begin{table}
\begin{center}
\renewcommand\arraystretch{1.5}
    \begin{tabular}{|c||c|c|}\hline
Killing vectors $\bm\xi$& Annihilation operators $a_{\omega,\bm k,\sigma}$&Creation operators $a_{\omega,\bm k,\sigma}^\dagger$\\\hline\hline 
$\bm\xi_X$&$a'_{\omega,\bm k,\sigma}=e^{-ik_x X_0}a_{\omega,\bm k,\sigma}$&$a'^\dagger_{\omega,\bm k,\sigma}=e^{ik_x X_0}a^\dagger_{\omega,\bm k,\sigma}$\\\hline 
$\bm\xi_Y$&$a'_{\omega,\bm k,\sigma}=e^{-ik_y Y_0}a_{\omega,\bm k,\sigma}$&$a'^\dagger_{\omega,\bm k,\sigma}=e^{ik_y Y_0}a^\dagger_{\omega,\bm k,\sigma}$\\\hline
$\bm\xi_{XY}$&$a'_{\omega,\bm k,\sigma}=a_{\omega,R^T\bm k,\sigma}$&$a'^\dagger_{\omega,\bm k,\sigma}=a^\dagger_{\omega,R^{T}\bm k,\sigma}$\\\hline
$\bm\xi_{TZ}$&$a'_{\omega,\bm k,\sigma}=e^{i\omega\gamma}a_{\omega,\bm k,\sigma}$&$a'^\dagger_{\omega,\bm k,\sigma}=e^{-i\omega\gamma}a^\dagger_{\omega,\bm k,\sigma}$\\\hline
\end{tabular}
\caption{Transformation law of annihilation and creation operators under residual global symmetry transformations. The symbol $R^T$ denotes the transpose of the rotation matrix $R$.}\label{ann}
\end{center}
\end{table} Regarding to the transformations generated by $\bm\xi_+,\ \bm\xi_{+X}$ and $\bm \xi_{+Y}$, they could leave the null hypersurface $\mathcal{H}^{+}$ invariant. However, since $\mathcal{H}^-$ is not preserved under these transformations, the Rindler vacuum is not invariant under these transformations. 

 From the conformal diagram of Figure \ref{conf}, there are four null hypersurfaces $\mathcal{H}^{--}\cup \mathcal{H}^{++}\cup \mathscr{I}^+_R\cup \mathscr{I}^-_R$ of  RRW. We have already discussed the boundary field on $\mathcal{H}^{--}$ and $\mathcal{H}^{++}$, it would be better to consider the field on $\mathscr{I}^{\pm}_R$. The Modified Bessel function $K_{i\omega}(\bar k \rho)$ decays exponentially near $\rho \to \infty$  
\bea
K_{i\omega}(\bar k \rho) \sim e^{-\bar k \rho}\sqrt{\frac{\pi}{2\bar k \rho}}.
\eea The fall-off behaviour is similar to a massive field in Minkowski spacetime. As a consequence, we may conclude that the boundary field is fixed to be zero at $\mathscr{I}_R^{\pm}$ since the bulk field decays exponentially, akin to the massive field at the null boundaries of Minkowski spacetime. Note that for $\bar k=0$, the field does not decay exponentially. The solution \eqref{solutionK} is not valid and one should discuss it separately. From $\bar k=0$, we can solve 
\be 
\bm k=0,\quad m=0
\ee and then the massless Klein-Gordon equation with zero transverse momentum becomes 
\be 
\rho^2\partial^2_\rho\Phi+\rho\partial_\rho\Phi-\partial_\tau^2\Phi=0,
\ee which is equivalent to a two dimensional massless Klein-Gordon equation. One can find the left moving and right moving modes and the general solution is 
\be 
\widetilde{\Phi}(u,v)=\int_0^\infty d\omega\left( c_\omega e^{-i\omega u}+c^\dagger_{\omega}e^{i\omega u}+\tilde{c}_\omega e^{-i\omega v}+\tilde{c}^\dagger_{\omega}e^{i\omega v}\right).\label{tildephi}
\ee We use $\widetilde{\Phi}$ to distinguish it from the solution \eqref{solutionK}. For this exceptional case, we may impose the  fall-off condition near $\mathscr{I}_R^{\pm}$ as 
\be 
\widetilde{\Phi}(\tau,\rho)=\left\{\begin{array}{cc}\widetilde{\Sigma}(u)+\cdots,& \text{at}\ \mathscr{I}_R^+,\\
\widetilde{\Xi}(v)+\cdots,& \text{at}\ \mathscr{I}_R^-.\end{array}\right.\label{falloffscr}
\ee We will show later that this is an independent branch and will be discarded in the following discussion.

\subsection{Incoming and outgoing states}
We consider a scalar theory with the action \be 
S[\Phi]=\int d^4 x [-\frac{1}{2}\partial_\mu\Phi \partial^\mu\Phi-m^2\Phi^2-\frac{\lambda_3}{6}\Phi^3-\frac{\lambda_4}{24}\Phi^4].
\ee Switching to the Rindler coordinates, the action becomes 
\bea 
S[\Phi]= \int_{-\infty}^{\infty} d \tau { \int_0^\infty}\rho d\rho \int_{-\infty}^{\infty} d\bm x [\frac{1}{2\rho^2}(\partial_\tau\Phi)^2-\frac{1}{2}(\partial_\rho\Phi)^2-\frac{1}{2}\partial_A\Phi \partial^A\Phi-m^2\Phi^2-\frac{\lambda_3}{6}\Phi^3-\frac{\lambda_4}{24}\Phi^4].
\eea We can regard  $\tau$ as the time of the Rindler wedge and the conjugate momentum of $\Phi$ is 
\be 
\Pi=\rho^{-1}\partial_\tau\Phi.
\ee As a consequence,  the Hamiltonian of the system is 
\bea 
H= \int_{-\infty}^{\infty}d\tau \int_0^\infty  \rho{ d \rho}  \int_{-\infty}^{\infty} d\bm x [\frac{1}{2\rho^2}(\partial_\tau\Phi)^2+\frac{1}{2}(\partial_\rho\Phi)^2+\frac{1}{2}\partial_A\Phi \partial^A\Phi+m^2\Phi^2+\frac{\lambda_3}{6}\Phi^3+\frac{\lambda_4}{24}\Phi^4].
\eea Notice that the 
 Rindler time is 
 \be 
 \tau=\left\{\begin{array}{cc}+\infty, &\text{at}\  \mathcal{H}^{++},\\ -\infty,&\text{at}\ \mathcal{H}^{--}.\end{array}\right.
 \ee 
 Since the evolution of the state is along the Rindler time direction, 
 we may define an incoming state 
 \bea 
 |\Sigma(u,\bm x_{},-)\rangle=\Sigma(u,\bm x_{},-)|0\rangle=\int_0^\infty \frac{d\omega}{\sqrt{4\pi\omega}}{ \int_{-\infty}^{\infty}}\frac{d\bm k}{\sqrt{(2\pi)^2}}a_{\omega,\bm k,-}^\dagger e^{i\omega u-i\bm k\cdot\bm x_{}} |0\rangle,\label{sigmamode}
 \eea whose Hermite conjugate is 
 \bea 
 \langle \Sigma(u,\bm x_{},-)|=\int_0^\infty \frac{d\omega}{\sqrt{4\pi\omega}}{\int_{-\infty}^{\infty}}\frac{d\bm k}{\sqrt{(2\pi)^2}} e^{-i\omega u+i\bm k\cdot\bm x_{}}\langle 0|a_{\omega,\bm k,-}.
 \eea The state $a_{\omega,\bm k,-}^\dagger |0\rangle$ is an incoming state with definite frequency $\omega$ and transverse momentum $\bm k$
 \be 
 |\omega,\bm k\rangle=c_{\omega,\bm k}^\dagger|0\rangle=\sqrt{\frac{\omega\sinh\pi\omega}{\pi}}(\frac{\bar{k}_{}}{2})^{i\omega}\Gamma(-i\omega)a^\dagger_{\omega,\bm k,-}|0\rangle.
 \ee
 
 Substituting it into \eqref{sigmamode}, we find that the state with definite position can be written as a superposition of the states with definite frequency and transverse momentum
 \bea 
 |\Sigma(u,\bm x_{},-)\rangle=\int_0^\infty \frac{d\omega}{\sqrt{4\pi\omega}}{\int_{-\infty}^{\infty}}\frac{d\bm k}{\sqrt{(2\pi)^2}} \sqrt{\frac{\omega\sinh\pi\omega}{\pi}}(\frac{\bar{k}}{2})^{-i\omega}\Gamma(i\omega) e^{i\omega u-i\bm k\cdot\bm x_{}}|\omega,\bm k\rangle.
 \eea The Hermite conjugate of the above state is \bea 
 \langle \Sigma(u,\bm x_{},-)|=\int_0^\infty \frac{d\omega}{\sqrt{4\pi\omega}}{\int_{-\infty}^{\infty}}\frac{d\bm k}{\sqrt{(2\pi)^2}} \sqrt{\frac{\omega\sinh\pi\omega}{\pi}}(\frac{\bar{k}}{2})^{i\omega}\Gamma(-i\omega) e^{-i\omega u+i\bm k\cdot\bm x_{}}\langle \omega,\bm k|.
 \eea Similarly, we can also define the asymptotic outgoing state and its Hermite conjugate
 \bs\begin{align}
 |\Sigma(u,\bm x_{},+)\rangle&=\int_0^\infty \frac{d\omega}{\sqrt{4\pi\omega}}{\int_{-\infty}^{\infty}}\frac{d\bm k}{\sqrt{(2\pi)^2}} \sqrt{\frac{\omega\sinh\pi\omega}{\pi}}(\frac{\bar{k}}{2})^{i\omega}\Gamma(-i\omega) e^{i\omega u-i\bm k\cdot\bm x_{}}| \omega,\bm k\rangle,\\
 \langle\Sigma(u,\bm x_{},+)|&=\int_0^\infty \frac{d\omega}{\sqrt{4\pi\omega}}{\int_{-\infty}^{\infty}}\frac{d\bm k}{\sqrt{(2\pi)^2}} \sqrt{\frac{\omega\sinh\pi\omega}{\pi}}(\frac{\bar{k}}{2})^{-i\omega}\Gamma(i\omega) e^{-i\omega u+i\bm k\cdot\bm x_{}}\langle  \omega,\bm k|.
 \end{align}\es
 \paragraph{Orthogonality relation.}
 The orthogonality relation of the states $|\omega,\bm k\rangle$ is 
 \be 
 \langle \omega,\bm k|\omega',\bm k'\rangle=\delta(\omega-\omega')\delta^{(2)}(\bm k-\bm k')
 \ee which comes from the commutation relations \eqref{com} and the definition of the Rindler vacuum. This relation can be transformed to the orthogonality relation of the asymptotic incoming/outgoing states 
\begin{align}
     \langle \Sigma(u,\bm x_{},\sigma)|\Sigma(u',\bm x'_{},\sigma)\rangle&=\frac{1}{4\pi}\int_0^\infty \frac{d\omega}{\omega}e^{-i\omega (u-u')}\delta^{(2)}(\bm x_{}-\bm x'_{})
 \end{align} which has been found in \cite{Liu:2022mne,Li:2023xrr} and it could be regularized by introducing an IR cutoff $\omega_0$ and utilizing $i\epsilon$ prescription
 \be 
  \langle \Sigma(u,\bm x_{},\sigma)|\Sigma(u',\bm x'_{},\sigma)\rangle=-\frac{1}{4\pi}I_0(\omega_0(u-u'-i\epsilon))\label{orthoS}\delta^{(2)}(\bm x_{}-\bm x'_{})
 \ee where 
 \be 
 I_0(\omega(u-i\epsilon))=\gamma_E+\log i\omega_0(u-i\epsilon)
 \ee with $\gamma_E$ the Euler constant.
 \paragraph{Completeness relation.} The completeness relation of the one-particle state $|\omega,\bm k\rangle$ is 
 \be 
 1=\int_0^\infty d\omega \int_{-\infty}^\infty d\bm k |\omega,\bm k\rangle\langle \omega,\bm k|
 \ee which could be transformed to the completeness relation of the incoming/outgoing states\begin{align}
     1&=2i\int du d\bm x_{} |\Sigma(u,\bm x_{},\sigma)\rangle\langle \dot{\Sigma}(u,\bm x_{},\sigma)|=-2i\int du d\bm x_{} |\dot\Sigma(u,\bm x_{},\sigma)\rangle\langle {\Sigma}(u,\bm x_{},\sigma)|\nn\\&=i\int du d\bm x_{}\left( |\Sigma(u,\bm x_{},\sigma\rangle\langle\dot\Sigma(u,\bm x_{},\sigma|-|\dot\Sigma(u,\bm x_{},\sigma\rangle\langle\Sigma(u,\bm x_{},\sigma|\right),
 \end{align} where $\sigma$ is either $+$ or $-$. Note that there is no summation on $\sigma$ since the incoming or outgoing states are already complete. Interestingly, the completeness relation is exactly the same as the one at future/past null infinity of Minkowski spacetime \cite{Liu:2024nfc}. 
One can also extend the completeness relation to multi-particle states
\bea 
1=\prod_{j} i\int du_j d\bm x_{ j}\left( |\Sigma(u_j,\bm x_{ j},\sigma\rangle\langle\dot\Sigma(u_j,\bm x_{ j},\sigma|-|\dot\Sigma(u_j,\bm x_{j},\sigma\rangle\langle\Sigma(u_j,\bm x_{ j},\sigma|\right).
\eea 
\subsection{Carrollian amplitude}
Given the asymptotic incoming and outgoing states in RRW, we can calculate the boundary correlator
\bea 
\langle\prod_{j=1}^n \Sigma(u_j,\bm  x_{j},\sigma_j)\rangle\label{cor}
\eea through the bulk scattering amplitude.
More explicitly, we consider a scattering process with $p$ incoming and $q$ outgoing particles 
\be 
n=p+q,\quad p,q\ge 0.
\ee Then the correlator \eqref{cor} becomes
\bea 
&&{}_{\text{out}}\langle \prod_{j=p+1}^n \Sigma(u_j,\bm x_{ j},+)|\prod_{j=1}^p\Sigma(u_j,\bm x_{ j},-)\rangle_{\text{in}} \nn\\&=&\prod_{j=1}^p \int_0^\infty \frac{d\omega_j}{\sqrt{4\pi\omega_j}}{\int_{-\infty}^{\infty}}\frac{d\bm k_j}{\sqrt{(2\pi)^2}}\sqrt{\frac{\omega_j\sinh\pi\omega_j}{\pi}} (\frac{{ \bar k_{j}}}{2})^{-i\omega_j}\Gamma(i\omega_j)e^{i\omega_j u_j-i\bm k_j\cdot\bm x_{ j}}\nn\\&&\times \prod_{j=p+1}^n \int_0^\infty \frac{d\omega_j}{\sqrt{4\pi\omega_j}}{\int_{-\infty}^{\infty}}\frac{d\bm k_j}{\sqrt{(2\pi)^2}}\sqrt{\frac{\omega_j\sinh\pi\omega_j}{\pi}}  (\frac{{\bar k_{j}}}{2})^{-i\omega_j}\Gamma(i\omega_j)e^{-i\omega_j u_j+i\bm k_j\cdot\bm x_{ j}}\mathcal{A}(1,2,\cdots,n).\nn\\\eea 
The expression in the last line is the scattering amplitude in momentum space which is defined  as 
\bea 
\mathcal{A}(1,2,\cdots,n)&=&{}_{\text{out}}\langle \omega_{p+1},\bm k_{p+1};\cdots;\omega_{n},\bm k_n|\omega_1,\bm k_1;\cdots;\omega_p,\bm k_p\rangle_{\text{in}}\nn\\&=&\langle \omega_{p+1},\bm k_{p+1};\cdots;\omega_{n},\bm k_n|\text{S}|\omega_1,\bm k_1;\cdots;\omega_p,\bm k_p\rangle
\eea where S is the scattering matrix.
Therefore, the $n$-point correlator is an integral transform of the scattering amplitude in momentum space
\bea \langle\prod_{j=1}^n \Sigma(u_j,\bm x_{j},\sigma_j)\rangle&=&\prod_{j=1}^n \int_0^\infty \frac{d\omega_j}{\sqrt{4\pi\omega_j}} { \int_{-\infty}^{\infty}}\frac{d\bm k_j}{\sqrt{(2\pi)^2}}\sqrt{\frac{\omega_j\sinh\pi\omega_j}{\pi}}(\frac{\bar k_j}{2})^{-i\omega_j}\Gamma(i\omega_j)e^{-i\sigma_j\omega_j u_j+i\sigma_j\bm k_j\cdot\bm x_{ j}}\nn\\&&\times \mathcal{A}(1,2,\cdots,n). \label{sansier}
\eea 
Note that the integral transform is different from the Fourier transform in four-dimensional Minkowski spacetime \cite{Donnay:2022wvx} and modified Fourier transform in higher dimensional Minkowski spacetime \cite{Liu:2024llk}. Similar to the scattering amplitude in Minkowski spacetime, one can divide the S-matrix into
\be 
\text{S}=1+i\text{T},
\ee where the T-matrix extracts the information of nontrivial interactions. Moreover, since the RRW is invariant under $\text{SO}(1,1)\times \text{ISO}(2)$, the frequency and the transverse momentum should be conserved \bea 
\sum_{j=1}^n \sigma_j\omega_j=0,\quad \sum_{j=1}^n \sigma_j\bm k_j=0.
\eea Therefore, one can always separate a $\mathcal{M}$ matrix from the T-matrix 
\be 
\langle \omega_{p+1},\bm k_{p+1};\cdots;\omega_{n},\bm k_n|i\text{T}|\omega_1,\bm k_1;\cdots;\omega_p,\bm k_p\rangle=\delta(\sum_{j=1}^n \sigma_j\omega_j)\delta(\sum_{j=1}^n \sigma_j\bm k_j)i\mathcal{M}(\omega_1,\bm k_1,\sigma_1;\cdots;\omega_n,\bm k_n,\sigma_n).
\ee 
We may throw out the identity and just write the correlator as 
\bea 
\langle\prod_{j=1}^n \Sigma(u_j,\bm x_{ j},\sigma_j)\rangle&=&\prod_{j=1}^n \int_0^\infty \frac{d\omega_j}{\sqrt{4\pi\omega_j}}{ \int_{-\infty}^{\infty}} \frac{d\bm k_j}{\sqrt{(2\pi)^2}}\sqrt{\frac{\omega_j\sinh\pi\omega_j}{\pi}}(\frac{\bar k_j}{2})^{-i\omega_j}\Gamma(i\omega_j)e^{-i\sigma_j\omega_j u_j+i\sigma_j\bm k_j\cdot\bm x_{ j}}\nn\\&&\times \delta(\sum_{j=1}^n \sigma_j\omega_j)\delta(\sum_{j=1}^n \sigma_j\bm k_j)i\mathcal{M}(1,2,\cdots,n),\nn\\
\eea where $\mathcal{M}(1,2,\cdots,n)$ is 
\bea 
\mathcal{M}(1,2,\cdots,n)=\mathcal{M}(\omega_1,\bm k_1,\sigma_1;\cdots;\omega_n,\bm k_n,\sigma_n).
\eea 
Note that the delta function and the  $\mathcal{M}$ matrix are invariant separately under $\text{SO}(1,1)\times \text{ISO}(2)$.

\paragraph{Ward identities}
Now we prove the transformation law of the Carrollian amplitude
\bea 
\langle\prod_{j=1}^n \Sigma(u'_j,\bm x'_{ j},\sigma_j)\rangle=\langle\prod_{j=1}^n \Sigma(u_j,\bm x_{ j},\sigma_j)\rangle\label{trans}
\eea  under general $\text{SO}(1,1)\times \text{ISO}(2)$ transformation. At first, we will consider the Lorentz transformation generated by $\bm\xi_{TZ}$, we have 
\be 
u'=u+\gamma,\quad \omega'=\omega,\quad \bm k'=\bm k,\quad \mathcal{M}(1',2',\cdots,n')=\mathcal{M}(1,2,\cdots,n).
\ee Then 
\bea 
&&\langle \prod_{j=1}^n \Sigma(u'_j,\bm x'_{j},\sigma_j)\rangle\nn\\&=&\prod_{j=1}^n {\int_{0}^{\infty}d\omega_j \int_{-\infty}^{+\infty}d\bm k_j} f(\omega_j,k_j) e^{-i\sigma_j \omega_j (u_j+\gamma)+i\sigma_j\bm k_j\cdot\bm x_{j}}\delta(\sum_{j=1}^n \sigma_j\omega_j)\delta(\sum_{j=1}^n \sigma_j\bm k_j)i\mathcal{M}(1',2',\cdots,n')\nn\\&=&\prod_{j=1}^n {\int_{0}^{\infty}d\omega_j \int_{-\infty}^{+\infty}d\bm k_j} f(\omega_j,k_j) e^{-i\sigma_j \omega_j u_j+i\sigma_j\bm k_j\cdot\bm x_{ j}}\delta(\sum_{j=1}^n \sigma_j\omega_j)\delta(\sum_{j=1}^n \sigma_j\bm k_j)i\mathcal{M}(1,2,\cdots,n)\nn\\&=&\langle \prod_{j=1}^n \Sigma(u_j,\bm x_{ j},\sigma_j)\rangle.
\eea In the second line, we have  defined the  function 
\be 
f(\omega,k)=\frac{1}{\sqrt{4\pi\omega}}\frac{1}{\sqrt{(2\pi)^2}}\sqrt{\frac{\omega\sinh\pi\omega}{\pi}}(\frac{\bar k}{2})^{-i\omega}\Gamma(i\omega)
\ee to simplify notation. Note that in the third line we have used the conservation of the frequency. In the same way, we can prove the translation invariance (transverse directions) of the Carrollian amplitude. For the spatial rotation 
\bea 
u'=u,\quad \bm x' =R\bm x,
\eea the frequency is invariant while the transverse momentum $\bm k$ is rotated 
\bea 
\omega'=\omega,\quad \bm k'=R^{T}\bm k.
\eea Therefore, 
\bea 
&&\langle \prod_{j=1}^n \Sigma(u'_j,\bm x'_{j},\sigma_j)\rangle\nn\\&=&\prod_{j=1}^n {\int_{0}^{\infty}d\omega'_j \int_{-\infty}^{+\infty}d\bm k'_j}f(\omega'_j,k'_j) e^{-i\sigma_j \omega'_j u'_j+i\sigma_j\bm k'_j\cdot\bm x'_{ j}}\delta(\sum_{j=1}^n \sigma_j\omega'_j)\delta(\sum_{j=1}^n \sigma_j\bm k'_j)i\mathcal{M}(1',2',\cdots,n')\nn\\&=&\prod_{j=1}^n {\int_{0}^{\infty}d\omega_j \int_{-\infty}^{+\infty}d\bm k_j} f(\omega_j,k_j)e^{-i\sigma_j\omega_ju_j+i\sigma_j\bm k_j\cdot\bm x_{ j}}\delta(\sum_{j=1}^n\sigma_j\omega_j)\delta(\sum_{j=1}^n\sigma_j\bm k_j)i\mathcal{M}(1,2,\cdots,n)\nn\\&=&\langle \prod_{j=1}^n \Sigma(u_j,\bm x_{ j},\sigma_j)\rangle.
\eea In the third line, we have used the fact that the $R$ is an orthogonal matrix. Therefore, the Jacobian matrix from $\bm k$ to $\bm k'$ is 1. Moreover, the integral measure and the Dirac delta function are invariant under this rotation. The function $f(\omega,k)$ only depends on the length of the transverse momentum and the frequency, both of them are invariant under rotation in the transverse plane. To obtain the Ward identities, we can expand the transformation law \eqref{trans} up to the first infinitesimal order 
\bs\label{Ward}\begin{align}
    \left(\sum_{j=1}^n \frac{\partial}{\partial u_j}\right)\langle \prod_{j=1}^n \Sigma(u_j,\bm x_{ j},\sigma_j)\rangle&=0,\label{timetrans}\\
    \left(\sum_{j=1}^n \frac{\partial}{\partial x_j}\right)\langle \prod_{j=1}^n \Sigma(u_j,\bm x_{ j},\sigma_j)\rangle&=0,\label{strans}\\
     \left(\sum_{j=1}^n \frac{\partial}{\partial y_j}\right)\langle \prod_{j=1}^n \Sigma(u_j,\bm x_{ j},\sigma_j)\rangle&=0,\label{ytrans}\\
     \left(\sum_{j=1}^n x_j\frac{\partial}{\partial y_j}-y_j\frac{\partial}{\partial x_j}\right)\langle \prod_{j=1}^n \Sigma(u_j,\bm x_{ j},\sigma_j)\rangle&=0.\label{xyrot}
\end{align}\es 
\paragraph{Fourier space}
The operator $\Sigma(u,\bm x,\sigma)$ is defined in the Carrollian space, we may transform it to the Fourier space 
\bs\label{Sf}\begin{align}
|\Sigma(\omega,\bm x_{},\sigma)\rangle&=\int_{-\infty}^\infty du e^{-i\omega u}|\Sigma(u,\bm x_{},\sigma)\rangle,\\
\langle \Sigma(\omega,\bm x_{},\sigma)|&=\int_{-\infty}^\infty du e^{i\omega u}\langle \Sigma(u,\bm x_{},\sigma)|.
\end{align}\es 
Note that this is actually the Fourier transform in the time direction. The transverse coordinates remain to be invariant. After Fourier transform, the Carrollian field $\Sigma(u,\bm x,\sigma)$, defined in three-dimensional null hypersurface, switches to an infinite tower of operators $\Sigma(\omega,\bm x,\sigma)$ defined in two-dimensional Euclidean plane. In this sense, the original dual Carrollian field theory is mapped to a two-dimensional Euclidean field theory, which is the analog of the putative celestial conformal field theory in celestial holography. In the following, we still call the dual field theory in the Fourier space the Carrollian field theory, though it is already reduced to a ``celestial field theory'' in two dimensions.
We could obtain the following amplitude in Fourier space 
\bea 
\langle\prod_{j=1}^n\Sigma(\omega_j,\bm x_{j},\sigma_j)\rangle=\left(\prod_{j=1}^n { \int_{-\infty}^{\infty}}du_j e^{i\sigma_j\omega_ju_j}\right)\langle\prod_{j=1}^n\Sigma(u_j,\bm x_{ j},\sigma_j)\rangle.
\eea Then the first Ward identity \eqref{timetrans} is solved by the conservation of the energy 
\be 
\sum_{j=1}^n \sigma_j\omega_j=0,
\ee which indicates that the amplitude in the Fourier space is always proportional to a Dirac delta function 
\bea 
\langle\prod_{j=1}^n\Sigma(\omega_j,\bm x_{j},\sigma_j)\rangle=\delta(\sum_{j=1}^n \sigma_j\omega_j)i\mathcal{T}(\omega_1,\bm x_{1},\sigma_1;\cdots;\omega_n,\bm x_{n},\sigma_n).
\eea The remaining three Ward identities \eqref{strans}-\eqref{xyrot} are transformed to the following constraints
\bs\begin{align}
     \left(\sum_{j=1}^n \frac{\partial}{\partial x_j}\right)\mathcal{T}(\omega_1,\bm x_{ 1},\sigma_1;\cdots;\omega_n,\bm x_{ n},\sigma_n)&=0,\label{stransT}\\
     \left(\sum_{j=1}^n \frac{\partial}{\partial y_j}\right)\mathcal{T}(\omega_1,\bm x_{ 1},\sigma_1;\cdots;\omega_n,\bm x_{ n},\sigma_n)&=0,\label{ytransT}\\
     \left(\sum_{j=1}^n x_j\frac{\partial}{\partial y_j}-y_j\frac{\partial}{\partial x_j}\right)\mathcal{T}(\omega_1,\bm x_{ 1},\sigma_1;\cdots;\omega_n,\bm x_{ n},\sigma_n)&=0.\label{xyrotT}
\end{align}\es Similarly, the orthogonality relation becomes
\be 
    \langle\Sigma(\omega,\bm x_{},\sigma)\Sigma(\omega',\bm x'_{ },\sigma)\rangle=\frac{\pi}{\omega}\delta(\omega-\omega')\delta^{(2)}(\bm x_{}-\bm x'_{}),
\ee and the completeness relation is transformed to 
\bea 
1=\frac{1}{\pi}\int_0^\infty\omega d\omega {\int_{-\infty}^{\infty}}d\bm x_{} |\Sigma(\omega,\bm x,\sigma)\rangle \langle \Sigma(\omega,\bm x,\sigma)|.
\eea 
The amplitude  $\mathcal{T}(\omega_1,\bm x_1,\sigma_1;\cdots;\omega_n,\bm x_n,\sigma_n)$ can be retarded as a correlator in the dual two-dimensional Euclidean field theory, which is the analog of the celestial amplitude. We will call it the Carrollian amplitude in the Fourier space. 

\section{Propagators}\label{pro}
To compute the Carrollian amplitude, the first step is to calculate various propagators. In this section, we will obtain the boundary-to-boundary, bulk-to-boundary and bulk-to-bulk propagators. 
\subsection{Boundary-to-boundary propagator}
\begin{figure}
  \centering
  \begin{tikzpicture}[scale=0.7]
\newcommand{\arrowIn}{
\tikz \draw[-stealth] (-2pt,0) -- (2pt,0);
}

     \draw[draw,thick]   (5,-5) node[below right]{\footnotesize $\mathcal{H}^{--}$}--(0,0) node[left]{\footnotesize $\mathcal B$}--  (5,5) node[above right]{\footnotesize $\mathcal{H}^{++}$};

    \draw[thick] (2.5,-2.5) node[left]{\footnotesize $(u',\bm x'_{ })$}--node[right]{\footnotesize $\langle \Sigma(v,\bm x_{ },+)\Sigma(u',\bm x'_{ },-)\rangle$}(3.5,3.5) node[ left]{\footnotesize $(v,\bm x_{ })$};

    \draw[<->] (-0.6,5) node[above left] {\footnotesize $u$} -- (0,4.4) -- (0.6,5) node[above right] {\footnotesize $v$};
\fill [fill,use as bounding box](2.5,-2.5) circle (1pt);
\fill [fill,use as bounding box](3.5,3.5) circle (1pt);

  \end{tikzpicture}
  \caption{Boundary-to-boundary propagator in RRW. An incoming state is inserted at $(u',\bm x')$  while an outgoing state is located at $(v,\bm x)$. The propagator from $\mathcal{H}^{--}$ to $\mathcal{H}^{++}$ is also the tree-level two-point Carrollian amplitude. According to the ``antipodal map'', the coordinate $v$ will be rewritten as $u$ in the propagator. We will always use the coordinate $u$ even when the state is outgoing in the following figures.}\label{bbdy}
\end{figure}
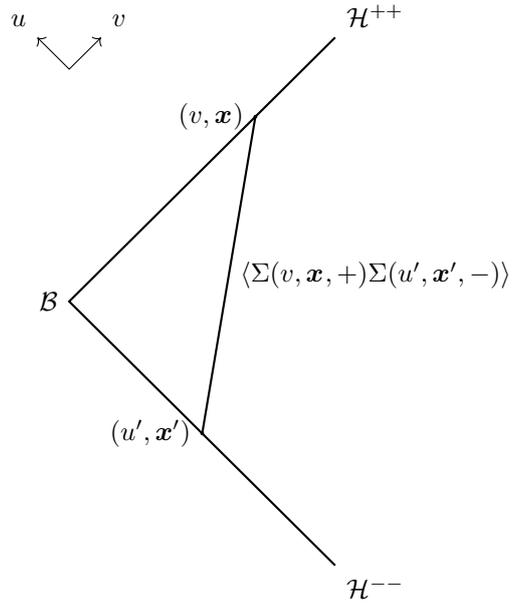

The boundary-to-boundary propagator is shown in Figure \ref{bbdy}. This is also the tree-level two-point Carrollian amplitude. 
We will consider massless theory at first.
\bea 
&&\langle \Sigma(u,\bm x_{ },+)\Sigma(u',\bm x'_{ },-)\rangle\nn\\&=& \int_0^\infty d\omega d\omega' { \int_{-\infty}^{\infty}}d\bm k d\bm k' f(\omega,k)f(\omega',k') e^{-i\omega u+i\omega'u'+i\bm k\cdot\bm x_{ }-i\bm k'\cdot\bm x'}\delta(\omega-\omega')\delta^{(2)}(\bm k-\bm k')\nn\\&=&\int_0^\infty d\omega \int_{-\infty}^\infty d\bm k f(\omega,k)^2 e^{-i\omega(u-u')+i\bm k\cdot(\bm x_{}-\bm x'_{ })}\nn\\&=&-\frac{i}{4\pi^2}\int_0^\infty d\omega  \frac{e^{-i\omega(u-u')}}{|\bm x_{}-\bm x'_{ }|^{2-2i\omega}}\nn\\&=&-\frac{1}{4\pi^2 |\bm x_{}-\bm x'_{ }|^2}\frac{1}{u-u'-\log|\bm x_{}-\bm x'_{}|^2-i\epsilon}.
\eea Note that this formula is completely different from the orthogonality relation \eqref{orthoS}. In Minkowski spacetime, the boundary-to-boundary propagator from $\mathscr{I}^-$ to $\mathscr{I}^+$ could be mapped to the orthogonality relation in $\mathscr{I}^-$ or $\mathscr{I}^+$ such that their forms are proportional to each other \cite{Liu:2024nfc}. In general spacetime, the orthogonality relation is not  necessarily equivalent to the boundary-to-boundary propagator. We could check that the boundary-to-boundary propagator is invariant under time translation $u\to u+\gamma$, spatial translation  $(x,y)\to (x+\beta,y+\delta)$ and rotation $\bm x\to R\bm x$ in transverse directions. In other words, the boundary-to-boundary propagator satisfies the Ward identities \eqref{Ward}.
We  notice that the propagator has a pole on the surface
\be 
u-u'=\log|\bm x-\bm x'|^2.\label{pole}
\ee Assuming a light is emitted from  $(T',X',Y',Z')$ and absorbed at $(T,X,Y,Z)$, then the classical trajectory of the light should obey the equation 
\be 
(T-Z-T'+Z')(T+Z-T'-Z')=(T-T')^2-(Z-Z')^2=(X-X')^2+(Y-Y')^2.
\ee Switching to the Rindler coordinates, we find 
\be 
(-\rho e^{-\tau}+\rho'e^{-\tau'})(\rho e^{\tau}-\rho' e^{\tau'})=|\bm x-\bm x'|^2.
\ee According to the definition of Carrollian amplitude, the light is emitted from $\mathcal{H}^{--}$ where $\rho'\to 0$ and $u'$ finite and absorbed at $\mathcal{H}^{++}$ where $\rho\to 0$ and $v$ finite. Therefore, the left hand side becomes 
\bea 
e^{v-u'}=|\bm x-\bm x'|^2\quad\Rightarrow\quad v-u'=\log|\bm x-\bm x'|^2,
\eea which matches with \eqref{pole} after taking the ``antipodal map'' at $\mathcal{H}^{++}$
\be 
v\to u.
\ee Therefore, the surface \eqref{pole} is actually composed by  light rays from $\mathcal{H}^{--}$ to $\mathcal{H}^{++}$.

The propagator can be transformed to the Fourier space  
\bea 
&&\langle\Sigma(\omega,\bm x_{},+)\Sigma(\omega',\bm x'_{ },-)\rangle\nn\\&=&\int_{-\infty}^\infty du\int_{-\infty}^\infty du' e^{i\omega u-i\omega' u'}\langle \Sigma(u,\bm x_{},+)\Sigma(u',\bm x'_{},-)\rangle\nn\\&=&-i\delta(\omega-\omega')\frac{1}{|\bm x_{}-\bm x'_{}|^{2-2i\omega}},
\eea from which we can read out the  $\mathcal{T}$ matrix 
\bea
\mathcal{T}(\omega,\bm x_{ },+;\omega',\bm x'_{ },-)=\left\{\begin{array}{cc}-|\bm x_{}-\bm x'_{ }|^{-2+2i\omega},&\omega=\omega',\\ 0,&\omega\not=\omega'.\end{array}\right.\label{Tmassless}
\eea This correlator has the same form of the two-point correlation function  of any conformal field theory by identifying the conformal weight $\Delta$ of the primary field $\Sigma(\omega,\bm x_{},\sigma)$ as\footnote{The minus sign can be absorbed into the redefinition of the field $\Sigma$.}
\be \Delta=1-i\omega.
\ee In general $d$ dimensions, the boundary-to-boundary propagator for massless scalar field can be found in \eqref{masslessd}
\bea 
\langle\Sigma(u,\bm x,+)\Sigma(u',\bm x',-)\rangle=-\frac{i}{4\pi^{d/2}}\int_0^\infty d\omega \frac{\Gamma(\frac{d}{2}-1-i\omega)}{\Gamma(1-i\omega)}|\bm x_{}-\bm x'_{}|^{-d+2+2 i \omega }e^{-i\omega(u-u')}.
\eea In Fourier space, the corresponding $\mathcal{T}$ matrix is 
\bea 
\mathcal{T}(\omega,\bm x_{},+;\omega',\bm x'_{},-)=\left\{\begin{array}{cc}-\frac{1}{\pi^{d/2-2}}\frac{\Gamma(\frac{d}{2}-1-i\omega)}{\Gamma(1-i\omega)}\left(|\bm x_{}-\bm x'_{ }|\right)^{-d+2+2i\omega},&\omega=\omega',\\ 0,&\omega\not=\omega'.\end{array}\right.
\eea 
We can also compute the boundary-to-boundary propagator in massive theory for later convenience. In this case,
\bea &&\langle \Sigma(u,\bm x_{ },+)\Sigma(u',\bm x'_{ },-)\rangle\nn\\&=&\int_0^\infty d\omega \int_{-\infty}^\infty d\bm k\frac{2^{2i\omega}\sinh\pi\omega \Gamma(i\omega)^2}{4\pi^2 (2\pi)^2}(\bm k^2+m^2)^{-i\omega} e^{-i\omega(u_1-u_2)+i\bm k\cdot(\bm x_{ }-\bm x'_{ })}\nn\\&=&-i\frac{1}{(2\pi)^{2}}\left(\frac{m}{|\bm x_{}-\bm x'_{ }|}\right)\int_0^\infty \frac{d\omega}{\Gamma(1-i\omega)}\left(\frac{2|\bm x_{}-\bm x'_{ }|}{m}\right)^{i\omega} e^{-i\omega(u-u')}K_{1-i\omega}(m|\bm x_{}-\bm x'_{}|).\nn\\\label{massivebb}
\eea The result can be extended to general dimensions. Interested reader can find the details in Appendix \ref{massived}. We can also transform it to the Fourier space to obtain the $\mathcal{T}$ matrix. It is non-zero only for $\omega=\omega'$
\bea 
&&\mathcal{T}(\omega,\bm x_{ },+;\omega,\bm x'_{ },-)=-\frac{2^{i\omega}}{\Gamma(1-i\omega)}\left(\frac{m}{|\bm x_{ }-\bm x'_{}|}\right)^{1-i\omega} K_{1-i\omega}(m|\bm x_{}-\bm x'_{}|).
\eea Obviously, it cannot be mapped to the two-point correlation function of any conformal field theory.
In the massless limit, we can find the asymptotic behaviour of the $\mathcal{T}$ matrix 
\bea 
\mathcal{T}(\omega,\bm x_{},+;\omega,\bm x'_{},-)&\sim& -|\bm x_{}-\bm x'_{}|^{-2+2 i \omega }+\frac{i m^2 |\bm x_{ }-\bm x'_{}|^{2 i \omega }}{4 \omega }+\mathcal{O}(m^4)\nn\\&&+\left(-\frac{m^{2-2i\omega} 2^{-2+2 i \omega } \Gamma (i \omega -1)}{\Gamma (1-i \omega )}+\mathcal{O}(m^{4-2i\omega})\right),\label{masscorrection}
\eea whose leading order is exactly the massless two-point correlation function  \eqref{Tmassless}. When the mass of the scalar particle is extremely heavy, 
\be 
m|\bm x-\bm x'|\gg 1, 
\ee the two-point correlation function decays exponentially 
\bea \mathcal{T}(\omega,\bm x_{ },+;\omega,\bm x'_{ },-)&\sim&-\frac{\sqrt{\pi } (m/2)^{1/2-i\omega}  }{|\bm x_{}-\bm x'|^{3/2-i\omega} \Gamma (1-i \omega )}e^{-m|\bm x-\bm x'|}+\cdots.
\eea 
\subsection{Bulk-to-boundary propagator}
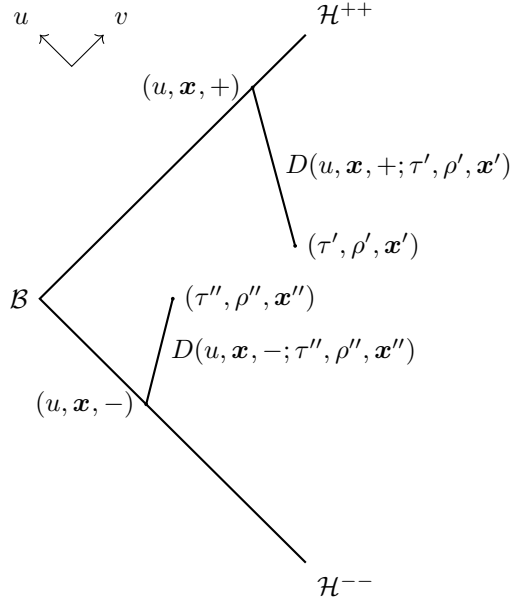
\begin{figure}
  \centering
   
  \begin{tikzpicture}[scale=0.7]
\newcommand{\arrowIn}{
\tikz \draw[-stealth] (-2pt,0) -- (2pt,0);
}

     \draw[draw,thick]   (5,-5) node[below right]{\footnotesize $\mathcal{H}^{--}$}--(0,0) node[left]{\footnotesize $\mathcal B$}--  (5,5) node[above right]{\footnotesize $\mathcal{H}^{++}$};
  \draw[thick] (2,-2) node[left]{\footnotesize $(u,\bm x_{},-)$}--node[right]{\footnotesize $D(u,\bm x_{},-;\tau'',\rho'',\bm x''_{})$}(2.5,0) node[ right]{\footnotesize $(\tau'',\rho'',\bm x''_{})$};
    
  \fill [fill,use as bounding box](2.5,0) circle (1pt);
    \fill [fill,use as bounding box](2,-2) circle (1pt);

    \draw[<->] (0,5) node[above left] {\footnotesize $u$} -- (0.6,4.4) -- (1.2,5) node[above right] {\footnotesize $v$};

  \draw[thick] (4.8,1) node[right]{\footnotesize $(\tau',\rho',\bm x'_{})$}--node[right]{\footnotesize $D(u,\bm x_{},+;\tau',\rho',\bm x'_{})$}(4,4) node[ left]{\footnotesize $(u,\bm x_{},+)$};

  \fill [fill,use as bounding box](4.8,1) circle (1pt);
   \fill [fill,use as bounding box](4,4) circle (1pt); 
   
  \end{tikzpicture}

  \caption{\centering{Bulk-to-boundary propagator in RRW.  }}\label{bulktoboundary}
\end{figure}

There are two bulk-to-boundary propagators which are shown in Figure \ref{bulktoboundary}. 

\begin{figure}
  \centering
  
  \begin{tikzpicture}[scale=0.7]

     \draw[draw,thick]   (0,-5) node[below right]{\footnotesize $\mathcal{H}^{--}$}--(-5,0) node[left]{\footnotesize $\mathcal B$}--  (0,5) node[above right]{\footnotesize $\mathcal{H}^{++}$};

      \draw[thick] (-2,0.5) node[right]{\footnotesize $|\Phi(\tau,\rho,\bm x_{})\rangle$}--node[right]{\footnotesize $D(u',\bm x'_{},+;\tau,\rho,\bm x_{})$}(-2.5,2.5) node[ left]{\footnotesize $(u',\bm x'_{},+)$};

    \draw[<->] (-5,5) node[above left] {\footnotesize $u$} -- (-4.4,4.4) -- (-3.8,5) node[above right] {\footnotesize $v$};

     \draw[draw,thick]   (9,-5) node[below right]{\footnotesize $\mathcal{H}^{--}$}--(4,0) node[left]{\footnotesize $\mathcal B$}--  (9,5) node[above right]{\footnotesize $\mathcal{H}^{++}$};
 
  \draw[thick] (6.5,-2.5) node[left]{\footnotesize $(u',\bm x'_{},-)$}--node[right]{\footnotesize $D^*(u',\bm x'_{},-;\tau,\rho,\bm x_{})$}(7,0.5) node[ right]{\footnotesize $|\Phi(\tau,\rho,\bm x_{})\rangle$};
    
  \node at (2,0) { =};
    
   \fill [fill,use as bounding box](-2,0.5) circle (1pt);
   \fill [fill,use as bounding box](-2.5,2.5) circle (1pt);
   \fill [fill,use as bounding box](6.5,-2.5) circle (1pt);
   \fill [fill,use as bounding box](7,0.5) circle (1pt);
  \end{tikzpicture}

  \caption{There are two equivalent ways to reconstruct the bulk state $|\Phi(\tau,\rho,\bm x)\rangle$. In the left figure, we integrate the states on $\mathcal{H}^{++}$ to obtain the bulk state. In the right figure, we integrate the states on $\mathcal{H}^{--}$ to reconstruct the bulk state. }\label{onsist}
\end{figure}
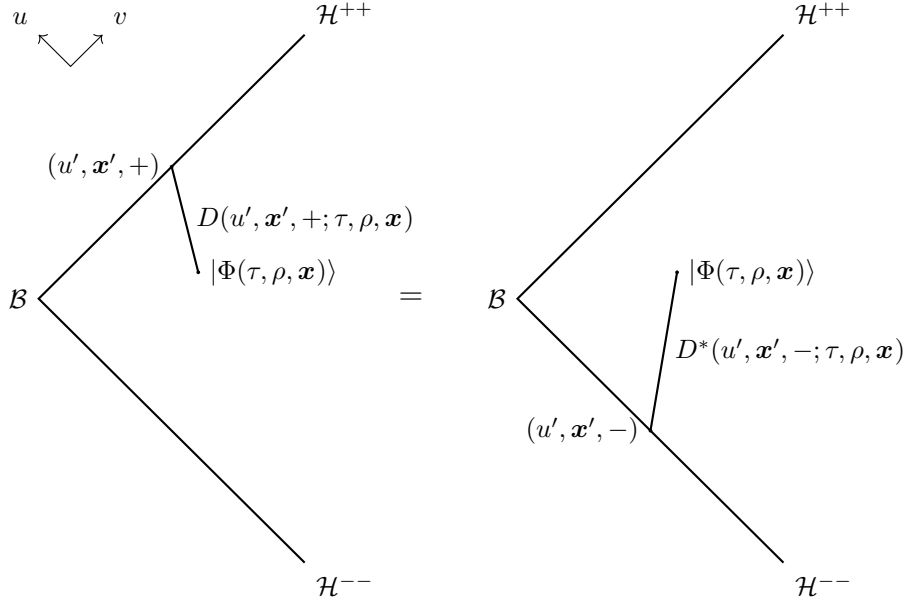

The propagator from bulk to $\mathcal{H}^{++}$ is \bea 
D(u,\bm x_{},+;\tau',\rho',\bm x_{}')=\langle \Sigma(u,\bm x_{},+)\Phi(\tau',\rho',\bm x_{}')\rangle
\eea while the one from bulk to $\mathcal{H}^{--}$ is 
\bea 
D(u,\bm x_{},-;\tau'',\rho'',\bm x''_{})=\langle\Phi(\tau'',\rho'',\bm x''_{})\Sigma(u,\bm x_{},-)\rangle.
\eea We still consider massless theory at first. Then the first bulk-to-boundary propagator is 
\bea 
&&D(u,\bm x_{},+;\tau',\rho',\bm x_{}')\nn\\&=&\frac{1}{8\pi^4}\int_0^\infty d\omega  \int_{-\infty}^{\infty}d\bm k (\frac{k}{2})^{-i\omega} \sinh\pi\omega\Gamma(i\omega)K_{i\omega}(k\rho')e^{-i\omega(u-\tau')+i\bm k\cdot(\bm x_{}-\bm x'_{})}\nn\\&=&-\frac{i}{4\pi^2}\int_0^\infty d\omega \frac{e^{-i\omega(u-u')}}{(\rho'^2+|\bm x_{}-\bm x'_{}|^2)^{1-i\omega}},
\eea where we have used the integral
\bea 
 \int_{-\infty}^{\infty} d\bm k k^{-i\omega} K_{i\omega}(k\rho')e^{i\bm k\cdot(\bm x_{}-\bm x'_{})}=\frac{2\pi\Gamma(1-i\omega)(2\rho')^{-i\omega}}{(\rho'^2+|\bm x_{}-\bm x'_{}|^2)^{1-i\omega}}.
\eea 
In the limit $\rho'\to0$ with $u'=\tau'-\log\rho'$ finite, the bulk point $x'$ approaches the null boundary $\mathcal{H}^{--}$ and the propagator becomes the boundary-to-boundary propagator
\bea 
\langle\Sigma(u,\bm x_{},+)\Sigma(u',\bm x'_{},-)\rangle=\lim_{\rho'\to0,\ u'\ \text{finite}}D(u,\bm x_{},+;\tau',\rho',\bm x'_{}).
\eea 
Similarly, we find the other bulk-to-boundary propagator
\bea 
D(u,\bm x_{},-;\tau',\rho',\bm x'_{})=-\frac{i}{4\pi^2}\int_0^\infty d\omega \frac{e^{-i\omega(v'-u)}}{(\rho'^2+|\bm x_{}-\bm x'_{}|^2)^{1-i\omega}}.
\eea We take the limit $\rho'\to 0$ and keep $v'=\tau'+\log\rho'$ finite, the propagator is reduced to the boundary-to-boundary propagator
\bea 
\langle \Sigma(v',\bm x'_{},+)\Sigma(u,\bm x_{},-)\rangle=\lim_{\rho'\to0,\ v'\ \text{finite}}D(u,\bm x_{},-;\tau',\rho',\bm x'_{}),
\eea where $v'$ could be written as $u'$ by mapping it to the ``antipodal'' point. The integral of $\omega$ can be worked out 
\bs\begin{align}
    D(u,\bm x,+;\tau',\rho',\bm x'_{})&=-\frac{1}{4\pi^2 (\rho'^2+|\bm x_{}-\bm x'_{}|^2)}\frac{1}{u-u'-\log[\rho'^2+|\bm x_{}-\bm x'_{}|^2]-i\epsilon},\\
    D(u,\bm x,-;\tau',\rho',\bm x'_{})&=-\frac{1}{4\pi^2 (\rho'^2+|\bm x_{}-\bm x'_{}|^2)}\frac{1}{v'-u-\log[\rho'^2+|\bm x_{}-\bm x'_{}|^2]-i\epsilon}.
\end{align}\es  We can also obtain the  Hermite conjugate of the bulk-to-boundary propagators
\bs\begin{align}
    D^*(u,\bm x,+;\tau',\rho',\bm x'_{})&=\langle\Phi(\tau',\rho',\bm x'_{})\Sigma(u,\bm x,+)\rangle=\frac{i}{4\pi^2}\int_0^\infty d\omega \frac{e^{i\omega(u-u')}}{(\rho'^2+|\bm x-\bm x'|^2)^{1+i\omega}},\\
    D^*(u,\bm x,-;\tau',\rho',\bm x')&=\langle\Sigma(u,\bm x,-)\Phi(\tau',\rho',\bm x')\rangle=\frac{i}{4\pi^2}\int_0^\infty d\omega \frac{e^{i\omega(v'-u)}}{(\rho'^2+|\bm x-\bm x'|^2)^{1+i\omega}}.
\end{align}\es 
For a state\be |\Phi(\tau,\rho,\bm x)\rangle=\Phi(\tau,\rho,\bm x)|0\rangle,\ee we can insert a set of complete basis at $\mathcal{H}^{++}$ to obtain 
\bea 
|\Phi(\tau,\rho,\bm x)\rangle&=&-2i\int du' d\bm x' D(u',\bm x',+;\tau,\rho,\bm x)|\dot\Sigma(u',\bm x',+)\rangle\nn\\&=&2i\int du' d\bm x' \partial_{u'}D(u',\bm x',+;\tau,\rho,\bm x)|\Sigma(u',\bm x',+)\rangle.\label{Phi1}
\eea Interestingly, we can also insert a set of complete basis at $\mathcal{H}^{--}$ to obtain  another identity
\bea 
|\Phi(\tau,\rho,\bm x)\rangle&=&-2i\int du' d\bm x' D^*(u',\bm x',-;\tau,\rho,\bm x)|\dot\Sigma(u',\bm x',-)\rangle\nn\\&=&2i\int du'd\bm x' \partial_{u'}D^*(u',\bm x',-;\tau,\rho,\bm x)|\Sigma(u',\bm x',-)\rangle.\label{Phi2}
\eea As shown in Figure \ref{onsist}, they should be consistent with each other. We have checked that  \eqref{Phi1} and \eqref{Phi2} are consistent in Appendix \ref{consistency}. Note that in classical physics, one can either use the retarded Green's function or advanced Green's function to solve the bulk field. In this sense, the propagator $D(u,\bm x,-;\tau',\rho',\bm x')$ is an advanced bulk-to-boundary propagator while $D(u,\bm x,+;\tau',\rho',\bm x')$ is a retarded bulk-to-boundary propagator. Note that these propagators indeed break the time reversal symmetry of the Feynman propagator. A time reversal that flips the arrow of time
\be 
T\to -T
\ee  leads to the reverse of the Rindler time
\be 
\tau\to-\tau.
\ee Therefore, the Feynman propagator is indeed invariant under time reversal. However, this transform will change the advanced and retarded time
\bea 
u\to -v,\quad v\to -u.
\eea As a consequence, it exchanges the two bulk-to-boundary propagators \footnote{One should also take care of the antipodal map.}
\bea 
D(u,\bm x,+;\tau',\rho',\bm x')\quad\leftrightarrow\quad D(u,\bm x,-;\tau',\rho',\bm x')
\eea 

\subsection{Bulk-to-bulk propagator}
The bulk-to-bulk propagator in Rindler vacuum  has been derived in \cite{Dowker:1978aza} by solving Green's function in RRW. In this subsection, we will study the split representation of the Feynman propagator using the product of bulk-to-boundary propagators. The Feynman propagator is defined as the time ordered two-point correlation function in the bulk
\bea 
G_F(\tau,\rho,\bm x;\tau',\rho',\bm x'_{})=\theta(\tau-\tau')\langle \Phi(\tau,\rho,\bm x)\Phi(\tau',\rho',\bm x')\rangle+\theta(\tau'-\tau)\langle \Phi(\tau',\rho',\bm x')\Phi(\tau,\rho,\bm x)\rangle.\nn\\
\eea For $\tau>\tau'$, we  insert the complete basis at $\mathcal{H}^{++}$ into the correlator 
\bea 
&&G_F(\tau,\rho,\bm x;\tau',\rho',\bm x'_{})\nn\\&=&2i\int du'' d\bm x'' \langle \Phi(\tau,\rho,\bm x)\Sigma(u'',\bm x''_{},+)\rangle\langle \dot{\Sigma}(u'',\bm x'',+)\Phi(\tau',\rho',\bm x'_{})\rangle \nn\\&=&2i \int du'' d\bm x'' D^*(u'',\bm x'',+;\tau,\rho,\bm x)\partial_{u''}D(u'',\bm x'',+;\tau',\rho',\bm x')\nn\\&=&\frac{2i}{(4\pi^2)^2}\int du'' d\bm x''\int_0^\infty d\omega  \frac{e^{i\omega(u''-u)}}{(\rho^2+|\bm x-\bm x''|^2)^{1+i\omega}}\int_0^\infty d\omega' (-i\omega')\frac{e^{-i\omega'(u''-u')}}{(\rho'^2+|\bm x'-\bm x''|^2)^{1-i\omega'}}\nn\\&=&\frac{1}{4\pi^3}\int_0^\infty d\omega \omega d\bm x'' \frac{e^{-i\omega(u-u')}}{(\rho^2+|\bm x-\bm x''|^2)^{1+i\omega}(\rho'^2+|\bm x'-\bm x''|^2)^{1-i\omega}}.\label{taubigger}
\eea We can use the Feynman integral formula \eqref{feynman} and the integral \eqref{twodim} to compute the integration of $\bm x''$
\bea 
&&G_F(\tau,\rho,\bm x;\tau',\rho',\bm x'_{})\nn\\&=&\frac{1}{4\pi^2}\int_0^\infty d\omega \frac{\omega e^{-i\omega(u-u')}}{\Gamma(1-i\omega)\Gamma(1+i\omega)}\int_0^1 dt \frac{t^{i\omega}(1-t)^{-i\omega}}{t\rho^2+(1-t)\rho'^2+t(1-t)|\bm x-\bm x'|^2}\nn\\&=&\frac{1}{4\pi^2}\int_0^\infty d\omega \frac{\omega e^{-i\omega(u-u')}}{\Gamma(1-i\omega)\Gamma(1+i\omega)}\int_0^\infty ds\frac{s^{i\omega}}{\rho^2 s^2+(\rho^2+\rho'^2+|\bm x-\bm x'|^2)s+\rho'^2}\nn\\
&=&\frac{1}{4\pi^2\rho\rho'}\int_0^\infty d\omega \frac{\omega e^{-i\omega(\tau-\tau')}}{\Gamma(1-i\omega)\Gamma(1+i\omega)}\int_0^\infty ds \frac{s^{i\omega}}{s^2+2\eta s+1}\nn\\&=&-\frac{i}{4\pi^2\rho\rho'}\int_0^\infty d\omega \frac{\zeta^{i\omega}-\zeta^{-i\omega}}{\zeta-\zeta^{-1}}e^{-i\omega(\tau-\tau')}\nn\\&=&-\frac{1}{4\pi^2\rho\rho'(\zeta-\zeta^{-1})}[\frac{1}{\tau-\tau'-\log\zeta-i\epsilon}-\frac{1}{\tau-\tau'+\log\zeta-i\epsilon}],
\eea where 
\bea 
\eta=\frac{\rho^2+\rho'^2+|\bm x-\bm x'|^2}{2\rho\rho'}\label{eta}
\eea and $-\zeta,-\zeta^{-1}$ are the two roots of the polynomial 
\be 
s^2+2\eta s+1=0\quad \Rightarrow\quad \zeta=\eta+\sqrt{\eta^2-1},\quad \zeta^{-1}=\eta-\sqrt{\eta^2-1}.\label{zeta}
\ee We may parameterize $\eta$ as 
\be 
\eta=\cosh\xi,\quad \xi>0,
\ee and then 
\bea 
\zeta=e^{\xi}.\label{zeta2}
\eea The Feynman propagator becomes 
\bea 
G_F(\tau,\rho,\bm x;\tau',\rho',\bm x')&=&\frac{\xi}{4\pi^2\rho\rho'\sinh\xi}\frac{1}{\xi^2-(\tau-\tau'-i\epsilon)^2}\theta(\tau-\tau')+(\tau\leftrightarrow\tau')\nn\\&=&\frac{\xi}{4\pi^2\rho\rho'\sinh\xi}\frac{1}{\xi^2-(\tau-\tau')^2+i\epsilon}.\label{feyn}
\eea 
We will discuss this propagator as follows.
\begin{enumerate}
    \item \textbf{Split representations.} The Feynman propagator \eqref{feyn} matches with the Green's function in \cite{Dowker:1978aza} up to a factor $i$ which comes from the convention. In \cite{Dowker:1978aza}, the Feynman propagator is found by solving Green's function in Rindler spacetime. On the other hand, we find the Feynman propagator by its split representation. Note that there are two equivalent split representations for the Feynman propagator, depending on the choice of the complete basis
    \bs\begin{align}
        G_F(\tau,\rho,\bm x;\tau',\rho',\bm x')=&\theta(\tau-\tau')2i\int du'' d\bm x'' D^*(u'',\bm x'',+;\tau,\rho,\bm x)\partial_{u''}D(u'',\bm x'',+;\tau',\rho',\bm x')\nn\\&+\theta(\tau'-\tau) 2i\int du'' d\bm x'' D^*(u'',\bm x'',+;\tau',\rho',\bm x')\partial_{u''}D(u'',\bm x'',+;\tau,\rho,\bm x),\\
        G_F(\tau,\rho,\bm x;\tau',\rho',\bm x')=&\theta(\tau-\tau')2i\int du'' d\bm x'' D(u'',\bm x'',-;\tau,\rho,\bm x)\partial_{u''}D^*(u'',\bm x'',-;\tau',\rho',\bm x')\nn\\&+\theta(\tau'-\tau) 2i\int du'' d\bm x'' D(u'',\bm x'',-;\tau',\rho',\bm x')\partial_{u''}D^*(u'',\bm x'',-;\tau,\rho,\bm x).
    \end{align}\es 
    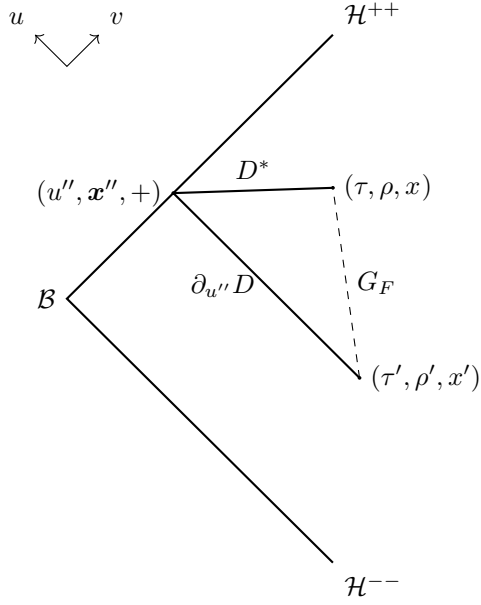
\begin{figure}
  \centering
  \begin{tikzpicture}[scale=0.7]
\newcommand{\arrowIn}{
\tikz \draw[-stealth] (-2pt,0) -- (2pt,0);
}

     \draw[draw,thick]   (5,-5) node[below right]{\footnotesize $\mathcal{H}^{--}$}--(0,0) node[left]{\footnotesize $\mathcal B$}--  (5,5) node[above right]{\footnotesize $\mathcal{H}^{++}$};

    \draw[dashed] (5.5,-1.5) node[right]{\footnotesize $(\tau',\rho',x'_{ })$}--node[right]{\footnotesize $G_F$}(5,2.1) node[ right]{\footnotesize $(\tau,\rho,x_{})$};
    
  \draw[thick] (5.5,-1.5)--node[left]{\footnotesize $\partial_{u''}D$}(2,2) node[ left]{\footnotesize $(u'',\bm x''_{ },+)$};
    
    \draw[thick] (5,2.1)--node[above]{\footnotesize $D^*$}(2,2) ;  
    
    \draw[<->] (-0.6,5) node[above left] {\footnotesize $u$} -- (0,4.4) -- (0.6,5) node[above right] {\footnotesize $v$};

   \fill [fill,use as bounding box](5.5,-1.5) circle (1pt);
   \fill [fill,use as bounding box](5,2.1) circle (1pt);
   \fill [fill,use as bounding box](2,2) circle (1pt);

  \end{tikzpicture}
  \caption{Split representation of the Feynman propagator in RRW for $\tau>\tau'$. The dashed line is the Feynman propagator and the black lines are bulk-to-boundary propagators. More explicitly, one is the bulk-to-boundary propagator while the other is the time derivative of the bulk-to-boundary propagator. We should integrate out all possible boundary positions. }\label{split}
\end{figure} 

Note that we have reproduced the Feynman propagator from the split representation which only consists the one from bulk to $\mathcal{H}^{--}$ or $\mathcal{H}^{++}$. This indicates that the possible propagator from bulk to $\mathscr{I}_R^{\pm}$ has no contribution to Feynman propagator.
\item \textbf{Bulk-to-boundary propagator from  bulk-to-bulk propagator.} Conversely, 
 we can reduce the Feynman propagator to the bulk-to-boundary propagator. As an illustration, the point $(\tau,\rho,\bm x)$ tends to the one on $\mathcal{H}^{--}$ when we take the limit
\be 
\rho\to0,\quad \tau\to-\infty \quad \text{with}\quad u=\tau-\log\rho\quad \text{finite}.
\ee Then the bulk-to-bulk propagator becomes the bulk to $\mathcal{H}^{--}$ propagator 
\bea 
\lim_{\rho\to0,\ u\ \text{finite}}G_F(\tau,\rho,\bm x;\tau',\rho',\bm x')&=&-\frac{1}{4\pi^2 (\rho'^2+|\bm x_{}-\bm x'_{}|^2)}\frac{1}{v'-u-\log[\rho'^2+|\bm x_{}-\bm x'_{}|^2]-i\epsilon}.\nn\\
\eea Similarly, the point $(\tau,\rho,\bm x)$ tends to $\mathscr{I}_R^+$ when we take the alternative limit \be \rho\to\infty,\ \tau\to\infty\quad \text{ with }\quad u=\tau-\log\rho\quad \text{finite}.\ee
In this limit, the bulk-to-bulk propagator becomes
\bea 
\lim_{\rho\to\infty,\ u\ \text{finite}}G_F(\tau,\rho,\bm x;\tau',\rho',\bm x')=\frac{1}{4\pi^2\rho^2(u'-u+i\epsilon)}.
\eea Since only the mode with zero transverse momentum may contribute to the bulk to $\mathscr{I}^+_R$ propagator, we use the condition \eqref{falloffscr} to extract it 
\be 
\langle \widetilde{\Sigma}(u)\Phi(\tau',\rho',\bm x')\rangle=\lim_{\rho\to\infty,\ u\ \text{finite}}G_F(\tau,\rho,\bm x;\tau',\rho',\bm x')=0.
\ee 
Note that there are other limits that approach $\mathscr{I}_R^+$, as have been shown in Appendix \ref{rindlerconformal}.  In these cases, the bulk-to-bulk propagator also falls off quickly and then the bulk to $\mathscr{I}_R^+$ propagator would be 0.
Therefore, the modes $\widetilde{\Sigma}(u)/\widetilde{\Xi}(v)$ have no effects on the bulk-to-boundary propagators. By taking a further limit, the propagators from the boundary to $\mathscr{I}_R^{\pm}$ are also vanishing.

    \item \textbf{Wightman functions.}  We can also define two Wightman functions
    \bs\begin{align}
    W^+(\tau,\rho,\bm x;\tau',\rho',\bm x')&=\langle \Phi(\tau,\rho,\bm x)\Phi(\tau',\rho',\bm x')\rangle,\\
    W^-(\tau,\rho,\bm x;\tau',\rho',\bm x')&=\langle \Phi(\tau',\rho',\bm x')\Phi(\tau,\rho,\bm x)\rangle
    \end{align}\es whose expressions are 
    \bs\begin{align}
        W^+(\tau,\rho,\bm x;\tau',\rho',\bm x')&=\frac{\xi}{4\pi^2\rho\rho'\sinh\xi}\frac{1}{\xi^2-(\tau-\tau'-i\epsilon)^2},\\
        W^-(\tau,\rho,\bm x;\tau',\rho',\bm x')&=\frac{\xi}{4\pi^2\rho\rho'\sinh\xi}\frac{1}{\xi^2-(\tau-\tau'+i\epsilon)^2}.
    \end{align}
    \es 
    The two Wightman functions are related to each other by complex conjugate 
    \be 
    W^+(\tau,\rho,\bm x;\tau',\rho',\bm x')=\left(W^-(\tau,\rho,\bm x;\tau',\rho',\bm x')\right)^*.
    \ee 
    Obviously, the Feynman propagator is related to the Wightman functions as follows 
    \be 
    G_F(\tau,\rho,\bm x;\tau',\rho',\bm x')=\theta(\tau-\tau')W^+(\tau,\rho,\bm x;\tau',\rho',\bm x')+\theta(\tau'-\tau)W^-(\tau,\rho,\bm x;\tau',\rho',\bm x').
    \ee 
    \item \textbf{General dimensions.} The split representation of the Feynman propagator is still valid in dimensions $d\not=4$. One can use the bulk-to-boundary propagator in \eqref{dbulk} or \eqref{dbulk2} to obtain an integral representation of the Wightman function in general dimensions
  \begin{align} 
    W^+(\tau,\rho,\bm x;\tau',\rho',\bm x')&=\frac{1}{4\pi^{d/2}(\rho\rho')^{d/2-1}}\int_0^\infty d\omega J(\eta;\omega;d)e^{-i\omega(\tau-\tau')},
    \end{align} where the function $J(\eta;\omega;d)$ is given in \eqref{Jfun} and becomes elementary function of $\eta$ in even dimension.
\end{enumerate}

\subsection{Propagators in Minkowski vacuum}
We have constructed various propagators in Rindler vacuum. The Rindler wedge is interesting due to its relation to the accelerated frame and black hole \cite{1966AmJPh..34.1174R}. The study on the vacuum in Rindler wedge leads to the famous Unruh effect \cite{Unruh:1976db}. This effect is also valid for theories with non-trivial interaction \cite{Sewell:1982zz,Unruh:1983ac} which satisfy the Wightman's axioms \cite{Bisognano:1975ih,Bisognano:1976za,Wightman:1956zz}. 
In this effect, an accelerating observer may detect a thermal state with temperature $\frac{1}{2\pi}$ in the Minkowski vacuum $|0\rangle_{M}$, which is distinguished from the Rindler vacuum studied in previous sections \cite{Takagi:1986kn,2008RvMP...80..787C}.  
Therefore, it would be better to study the propagators in Minkowski vacuum. The bulk-to-bulk  propagator in Minkowski vacuum is 
\bea 
G^{\text{Mink}}_{F}(\tau,\rho,\bm x;\tau',\rho',\bm x')={}_M\langle 0|\Phi(\tau,\rho,\bm x)\Phi(\tau',\rho',\bm x')|0\rangle_M, 
\eea which could be found by using the mode expansion \eqref{modePhi} in RRW and taking into account the Bogoliubov coefficients. Equivalently, we can also sum over all possible Feynman propagators with different winding numbers in the RRW \cite{Israel:1976ur,Gibbons:1976pt,Christensen:1978tw,Troost:1978yk,Linet:1995mq}
\bea 
G^{\text{Mink}}_{F}(\tau,\rho,\bm x;\tau',\rho',\bm x')=\sum_{n=-\infty}^\infty G_F(\tau+2\pi ni,\rho,\bm x;\tau',\rho',\bm x'),
\eea where $2\pi$ is the inverse temperature, the period of the imaginary time. Using the summation formula 
\bea 
\sum_{n=-\infty}^\infty \frac{1}{a^2-(t+2\pi i n)^2}=\frac{\coth \left(\frac{a-t}{2}\right)+\coth \left(\frac{a+t}{2}\right)}{4 a},
\eea we find the bulk-to-bulk propagator 
\bea 
G^{\text{Mink}}_{F}(\tau,\rho,\bm x;\tau',\rho',\bm x')=\frac{1}{16\pi^2\rho\rho'\sinh\frac{\xi+(\tau-\tau')}{2}\sinh\frac{\xi-(\tau-\tau')}{2}}
\eea which is exactly the standard Feynman propagator in Minkowski spacetime after switching to the Cartesian coordinates. We can extend the result to boundary-to-boundary propagator 
\bea 
   {}_M\langle  0|\Sigma(u,\bm x,+)\Sigma(u',\bm x',-)|0\rangle_{M}&=&\sum_{n=0}^\infty \langle \Sigma(u+2\pi n i,\bm x,+)\Sigma(u',\bm x',-)\rangle\nn\\&=&-\frac{1}{4\pi^2|\bm x-\bm x'|^2}\sum_{n=-\infty}^\infty\frac{1}{u-u'+2\pi n i-\log|\bm x-\bm x'|^2-i\epsilon}.\nn
\eea 
Note that the summation 
\bea 
\sum_{n=-\infty}^\infty \frac{1}{u+2\pi n i}
\eea is divergent. To regularize it, we may modify the summation to 
\bea 
\sum_{n=-\infty}^\infty \frac{1}{(u+2\pi n i)^s}. 
\eea For $\text{Re}(s)>1$, the summation is convergent 
\bea 
\sum_{n=-\infty}^\infty \frac{1}{(u+2\pi n i)^s}=\left(\frac{i}{2 \pi }\right)^s \zeta \left(s,\frac{i u}{2 \pi }+1\right)+\left(-\frac{i}{2 \pi }\right)^s \zeta \left(s,-\frac{i u}{2 \pi }\right)
\eea where $\zeta(s,x)$ is the Hurwitz zeta function \footnote{Please find more details on  Hurwitz zeta function in Appendix \ref{special}.}
\bea 
\zeta (s,x)=\sum _{n=0}^{\infty } (x+n)^{-s}.
\eea We expand the result near $s=1$ 
\bea 
\sum_{n=-\infty}^\infty \frac{1}{(u+2\pi n i)^s}=\frac{1}{e^u-1}+\mathcal{O}(s-1),\label{hur}
\eea and then the regularized  boundary-to-boundary propagator becomes
\bea 
{}_M\langle0|\Sigma(u,\bm x,+)\Sigma(u',\bm x',-)|0\rangle_{M}&=&\frac{1}{4\pi^2}\frac{1}{|\bm x-\bm x'|^2-e^{u-u'}+i\epsilon}.
\eea Similarly, we find the  bulk-to-boundary propagators 
\bs\begin{align}
    D^{\text{Mink}}(u,\bm x,+;\tau',\rho',\bm x')&=\frac{1}{4\pi^2}\frac{1}{\rho'^2+|\bm x-\bm x'|^2-e^{u-u'}+i\epsilon},\\
    D^{\text{Mink}}(u,\bm x,-;\tau',\rho',\bm x')&=\frac{1}{4\pi^2}\frac{1}{\rho'^2+|\bm x-\bm x'|^2-e^{v'-u}+i\epsilon}.
\end{align}\es 
Note that the bulk-to-boundary propagator is the $\rho\to 0$ limit of the   bulk-to-bulk propagator in Minkowski vacuum
\bea 
D^{\text{Mink}}(u,\bm x,+;\tau',\rho',\bm x')=\lim_{\rho\to0,\ u\ \text{finite}}G_F(\tau,\rho,\bm x;\tau',\rho',\bm x').
\eea Similarly, the boundary-to-boundary propagator could be obtained from the  bulk-to-boundary propagator by setting $\rho'\to 0$. All the propagators have a periodic $2\pi$ in the imaginary time. As has been stated, the boundary-to-boundary and bulk-to-boundary propagators are asymmetric under time reversal. Interestingly, we find the following behaviour
\bea 
{}_M\langle 0| \Sigma(u,\bm x,+)\Sigma(u',\bm x',-)|0\rangle_M=\left\{\begin{array}{cc}0,&u\gg u',\\
\frac{1}{4\pi^2|\bm x-\bm x'|^2},&u\ll u'\end{array}\right.
\eea for boundary-to-boundary propagator and 
\bs\begin{align}
D^{\text{Mink}}(u,\bm x,+;\tau',\rho',\bm x')&=\left\{\begin{array}{cc}0,&u\gg u',\\
\frac{1}{4\pi^2(\rho'^2+|\bm x-\bm x'|^2)},&u\ll u',\end{array}\right.\\
D^{\text{Mink}}(u,\bm x, -;\tau',\rho',\bm x')&=\left\{\begin{array}{cc}0,&v'\gg u,\\
\frac{1}{4\pi^2(\rho'^2+|\bm x-\bm x'|^2)},&v'\ll u.\end{array}\right.
\end{align}\es for bulk-to-boundary propagators.

\section{Amplitudes}\label{app}
In this section, we will derive several Carrollian amplitudes in RRW using the propagators obtained in previous section. The Feynman rule in RRW is almost the same as the one in Minkowski spacetime, except that one should use the newly found bulk-to-boundary propagators. Note that one should only integrate out the bulk points in RRW.
\subsection{Two-point Carrollian amplitude}
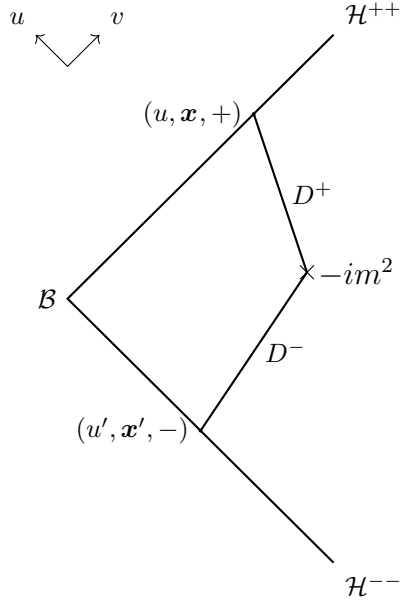
\begin{figure}
  \centering
  \usetikzlibrary {shapes.misc}
  \tikzset{cross/.style={cross out, draw=black, fill=none, minimum size=2*(#1-\pgflinewidth), inner sep=0pt, outer sep=0pt}, cross/.default={3pt}}
  \begin{tikzpicture}[scale=0.7]
  
\newcommand{\arrowIn}{
\tikz \draw[-stealth] (-2pt,0) -- (2pt,0);
}

     \draw[draw,thick]   (5,-5) node[below right]{\footnotesize $\mathcal{H}^{--}$}--(0,0) node[left]{\footnotesize $\mathcal B$}--  (5,5) node[above right]{\footnotesize $\mathcal{H}^{++}$};

    \draw[thick] (2.5,-2.5)node[left]{\footnotesize $(u',\bm x' ,-)$}--node[right]{\footnotesize $D^-$}(4.5,0.5) node[right]{$-im^2$}node[left]{$$}--node[right]{\footnotesize $D^+$}(3.5,3.5) node[ left]{\footnotesize $(u,\bm x_{ },+)$};

    \draw[<->] (-0.6,5) node[above left] {\footnotesize $u$} -- (0,4.4) -- (0.6,5) node[above right] {\footnotesize $v$};
\fill [fill,use as bounding box](2.5,-2.5) circle (1pt);
\fill [fill,use as bounding box](3.5,3.5) circle (1pt);
 \coordinate (p1) at (4.5,0.5);

    \draw (p1) node[cross] {} ;

  \end{tikzpicture}
  \caption{The leading order mass correction to the two-point Carrollian amplitude of massless scalar field theory. A vertex $-im^2$ is inserted in the bulk. The retarded and advanced bulk-to-boundary propagators are abbreviated as $D^+$ and $D^-$,respectively.}\label{mass}

\end{figure}

In this subsection, we will compute the two-point Carrollian amplitude in massive scalar theory. The result has been given in \eqref{massivebb}. However, by assuming that the mass term is a perturbation of the massless theory, we can use perturbation theory to check this result. This is also a consistency check for the Feynman rule in RRW. We will consider the leading correction whose Feynman diagram is shown in Figure \ref{mass}.
\bea 
&&\langle\Sigma(u,\bm x,+)\Sigma(u',\bm x',-)\rangle\nn\\&=&-im^2 \int_{-\infty}^\infty d\tau'' \int_0^\infty \rho''d\rho'' \int d\bm x'' D(u,\bm x,+;\tau'',\rho'',\bm x'')D(u',\bm x',-;\tau'',\rho'',\bm x'')\nn\\&=&-im^2(-\frac{i}{4\pi^2})^2 \int_{-\infty}^\infty d\tau''\int_0^\infty \rho''d\rho''\int d\bm x'' \int_0^\infty d\omega  \frac{e^{-i\omega(u-u'')}}{(\rho''^2+|\bm x-\bm x''|^2)^{1-i\omega}}\int_0^\infty d\omega' \frac{e^{-i\omega'(v''-u')}}{(\rho''^2+|\bm x'-\bm x''|^2)^{1-i\omega'}}\nn\\&=&\frac{im^2}{8\pi^3}\int_0^\infty \rho''d\rho''\int d\bm x''
\int_0^\infty d\omega \frac{e^{-i\omega(u-u')}\rho''^{-2i\omega}}{(\rho''^2+|\bm x-\bm x''|^2)^{1-i\omega}(\rho''^2+|\bm x'-\bm x''|^2)^{1-i\omega}}\nn\\&=&
\frac{im^2}{8\pi^3}\int_0^\infty d\omega \frac{\Gamma(2-2i\omega)}{\Gamma(1-i\omega)^2}e^{-i\omega(u-u')}\int_0^\infty d\rho''\int d\bm x''\int_0^1 dt \frac{\rho''^{1-2i\omega}t^{-i\omega}(1-t)^{-i\omega}}{(\rho''^2+t|\bm x''-\bm x|^2+(1-t)|\bm x''-\bm x'|^2)^{2-2i\omega}}\nn\\&=&\frac{im^2}{8\pi^2}\int_0^\infty d\omega \frac{\Gamma(1-2i\omega)}{\Gamma(1-i\omega)^2}e^{-i\omega(u-u')}\int_0^\infty d\rho''\rho''^{1-2i\omega}\int_0^1 dt \frac{t^{-i\omega}(1-t)^{-i\omega}}{(\rho''^2+t(1-t)|\bm x-\bm x'|^2)^{1-2i\omega}}\nn\\&=&-\frac{m^2}{16\pi^2}\int_0^\infty d\omega \frac{e^{-i\omega(u-u')}|\bm x-\bm x'|^{2i\omega}}{\omega}.
\eea 
Now we can transform it to the Fourier space and find the $\mathcal{O}(m^2)$ correction of the boundary-to-boundary propagator 
\bea 
\mathcal{T}(\omega,\bm x,+;\omega',\bm x',-)=\frac{im^2}{4\omega}|\bm x-\bm x'|^{2i\omega},
\eea which matches with the $\mathcal{O}(m^2)$ correction of \eqref{masscorrection}.
\subsection{Three-point Carrollian amplitude}
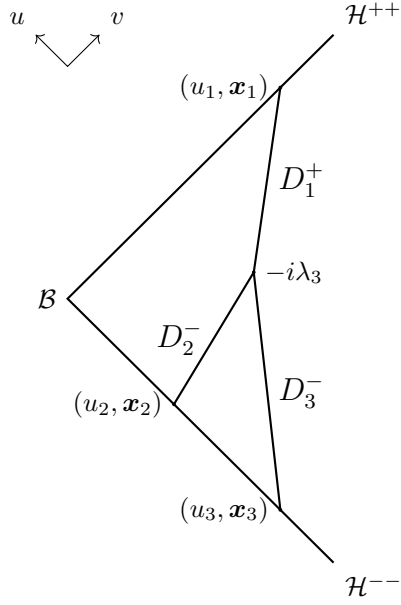
\begin{figure}
  \centering
  \usetikzlibrary {shapes.misc}
  \tikzset{cross/.style={cross out, draw=black, fill=none, minimum size=2*(#1-\pgflinewidth), inner sep=0pt, outer sep=0pt}, cross/.default={3pt}}
  \begin{tikzpicture}[scale=0.7]
  
\newcommand{\arrowIn}{
\tikz \draw[-stealth] (-2pt,0) -- (2pt,0);
}

     \draw[draw,thick]   (5,-5) node[below right]{\footnotesize $\mathcal{H}^{--}$}--(0,0) node[left]{\footnotesize $\mathcal B$}--  (5,5) node[above right]{\footnotesize $\mathcal{H}^{++}$};

    \draw[thick] (2,-2)node[left]{\footnotesize $(u_2,\bm x_{ 2})$}--node[left]{$D_2^-$}(3.5,0.5) node[right]{\footnotesize $-i\lambda_3$}--node[right]{$D_1^+$}(4,4) node[ left]{\footnotesize $(u_1,\bm x_{1})$};
    
    \draw[thick] (3.5,0.5) --node[right]{$D_3^-$}(4,-4) node[ left]{\footnotesize $(u_3,\bm x_{ 3})$};
    
    \draw[<->] (-0.6,5) node[above left] {\footnotesize $u$} -- (0,4.4) -- (0.6,5) node[above right] {\footnotesize $v$};
\fill [fill,use as bounding box](2,-2) circle (1pt);
\fill [fill,use as bounding box](4,4) circle (1pt);
\fill [fill,use as bounding box](4,-4) circle (1pt);
\fill [fill,use as bounding box](3.5,0.5) circle (1pt);
  \end{tikzpicture}
  \caption{\centering{Three-point Carrollian amplitude at the tree-level.} }\label{threepoint}
\end{figure}

In this subsection, we will compute the three-point Carrollian amplitude in $\Phi^3$ theory with two incoming and one outgoing states. The outgoing state is located at $(u_1,\bm x_1)$ and the incoming states are inserted at $(u_2,\bm x_2)$ and $(u_3,\bm x_3)$, respectively.  The Feynman diagram is shown in Figure \ref{threepoint} and the three-point Carrollian amplitude is
\bea 
&&\langle \Sigma(u_1,\bm x_{ 1},+)\Sigma(u_2,\bm x_{ 2},-)\Sigma(u_3,\bm x_{ 3},-)\rangle\nn\\&=&-i\lambda_3\int_{-\infty}^\infty d\tau \int_0^\infty d\rho \rho \int d\bm y D(u_1,\bm x_{ 1},+;\tau,\rho,\bm y)D(u_2,\bm x_{ 2},-;\tau,\rho,\bm y)D(u_3,\bm x_{ 3},-;\tau,\rho,\bm y)\nn\\&=& \frac{\lambda_3}{(4\pi^2)^3}\int_{-\infty}^\infty d\tau\int_0^\infty d\rho \rho \int d\bm y \prod_{j=1}^3 \int_0^\infty d\omega_j
\frac{e^{-i\omega_1(u_1-u)-i\omega_2(v-u_2)-i\omega_3(v-u_3)}}{\prod_{j=1}^3(\rho^2+|\bm x_j-\bm y|^2)^{1-i\omega_j}}\nn\\&=&\frac{\lambda_3}{32\pi^5}\int_0^\infty d\rho \rho \int d\bm y \prod_{j=1}^3\int_0^\infty d\omega_j\frac{\delta(\omega_1-\omega_2-\omega_3)e^{{ -}i\omega_1 u_1+i\omega_2u_2+i\omega_3 u_3}\rho^{-i\omega_1-i\omega_2-i\omega_3}}{\prod_{j=1}^3(\rho^2+|\bm x_j-\bm y|^2)^{1-i\omega_j}}.\nn\\
\eea We switch it to the Fourier space ($\omega_1=\omega_2+\omega_3$)
\bea 
&&\mathcal{T}(\omega_1,\bm x_1,+;\omega_2,\bm x_2,-;\omega_3,\bm x_3,-)\nn\\&=&-i\frac{\lambda_3}{4\pi^2}\int_0^\infty \rho d\rho \int d\bm y \frac{\rho^{-iw}}{\prod_{j=1}^3(\rho^2+|\bm x_j-\bm y|^2)^{1-i\omega_j}}\nn\\&=&-\frac{i\lambda_3}{4\pi^2}\frac{\Gamma(3-iw)}{\prod_{j=1}^3 \Gamma(1-i\omega_j)}\int_0^\infty d\rho \int d\bm y \int_0^1 dt_1 dt_2 dt_3  \frac{\delta(t_1+t_2+t_3-1)t_1^{-i\omega_1}t_2^{-i\omega_2}t_3^{-i\omega_3}\rho^{1-iw}}{[\sum_{j=1}^3 t_j(\rho^2+|\bm x_j-\bm y|^2)]^{3-iw}}\nn\\&=&-\frac{i\lambda_3}{4\pi^2}\frac{\sqrt{\pi } 2^{-3+i w}  \Gamma \left(1-\frac{i w}{2}\right)}{\Gamma \left(\frac{3}{2}-\frac{i w}{2}\right)}\frac{\Gamma(3-iw)}{\prod_{j=1}^3 \Gamma(1-i\omega_j)}\int d\bm y \int_0^1 dt_1 dt_2 dt_3 \frac{\delta(t_1+t_2+t_3-1)t_1^{-i\omega_1}t_2^{-i\omega_2}t_3^{-i\omega_3}}{[\sum_{j=1}^3 t_j|\bm x_j-\bm y|^2]^{2-iw/2}}\nn\\&=&-\frac{i\lambda_3}{4\pi}\frac{\sqrt{\pi } 2^{-3+i w}  \Gamma \left(1-\frac{i w}{2}\right)}{\Gamma \left(\frac{3}{2}-\frac{i w}{2}\right)(1-\frac{iw}{2})}\frac{\Gamma(3-iw)}{\prod_{j=1}^3 \Gamma(1-i\omega_j)}\int_0^1 dt_1 dt_2 dt_3 \frac{\delta(t_1+t_2+t_3-1)t_1^{-i\omega_1}t_2^{-i\omega_2}t_3^{-i\omega_3}}{S_3^{1-iw/2}}\label{3pt}
\eea with
\bea 
S_3&=&S_3(\bm x_1,\bm x_2,\bm x_3;t_1,t_2,t_3)\nn\\&=&t_1(1-t_1) \bm x_1^2+t_2(1-t_2)\bm x_2^2+t_3(1-t_3)\bm x_3^2-2t_1t_2\bm x_1\cdot\bm x_2-2t_1t_3\bm x_1\cdot\bm x_3-2t_2t_3\bm x_2\cdot\bm x_3\nn\\
\eea and 
\bea 
w=\omega_1+\omega_2+\omega_3=2\omega_1.
\eea
Utilizing the translation invariance, we may shift $\bm x_1=0$ and $S_3$ becomes 
\be 
S_3=t_2(1-t_2)\bm x_2^2+t_3(1-t_3)\bm x_3^2-2t_2 t_3 \bm x_2\cdot\bm x_3.
\ee We may change the variables
\be 
t_2=\frac{z_2}{1+z_2+z_3},\quad t_3=\frac{z_3}{1+z_2+z_3},
\ee then 
\bea 
&&\mathcal{T}(\omega_1,0,+;\omega_2,\bm x_2,-;\omega_3,\bm x_3,-)\nn\\&=&\tilde{\lambda}_3\int_0^\infty dz_2\int_0^\infty dz_3 \frac{z_2^{-i\omega_2}z_3^{-i\omega_3}}{(1+z_2+z_3)[z_2\bm x_2^2+z_3\bm x_3^2+z_2z_3\bm x_{23}^2]^{1-iw/2}}\label{intT}
\eea where 
\bea 
\tilde{\lambda}_3=-\frac{i\lambda_3}{4\pi}\frac{\sqrt{\pi } 2^{-3+i w}  \Gamma \left(1-\frac{i w}{2}\right)}{\Gamma \left(\frac{3}{2}-\frac{i w}{2}\right)(1-\frac{iw}{2})}\frac{\Gamma(3-iw)}{\prod_{j=1}^3 \Gamma(1-i\omega_j)}
\eea and 
\bea 
\bm x_{23}=\bm x_2-\bm x_3.
\eea
\subsubsection{Zero-energy Carrollian amplitude(ZECA)}
 In the limit $\omega_1=\omega_2=\omega_3=0$, we find 
\bea 
&&\mathcal{T}(0,0,+;0,\bm x_2,-;0,\bm x_3,-)\nn\\&=&\tilde{\lambda}_3\int_0^\infty dz_2\int_0^\infty dz_3 \frac{1}{(1+z_2+z_3)[z_2\bm x_2^2+z_3\bm x_3^2+z_2z_3\bm x_{23}^2]^{}}\nn\\&=&\tilde{\lambda}_3 \int_0^\infty dz_3\frac{\log(1+z_3)-\log z_3-\log \bm x_3^2+\log[\bm x_2^2+z_3\bm x^2_{23}]}{\bm x_2^2+2\bm x_2\cdot\bm x_{23}z_3+\bm x_{23}^2z_3^2}.\label{inv}
\eea Introducing a new variable 
\bea 
t=z_3 \frac{|\bm x_{23}|}{|\bm x_2|}
\eea and the normal vectors 
\bea 
\bm n_2=\frac{\bm x_2}{|\bm x_2|},\quad \bm n_{3}=\frac{\bm x_3}{|\bm x_{3}|},\quad \bm n_{23}=\frac{\bm x_{23}}{|\bm x_{23}|},
\eea the three-point Carrollian amplitude becomes 
\bea 
&&\mathcal{T}(0,0,+;0,\bm x_2,-;0,\bm x_3,-)\nn\\&=&\tilde{\lambda}_3\frac{1}{|\bm x_2||\bm x_{23}| }\int_0^\infty dt \frac{\log(1+\frac{|\bm x_2|}{|\bm x_{23}|}t)+\log(1+\frac{|\bm x_{23}|}{|\bm x_2|}t)-\log t+\log\frac{|\bm x_2||\bm x_{23}|}{|\bm x_3|^2}}{1+2\bm n_2\cdot\bm n_{23}t+t^2}\nn\\&=&\tilde{\lambda}_3\frac{1}{|\bm x_2||\bm x_{23}|}\int_0^\infty dt \frac{\log(t+a)+\log(t+a^{-1})-\log t+\log c}{(t+b)(t+b^{-1})}
\eea where the constants $a,b,c$ are
\bea 
a=\frac{|\bm x_2|}{|\bm x_{23}|},\quad b=e^{i\psi},\quad c=\frac{|\bm x_2||\bm x_{23}|}{|\bm x_3|^2}
\eea with $\psi$ the angle between $\bm n_2$ and $\bm n_{23}$
\be 
\cos\psi=\bm n_2\cdot\bm n_{23}.
\ee The integral can be expressed as polylogarithm function
\bea 
&&\mathcal{T}(0,0,+;0,\bm x_2,-;0,\bm x_3,-)\nn\\&=&\tilde{\lambda}_3\frac{1}{|\bm x_2||\bm x_{23}|}\frac{1}{b^{-1}-b}[\text{Li}_2(1-\frac{b}{a})-\text{Li}_2(1-\frac{1}{ab})+\text{Li}_2(1-ab)-\text{Li}_2(1-\frac{a}{b})-2\log c\log b].\nn\\\label{T}
\eea Interested reader can find more details in Appendix \ref{int}.

   \paragraph{$S_3$ symmetry.} To restore $\bm x_1$, one can just replace \be 
   \bm x_2\to \bm x_{21},\quad \bm x_3\to \bm x_{31}.
   \ee We have checked the invariance of the three-point Carrollian amplitude \eqref{T} under permutation group $S_3$ numerically
   \bea 
   &&\mathcal{T}(0,\bm x_1,+;0,\bm x_2,-;0,\bm x_3,-)= \mathcal{T}(0,\bm x_2,+;0,\bm x_1,-;0,\bm x_3,-)\nn\\&=& \mathcal{T}(0,\bm x_3,+;0,\bm x_1,-;0,\bm x_2,-)=\mathcal{T}(0,\bm x_1,+;0,\bm x_3,-;0,\bm x_2,-)\nn\\&=& \mathcal{T}(0,\bm x_2,+;0,\bm x_3,-;0,\bm x_1,-)= \mathcal{T}(0,\bm x_3,+;0,\bm x_2,-;0,\bm x_1,-).
   \eea 
   \paragraph{Three mass triangle integrals.} Unlike the two-point Carrollian amplitude, the three-point Carrollian amplitude does not correspond to three-point correlator of any conformal field theory. However, the integral representation of the tree-level three-point ZECA \eqref{inv} for general frequencies is akin to the three mass triangle loop integrals in particle physics\cite{Usyukina:1992jd,Usyukina:1994iw}. To see this point, we restore $\bm x_1$ and the three-point ZECA becomes 
   \bea 
   &&\mathcal{T}(0,\bm x_1,+;0,\bm x_2,-;0,\bm x_3,-)\nn\\&=&\tilde{\lambda}_3|\bm x_{23}|^{-2}\int_0^\infty dz_2\int_0^\infty dz_3 \frac{1}{(1+z_2+z_3)(z_2 z_3+{\tt{u}} z_2+{\tt{v}} z_3)},
   \eea where $\tt{u},\tt{v}$ are 
   \bea 
   {\tt{u}}=\frac{\bm x_{12}^2}{\bm x_{23}^2},\quad {\tt{v}}=\frac{\bm x_{13}^2}{\bm x_{23}^2}.
   \eea 
   One may define two variables 
   \bs\begin{align}
       z&=\frac{1}{2}(1+{\tt u}-{\tt v}+\sqrt{K(1,{\tt u},{\tt v})}),\\
       \bar z&=\frac{1}{2}(1+{\tt u}-{\tt v}-\sqrt{K(1,{\tt u},{\tt v})})
   \end{align}\es where the K$\ddot{a}$llen function is 
   \bea
   K(d_1,d_2,d_3)=d_1^2+d_2^2+d_3^2-2d_1 d_2-2d_2 d_3-2d_3 d_1.
   \eea In general, the  K$\ddot{a}$llen function $K(1,{\tt u},{\tt v})$ divides the $u$-$v$ plane into four parts according to  the sign of the parabola 
   \be 
   K(1,{\tt u},{\tt v})=1+{\tt u}^2+{\tt v}^2-2{\tt u} {\tt v}-2{\tt u}-2{\tt v}
   \ee and the domain of $z,\bar z$. This is shown in Figure \ref{kallen}. The red line is the K$\ddot{a}$llen function in the ${\tt u}$-${\tt v}$ plane and it separates the complex  and real $z,{ \bar z}$ region in the plane. In the shaded region  I, the variables $z,\bar z$ are complex. In region II-IV, the variables $z,\bar z$ are real. More explicitly, the domain of $z$ and $\bar z$ in the regions II,III and IV are 
    \bea 
   \left\{\begin{array}{cc} 0<z,\bar z<1,&\ \text{II},\\
   z,\bar z<0,& \text{III},\\
   z,\bar z>1,& \text{IV}.\end{array}\right.
   \eea 
   
   \begin{figure}
\usetikzlibrary{shapes.geometric,backgrounds}
  \centering
\begin{tikzpicture} [scale=1.8]
\draw[-] (0,0)--node[below=10pt]{\footnotesize ${\tt u}$}(4,0);
\draw[-] (0,0)--node[left=10pt]{\footnotesize ${\tt v}$}(0,4);
\foreach \x in {0,0.5,...,4}
{
    \draw[xshift=\x cm] (0,0) -- (0,0.1);
    \draw[yshift=\x cm] (0,0) -- (0.1,0);
};  
\node[below] at (0,0){0};
\foreach \y in {1,2,...,4}
    \node[below] at(\y,0){\footnotesize \y};
\foreach \y in {1,2,...,4}
    \node[left] at(0,\y){\footnotesize \y};
 \draw[smooth,domain=0:2,color=red]    plot ({(\x)^2},{1-2*(\x)+(\x)^2})        ;
    \draw[smooth,domain=0:1,color=red]    plot ({(\x)^2},{1+2*(\x)+(\x)^2})          ;
\draw[dotted] (1,0)--(1,4);
 \draw[dotted] (0,1)--(4,1);   
  \draw[dotted] (0,0)--(4,4);   
\node at (0.5,3.9) {\footnotesize ${\rm \uppercase\expandafter{\romannumeral3}}$};
\node at (0.5,3.7) {\footnotesize $$};

  \node at (2,2) {\footnotesize ${\rm \uppercase\expandafter{\romannumeral1}}$};
  \node at (2,1.8) {\footnotesize $z,\bar{z}\quad complex$ };
  
 \node at (0.15,0.15) {\footnotesize ${\rm \uppercase\expandafter{\romannumeral2}}$};

 \node at (3.7,0.5) {\footnotesize ${\rm \uppercase\expandafter{\romannumeral4}}$};
 \node at (3.7,0.3){\footnotesize $$};
 \begin{scope}[on background layer]
\fill [lightgray, domain=0:1, variable=\x]
 plot ({(\x)^2},{1+2*(\x)+(\x)^2})--(4,4)--(4,1);
 \fill[lightgray,domain=0:2, variable=\x]    plot ({(\x)^2},{1-2*(\x)+(\x)^2})--(4,4);
\end{scope}
\end{tikzpicture}
  \caption{K$\ddot{a}$llen function in the ${\tt u}$-${\tt v}$ plane. Similar  figure can be found in \cite{Chavez:2012kn}. However, we are in the real position space and the unshaded region are ruled out. }\label{kallen}
\end{figure}
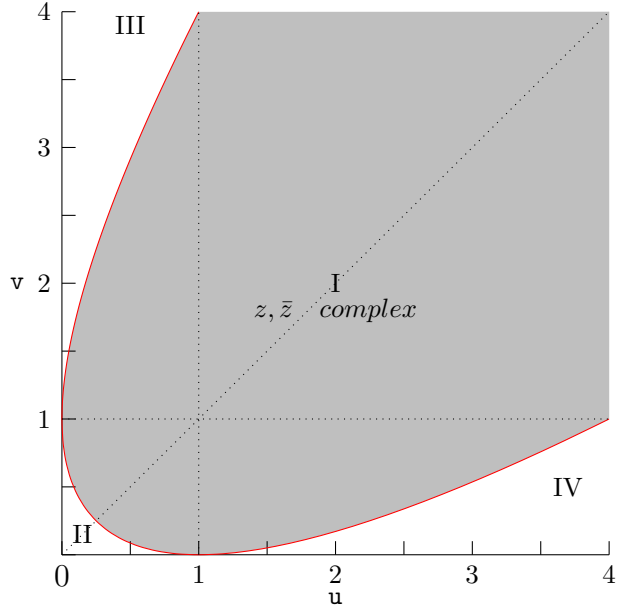
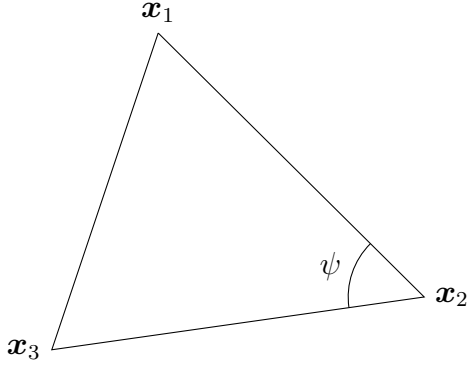
\begin{figure}
  \centering
  \usetikzlibrary{quotes,angles}
\begin{tikzpicture}[scale=0.7]
\coordinate (o) at (-2,-3);
\coordinate (a) at (0,3);
\coordinate (b) at (5,-2);
\draw (a)node[above]{$\bm x_1$}--(b)node[right]{$\bm x_2$}--(o)node[left]{$\bm x_3$}--(a);

\pic[draw,"$\psi$",  -, angle eccentricity=1.3, angle radius=1cm]
    {angle=a--b--o};
\end{tikzpicture}
  \caption{Three points $\bm x_1,\bm x_2,\bm x_3$ form a triangular in the transverse plane. The symbol $\psi$ is the angle between the vector $\bm x_{21}$ and $\bm x_{23}$.}\label{triangular}
\end{figure}

 In our case, we find 
   \bea 
   K(1,{\tt u},{\tt v})=\frac{\bm x_{12}^4+\bm x_{13}^4+\bm x_{23}^4-2\bm x_{12}^2\bm x_{13}^2-2\bm x_{12}^2\bm x_{23}^2-2\bm x_{13}^2\bm x_{23}^2}{\bm x_{23}^4}=|z-\bar z|^2.
   \eea Interestingly, the numerator is related to the famous Heron's formula for the area $\text{A}$ of a triangular where the length of the three sides are $|\bm x_{12}|, |\bm x_{13}|, |\bm x_{23}|$ ,respectively
   \bea 
   |\text{A}_{\Delta}|=\frac{1}{4}\sqrt{2\bm x_{12}^2\bm x_{23}^2+2\bm x_{13}^2\bm x_{23}^2+2\bm x_{12}^2\bm x_{13}^2-\bm x_{12}^4-\bm x_{13}^4-\bm x_{23}^4}.
   \eea The triangular is exactly determined by  the three points $\bm x_1,\bm x_2,\bm x_3$ in the transverse plane, as shown in Figure \ref{triangular}. 
   
   The area of the triangular can also be expressed through the Law of  Sines 
   \be 
   \text{A}_{\Delta}=\frac{1}{2}|\bm x_{21}||\bm x_{23}|\sin\psi,\label{area}
   \ee where $\psi$ is the angle between $\bm x_{21}$ and $\bm x_{23}$.
   Since the area $\text{A}$ is always non-negative, we conclude that the variable $z,\ \bar z$ are complex numbers that conjugate to each other. Therefore, region II-IV are ruled out in Figure \ref{kallen} and we can only consider the shaded region I and its boundary. Note that 
   \bea 
   \frac{1}{2}(1+{\tt u}-{\tt v})=\frac{\bm x_{23}^2+\bm x_{12}^2-\bm x_{13}^2}{2\bm x_{23}^2}=-\frac{\bm x_{23}\cdot\bm x_{12}}{\bm x_{23}^2},
   \eea where we have used the Law of Cosines. Therefore, the variables $z,\bar z$ can be written as 
   \bea 
   z=\frac{|\bm x_{12}|}{|\bm x_{23}|}e^{i\psi},\quad \bar z=\frac{|\bm x_{12}|}{|\bm x_{23}|}e^{-i\psi}.
   \eea 
 Using the result of \cite{Chavez:2012kn}, the integral is
   \bea 
   &&\mathcal{T}(0,\bm x_1,+;0,\bm x_2,-;0,\bm x_3,-)=\tilde{\lambda}_3|\bm x_{23}|^{-2}\frac{4i}{ z-\bar z}P_2(z),\label{block}
   \eea with $P_2(z)$ the Bloch-Wigner dilogarithm which is the  single-valued analog of the classical polylogarithms whose properties are given in Appendix \ref{special}. We also checked the equivalence of the result \eqref{T} and \eqref{block}. 
   Note  the identity 
   \be 
   \frac{4i}{z-\bar z}|\bm x_{23}|^{-2}=\frac{2}{|\bm x_{12}||\bm x_{23}|\sin\psi}=\frac{1}{\text{A}_{\Delta}},
   \ee the three-point ZECA may be simplified further 
   \bea 
   &&\mathcal{T}(0,\bm x_1,+;0,\bm x_2,-;0,\bm x_3,-)=\tilde{\lambda}_3\frac{P_2(z)}{\text{A}_\Delta}.
   \eea 
We list the correspondence between the $S_3$ symmetry on $z,\bar z$ variables and the positions $\bm x_j,\ j=1,2,3$ in Table \ref{zbarz}. 
 \begin{table}
\begin{center}
\renewcommand\arraystretch{1.5}
    \begin{tabular}{|c||c|}\hline
$S_3$ transformation on positions& $S_3$ transformation on $z$\\\hline\hline
$123$&$z\to z$\\\hline
$132$&$z\to 1-\bar z$\\\hline
$213$&$z\to \frac{\bar z}{\bar z-1}$\\\hline
$321$&$z\to \frac{1}{\bar z}$\\\hline
$231$&$z\to \frac{1}{1-z}$\\\hline
$312$&$z\to \frac{z-1}{z}$\\\hline
\end{tabular}
\caption{$S_3$ transformation. In the first column, the $S_3$ is represented by the number $ijk$ which means the permutation $\bm x_1\to \bm x_i,\ \bm x_2\to \bm x_j,\ \bm x_3\to \bm x_k$. In the second column, we only write down the transform of the $z$ variable.}\label{zbarz}
\end{center}
\end{table}
Note that the area $\text{A}_{\Delta}$ is invariant under the $S_3$ transformation of the positions. Combining with the 6-fold symmetry of Bloch-Wigner dilograithm in Appendix \eqref{six}, we can prove that the three-point ZECA \eqref{block} is invariant under $S_3$.
   \paragraph{Collinear points.}
   When the three points $\bm x_j$ are collinear, both of the area of the triangular and the Bloch-Wigner dilograithm vanish. This corresponds to the red line in Figure \ref{kallen} and we find
   \bea 
   \bar z=z,\quad {\tt u}=z^2,\quad {\tt v}=(1-z)^2,
   \eea and the three-point ZECA can be integrated out straightforwardly
   \bea 
 \mathcal{T}(0,\bm x_1,+;0,\bm x_2,-;0,\bm x_3,-)=\tilde{\lambda}_3|\bm x_{23}|^{-2}\left(\frac{\log z^2}{z-1}-\frac{\log(z-1)^2}{z}\right).
   \eea 
   The result is finite except at $z=0,1,\infty$, which corresponds to $\bm x_1=\bm x_2,\ \bm x_1=\bm x_3,\ \bm x_2=\bm x_3$, respectively. This is reasonable since in these cases two of the operators are  inserted at the same point. It follows that the Bloch-Wigner dilogarithm can be expanded near the collinear limit as 
   \bea 
   4iP_2(z)\sim \left(\frac{\log z^2}{z-1}-\frac{\log(z-1)^2}{z}\right)(z-\bar z)+\cdots.\label{collinear}
   \eea 
   \paragraph{Isosceles triangle.} In this case, we may assume 
   \be 
  |\bm x_{12}|=|\bm x_{23}|\quad\Rightarrow\quad z=e^{i\psi}
   \ee without loss of generality. The Bloch-Wigner dilogarithm becomes 
   \bea 
   P_2(e^{i\psi})=\sum_{n=1}^\infty \frac{\sin n\psi}{n^2}=\text{Im}\left(\text{Li}_2(e^{i\psi})\right)
   \eea and the three-point ZECA is 
   \bea 
    \mathcal{T}(0,\bm x_1,+;0,\bm x_2,-;0,\bm x_3,-)=2\tilde{\lambda}_3 \frac{\text{Im}\left(\text{Li}_2(e^{i\psi})\right)}{\sin \psi}\frac{1}{\bm x_{23}^2 }.
   \eea Note that $P_2(e^{i\psi})$ is the Clausen function $\text{Cl}_n(\psi)$ with $n=2$.

 \subsubsection{Non-zero energy Carrollian amplitude (NECA)}  
   We may introduce two new variables 
   \bea 
   t=z_2,\quad t'=\frac{\bm x_{12}^2}{\bm x_{23}^2}\frac{1}{z_3},
   \eea then the three-point Carrollian amplitude in Fourier space can be found as 
   \bea 
   &&\mathcal{T}(\omega_1,\bm x_1,+;\omega_2,\bm x_2,-;\omega_3,\bm x_3,-)\nn\\&=&\tilde{\lambda}_3|\bm x_{12}|^{2i\omega_2}|\bm x_{23}|^{-2+2i\omega_3}\int_0^\infty dt \int_0^\infty dt' \frac{t^{-i\omega_2}t'^{-i\omega_2}}{(tt'+t'+{\tt u})(tt'+t+{\tt v})^{1-i\omega_1}}.
   \eea We define a function by the integral with two variables ${\tt u},{\tt v}$ and four parameters $a_i,\ i=1,2,3,4$,
   \bea 
   \mathcal{I}(a_1,a_2,a_3,a_4;{\tt u},{\tt v})=\int_0^\infty dt \int_0^\infty dt' \frac{t^{a_1}t'^{a_2}}{(tt'+t'+{\tt u})^{a_3}(t t'+t+{\tt v})^{a_4}},\label{funI}
   \eea then the three-point Carrollian amplitude is 
   \bea 
   &&\mathcal{T}(\omega_1,\bm x_1,+;\omega_2,\bm x_2,-;\omega_3,\bm x_3,-)=\tilde{\lambda}_3|\bm x_{12}|^{2i\omega_2}|\bm x_{23}|^{-2+2i\omega_3}\mathcal{I}(-i\omega_2,-i\omega_2,1,1-i\omega_1;{\tt u},{\tt v}).\nn\\
   \eea The same integral will appear in the four-point Carrollian amplitude of $\Phi^4$ theory, we will discuss it later.

\subsection{Four-point Carrollian amplitude}
In this subsection, we will compute the four-point Carrollian amplitude in $\Phi^4$ theory with two incoming and two outgoing states. The Feynman diagram is shown in Figure \ref{phi4}. The outgoing states are located at $(u_1,\bm x_1)$ and $(u_2,\bm x_2)$ and the incoming states are inserted at $(u_3,\bm x_3)$ and $(u_4,\bm x_4)$, respectively. 
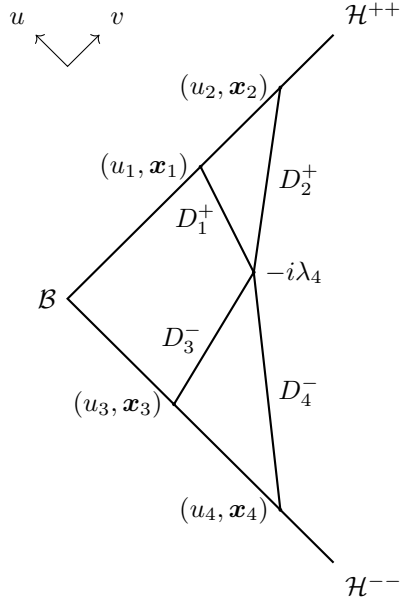
\begin{figure}
  \centering
  \usetikzlibrary {shapes.misc}
  \tikzset{cross/.style={cross out, draw=black, fill=none, minimum size=2*(#1-\pgflinewidth), inner sep=0pt, outer sep=0pt}, cross/.default={3pt}}
  \begin{tikzpicture}[scale=0.7]
  
\newcommand{\arrowIn}{
\tikz \draw[-stealth] (-2pt,0) -- (2pt,0);
}

     \draw[draw,thick]   (5,-5) node[below right]{\footnotesize $\mathcal{H}^{--}$}--(0,0) node[left]{\footnotesize $\mathcal B$}--  (5,5) node[above right]{\footnotesize $\mathcal{H}^{++}$};

    \draw[thick] (2,-2)node[left]{\footnotesize $(u_3,\bm x_{ 3})$}--node[left]{\footnotesize $D_3^-$}(3.5,0.5) node[right]{\footnotesize $-i\lambda_4$}--node[right]{\footnotesize $D_2^+$}(4,4) node[ left]{\footnotesize $(u_2,\bm x_{2})$};
    
    \draw[thick] (2.5,2.5) node[ left]{\footnotesize $(u_1,\bm x_{ 1})$}--node[left]{\footnotesize $D_1^+$}(3.5,0.5) --node[right]{\footnotesize $D_4^-$}(4,-4) node[ left]{\footnotesize $(u_4,\bm x_{4})$};
    
    \draw[<->] (-0.6,5) node[above left] {\footnotesize $u$} -- (0,4.4) -- (0.6,5) node[above right] {\footnotesize $v$};
\fill [fill,use as bounding box](2,-2) circle (1pt);
\fill [fill,use as bounding box](4,4) circle (1pt);
\fill [fill,use as bounding box](4,-4) circle (1pt);
\fill [fill,use as bounding box](3.5,0.5) circle (1pt);
\fill [fill,use as bounding box](2.5,2.5) circle (1pt);

  \end{tikzpicture}
  \caption{\centering{Four-point Carrollian amplitude at tree-level in $\Phi^4$ theory.} }\label{phi4}
\end{figure}
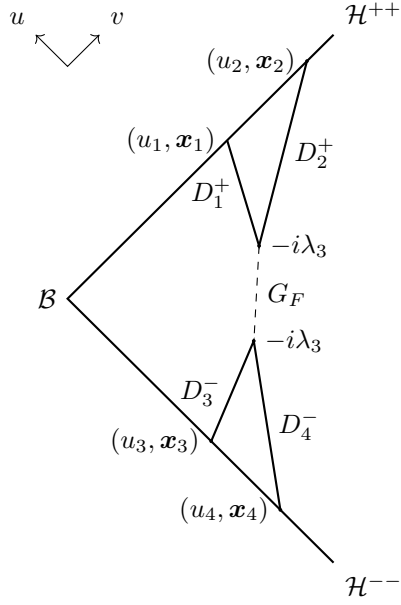
\begin{figure}
  \centering
  \usetikzlibrary {shapes.misc}
  \tikzset{cross/.style={cross out, draw=black, fill=none, minimum size=2*(#1-\pgflinewidth), inner sep=0pt, outer sep=0pt}, cross/.default={3pt}}
  \begin{tikzpicture}[scale=0.7]
  
\newcommand{\arrowIn}{
\tikz \draw[-stealth] (-2pt,0) -- (2pt,0);
}

     \draw[draw,thick]   (5,-5) node[below right]{\footnotesize $\mathcal{H}^{--}$}--(0,0) node[left]{\footnotesize $\mathcal B$}--  (5,5) node[above right]{\footnotesize $\mathcal{H}^{++}$};

    \draw[thick] (2.7,-2.7)node[left]{\footnotesize $(u_3,\bm x_{ 3})$}--node[left]{\footnotesize $D_3^-$}(3.5,-0.8) node[right]{\footnotesize $-i\lambda_3$}--node[right]{\footnotesize $D_4^-$}(4,-4) node[ left]{\footnotesize $(u_4,\bm x_{ 4})$};
    
    \draw[thick] (3,3) node[ left]{\footnotesize $(u_1,\bm x_{ 1})$}--node[left]{\footnotesize $D_1^+$}(3.6,1)node[right]{\footnotesize $-i\lambda_3$} --node[right]{\footnotesize $D_2^+$}(4.5,4.5) node[ left]{\footnotesize $(u_2,\bm x_{2})$};
    
    \draw[dashed] (3.6,1) --node[right]{\footnotesize $G_F$}(3.5,-0.85) ;
    
    \draw[<->] (-0.6,5) node[above left] {\footnotesize $u$} -- (0,4.4) -- (0.6,5) node[above right] {\footnotesize $v$};
\fill [fill,use as bounding box](2.7,-2.7) circle (1pt);
\fill [fill,use as bounding box](4.5,4.5) circle (1pt);
\fill [fill,use as bounding box](4,-4) circle (1pt);
\fill [fill,use as bounding box](3.6,1) circle (1pt);
\fill [fill,use as bounding box](3.5,-0.8) circle (1pt);

  \end{tikzpicture}
  \caption{Four-point Carrollian amplitude at tree-level in $\Phi^3$ theory. This is the ``s'' channel diagram. The ``u'' and ``t'' channels have been omitted.}\label{fourpoint}
\end{figure}
The four-point Carrollian amplitude is\footnote{When there is a cubic interaction, there should be more Feynman diagrams which  are shown in Figure \ref{fourpoint}. These diagrams contain double bulk integrals which are much more involved.  We will not study them in this paper.}
\bea 
&&\langle \Sigma(u_1,\bm x_{ 1},+)\Sigma(u_2,\bm x_{ 2},+)\Sigma(u_3,\bm x_{ 3},-)\Sigma(u_4,\bm x_{ 4},-)\rangle\nn\\&=&-i\lambda_4\int_{-\infty}^\infty d\tau \int_0^\infty \rho d\rho \int d\bm y \prod_{j=1}^4 D(u_j,\bm x_{ j},\sigma_j;\tau,\rho,\bm y)\nn\\&=& -i\frac{\lambda_4}{(4\pi^2)^4}\int_{-\infty}^\infty d\tau\int_0^\infty d\rho \rho \int d\bm y \prod_{j=1}^4 \int_0^\infty d\omega_j
\frac{e^{-i\omega_1(u_1-u)-i\omega_2(u_2-u)-i\omega_3(v-u_3)-i\omega_4(v-u_4)}}{\prod_{j=1}^4(\rho^2+|\bm x_j-\bm y|^2)^{1-i\omega_j}}\nn\\&=&-i\frac{\lambda_4}{128\pi^7}\int_0^\infty d\rho \rho \int d\bm y \prod_{j=1}^4\int_0^\infty d\omega_j\frac{\delta(\omega_1+\omega_2-\omega_3-\omega_4)e^{{ -}i\omega_1 u_1-i\omega_2u_2+i\omega_3 u_3+i\omega_4 u_4}\rho^{-iw}}{\prod_{j=1}^4(\rho^2+|\bm x_j-\bm y|^2)^{1-i\omega_j}}.\nn\\
\eea Switching to the Fourier space, we find 
\bea 
&&\mathcal{T}(\omega_1,\bm x_1,+;\omega_2,\bm x_2,+;\omega_3,\bm x_3,-;\omega_4,\bm x_4,-)\nn\\&=&-\frac{\lambda_4}{8\pi^3}\int_0^\infty d\rho \rho \int d\bm y \frac{\rho^{-i\omega}}{\prod_{j=1}^4(\rho^2+|\bm x_j-\bm y|^2)^{1-i\omega_j}}\nn\\&=&-\frac{\lambda_4}{8\pi^3}\frac{\Gamma(4-iw)}{\prod_{j=1}^4\Gamma(1-i\omega_j)}\int_0^\infty d\rho \rho \int d\bm y\int_0^1 dt_1dt_2dt_3dt_4\frac{\delta(t_1+t_2+t_3+t_4-1)\rho^{-iw} t_1^{-i\omega_1}t_2^{-i\omega_2}t_3^{-i\omega_3}t_4^{{ -}i\omega_4}}{[\sum_{j=1}^4 t_j(\rho^2+|\bm x_j-\bm y|^2)]^{4-iw}}\nn\\&=&-\frac{\lambda_4}{16\pi^3}\frac{\Gamma \left(1-\frac{i w}{2}\right) \Gamma \left(3-\frac{i w}{2}\right)}{\prod_{j=1}^4\Gamma(1-i\omega_j)}\int d\bm y \int_0^1 dt_1 dt_2 dt_3 dt_4 \frac{\delta(t_1+t_2+t_3+t_4-1)\prod_{j=1}^4 t_j^{-i\omega_j}}{(\sum_{j=1}^4 t_j|\bm x_j-\bm y|^2)^{3-\frac{i w}{2}}}
\nn\\&=&\tilde{\lambda}_4\int_0^1 dt_1dt_2dt_3dt_4\frac{\delta(t_1+t_2+t_3+t_4-1)\prod_{j=1}^4 t_j^{-i\omega_j}}{S_4^{2-iw/2}},\label{TT}
\eea where
\bea 
S_4=\sum_{j=1}^4 t_j(1-t_j)\bm x_j^2-2\sum_{j_1<j_2}t_{j_1}t_{j_2}\bm x_{j_1}\cdot\bm x_{j_2}
\eea and 
\bea w=\omega_1+\omega_2+\omega_3+\omega_4,\quad \tilde{\lambda}_4=-\frac{\lambda_4}{16\pi^2}\frac{\Gamma \left(1-\frac{i w}{2}\right) \Gamma \left(2-\frac{i w}{2}\right)}{\prod_{j=1}^4\Gamma(1-i\omega_j)}.
\eea 
\subsubsection{Zero-energy Carrollian amplitude(ZECA)}
We can shift $\bm x_1=0$ by translation invariance. The resulting integral is still hard and we may explore the limit $\omega_1=\omega_2=\omega_3=\omega_4=0$
\bea 
&&\mathcal{T}(0,0,+;0,\bm x_2,+;0,\bm x_3,-;0,\bm x_4,-)\nn\\&=&\tilde{\lambda}_4\int_0^1 dt_1dt_2dt_3dt_4 \frac{\delta(t_1+t_2+t_3+t_4-1)}{S_4^2}\nn\\&=&\tilde{\lambda}_4\int_0^\infty dz_2\int_0^\infty dz_3\int_0^\infty dz_4 \frac{1}{\tilde{S}_4^2}\label{FeynT}
\eea where 
\bea 
\tilde{S}_4&=&z_2(1+z_3+z_4)\bm x_2^2+z_3(1+z_2+z_4)\bm x_3^2+z_4(1+z_2+z_3)\bm x_4^2\nn\\&&-2z_2 z_3\bm x_2\cdot\bm x_3-2z_2 z_4\bm x_2\cdot\bm x_4-2z_3 z_4\bm x_3\cdot\bm x_4.
\eea Note that we have changed the variable $t_j$ to $z_j$
\be 
t_j=\frac{z_j}{1+z_2+z_3+z_4},\quad j=2,3,4\label{changevariable}
\ee in the last step. The integral can be found after some efforts, 
\bea 
&&\mathcal{T}(0,\bm x_1,+;0,\bm x_2,+;0,\bm x_3,-;0,\bm x_4,-)\nn\\&=&\tilde{\lambda}_4\frac{1}{|\bm x_{21}||\bm x_{31}||\bm x_{24}||\bm x_{34}|}\frac{1}{\overline{b}^{-1}-\overline{b}}[\text{Li}_2(1-\frac{\overline{b}}{\overline{a}})-\text{Li}_2(1-\frac{1}{\overline{a}\overline{b}})+\text{Li}_2(1-\overline{a}\overline{b})-\text{Li}_2(1-\frac{\overline{a}}{\overline{b}})-2\log \overline{c}\log \overline{b}]\nn\\
&=&\tilde{\lambda}_4 \frac{1}{|\bm x_{21}||\bm x_{31}||\bm x_{24}||\bm x_{34}|}\frac{4i}{z-\bar  z}\sqrt{z\bar z}P_2(z)\label{4pt}
\eea where we have inserted back $\bm x_1$ by replacing $\bm x_{j}$ to $\bm   x_{j1},\ j=2,3,4$. we have used the complex coordinates 
    \be 
    z_i=x_i+iy_i,\quad \bar z_i=x_i-iy_i
    \ee in the last line and 
    \bea 
    z=\frac{z_{12}z_{34}}{z_{13}z_{24}},\quad \bar z=\frac{\bar z_{12}\bar{z}_{34}}{\bar z_{13}\bar z_{24}}
    \eea are cross ratios. Interestingly, we still find the Bloch-Wigner dilogarithm which is the same as the three-point ZECA, except that the parameters $\overline{a},\overline{b},\overline{c}$ are slightly different
\bea 
\overline{b}+\overline{b}^{-1}=2\frac{\bm x_{21}^2\bm x_{31}^2-\bm x_{21}^2\bm x_{31}\cdot\bm x_{41}-\bm x_{31}^2\bm x_{21}\cdot\bm x_{41}+\bm x_{41}^2\bm x_{21}\cdot\bm x_{31}}{|\bm x_{21}||\bm x_{31}||\bm x_{24}||\bm x_{34}|}=\frac{\bm x_{12}^2\bm x_{34}^2+\bm x_{13}^2\bm x_{24}^2-\bm x_{14}^2\bm x_{23}^2}{|\bm x_{21}||\bm x_{31}||\bm x_{24}||\bm x_{34}|}\nn\\
\eea and 
\bs\begin{align}
\overline{a}&=\frac{|\bm x_{21}||\bm x_{34}|}{|\bm x_{31}||\bm x_{24}|},\\
\overline{c}&=\frac{|\bm x_{21}||\bm x_{31}||\bm x_{24}||\bm x_{34}|}{|\bm x_{41}|^2|\bm x_{23}|^2}.
\end{align}\es 
The three parameters can be expressed as elementary functions of the cross ratios 
    \bea 
    \bar a=\sqrt{z\bar z},\quad \bar b=\sqrt{\frac{z}{\bar z}},\quad \bar c=\frac{\sqrt{z\bar z}}{(1-z)(1-\bar z)}.
    \eea As before, $\bar a$ is the magnitude of the cross ratio while $\bar b$ is the phase of the cross ratio. 
One can find more details  in Appendix \ref{int}.

\paragraph{$S_4$ symmetry.} We could check numerically that the four-point Carrollian amplitude is invariant under the permutation of  the four coordinates. To prove this point, we may define a function 
\bea 
Q(\bm x_1,\bm x_2,\bm x_3,\bm x_4)=\frac{z-\bar z}{4i\sqrt{z\bar z}}|\bm x_{21}||\bm x_{31}||\bm x_{24}||\bm x_{34}|=\frac{1}{4i}(z_{12}z_{34}\bar{z}_{13}\bar{z}_{24}-\bar{z}_{12}\bar{z}_{34}z_{13}z_{24}).
\eea 
Then the four-point ZECA is 
\be 
\mathcal{T}(0,\bm x_1,+;0,\bm x_2,+;0,\bm x_3,-;0,\bm x_4,-)=\tilde{\lambda}_4 \frac{P_2(z)}{Q(\bm x_1,\bm x_2,\bm x_3,\bm x_4)}.\label{T4}
\ee 
This is a signed function which can only flip sign under $S_4$ transformations that is shown in Table \ref{Qua}. Since the Bloch-Wigner dilogarithm transforms in the same way, \eqref{T4} should be invariant under any permutation of the four positions.
\begin{table}
\begin{center}
\renewcommand\arraystretch{1.5}
    \begin{tabular}{|c||c||c|}\hline
$S_4$ transformation on positions& $S_4$ transformation on $z$&Transformation law of $Q$\\\hline\hline
1234,2143,3412,4321&$z$&$+Q$\\\hline
1243,2134,3421,4312&$\frac{z}{z-1}$&$-Q$\\\hline
1324,2413,3142,4231&$\frac{1}{z}$&$-Q$\\\hline
1342,2431,3124,4213&$\frac{1}{1-z}$&$+Q$\\\hline
1423,2314,3241,4132&$\frac{z-1}{z}$&$+Q$\\\hline
1432,2341,3214,4123&$1-z$&$-Q$\\\hline
\end{tabular}
\caption{$S_4$ transformation. In the first column, we use $ijkl$ to represent the transformation $\bm x_1\to \bm x_i,\ \bm x_2\to \bm x_j,\ \bm x_3\to\bm x_k,\ \bm x_4\to\bm x_l$. The transformations that change the variable $z$ in the same way are placed in the same row.}\label{Qua}
\end{center}
\end{table}
\paragraph{Concyclic points.} We assume that any two of the four points do not coincide. The four-point ZECA is ill defined superficially when the  function 
$Q(\bm x_1,\bm x_2,\bm x_3,\bm x_4)=0$ which is also equivalent to  $z=\bar z$. When four points are collinear, then it is easy to show that $Q=0$. Actually, the condition $Q=0$ may also be satisfied  when  the four points are concyclic as a consequence of the (converse of) Ptolemy's theorem\footnote{Please find more details in Appendix \ref{coll}.}, as has been shown in Figure \ref{colic}. In this case, the four-point ZECA becomes 
\bea 
\mathcal{T}(0,\bm x_1,+;0,\bm x_2,+;0,\bm x_3,-;0,\bm x_4,-)=\tilde{\lambda}_4\ |\bm x_{13}|^{-2}|\bm x_{24}|^{-2}\left(\frac{\log z^2}{z-1}-\frac{\log(z-1)^2}{z}\right),
\eea 
where we have used the asymptotic expansion \eqref{collinear}.
As a consistency check, it diverges when any two of the points coincide.
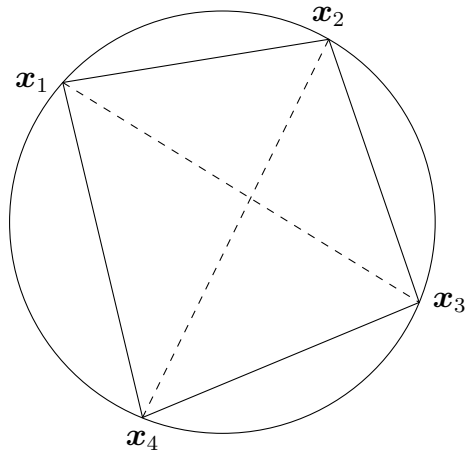
\begin{figure}
  \centering
\begin{tikzpicture}[scale=0.7]
\draw(0,0) circle (4);
\draw (-3,2.646)node[left=1pt]{$\bm x_1$}--(2,3.464)node[above=1pt]{$\bm x_2$}--(3.7,-1.5198)node[right=1pt]{$\bm x_3$}--(-1.5,-3.7)node[below=1pt]{$\bm x_4$}--(-3,2.646);
\draw [dashed](-3,2.646)--(3.7,-1.5198);
\draw [dashed](2,3.464)--(-1.5,-3.7);
\end{tikzpicture}
  \caption{\centering{Conclyclic points.} }\label{colic}
\end{figure}
    \subsubsection{Conformal invariance}  We can find the striking resemblance between the tree-level Carrollian amplitude of massless $\Phi^4$ theory and the correlators of two-dimensional conformal field theory. 
    \begin{itemize}
        \item The two-point Carrollian amplitude \eqref{Tmassless} is the same as the two-point correlation function of a primary operator with dimension $\Delta=1-i\omega$.
        \item The three-point Carrollian amplitude of  $\Phi^4$ theory vanishes identically for the operator $\Sigma(u,\bm x)$.
        \item The four-point Carrollian amplitude \eqref{4pt} satisfies the transformation law 
        \bea 
       && \mathcal{T}(0,\bm x'_1,+;0,\bm x'_2,+;0,\bm x'_3,-;0,\bm x'_4,-)\nn\\&=&\left(\prod_{j=1}^4\Omega(\bm x_j)^{\Delta_j}\right)\mathcal{T}(0,\bm x_1,+;0,\bm x_2,+;0,\bm x_3,-;0,\bm x_4,-),\label{ward}
        \eea where $\Omega(\bm x)$ is the conformal factor which is related to the Jacobian of the conformal transformation of the coordinates 
        \be 
        \Omega(\bm x)=\Big|\frac{\partial\bm x'}{\partial\bm x}\Big|^{-1/2}.
        \ee The primary field $\Sigma(\omega,\bm x)$ has a conformal dimension  $\Delta=1$ in the zero frequency  limit. When $\omega_j\not=0$, one should study the  integral \eqref{TT}. 
        For scaling transformation, 
        \bea 
        \bm x\to \bm x'=\lambda \bm x,
        \eea
        The integral obeys the transformation law 
        \bea 
        &&\mathcal{T}(\omega_1,\bm x'_1,+;\omega_2,\bm x'_2,+;\omega_3,\bm x'_3,-;\omega_4,\bm x'_4,-)\nn\\&=&\left(\prod_{j=1}^4\lambda^{-1+i\omega_j}\right)\mathcal{T}(\omega_1,\bm x_1,+;\omega_2,\bm x_2,+;\omega_3,\bm x_3,-;\omega_4,\bm x_4,-)
        \eea which matches with \eqref{ward}. For special conformal transformation, one can not obtain the transformation law at first sight.
       However,  we can still change the variables $t_j$ to $z_j$ using \eqref{changevariable} and integrate out  $z_2$
        \bea 
        &&\mathcal{T}(\omega_1,\bm x_1,+;\omega_2,\bm x_2,+;\omega_3,\bm x_3,-;\omega_4,\bm x_4,-)\nn\\&=&\tilde{\lambda}_4\int_0^\infty dz_2\int_0^\infty dz_3\int_0^\infty dz_4\frac{z_2^{-i\omega_2}z_3^{-i\omega_3}z_4^{-i\omega_4}}{\tilde{S}_4^{2-iw/2}}\nn\\&=&\tilde{\lambda}'_4\int_0^\infty dz_3\int_0^\infty dz_4 \frac{z_3^{-i\omega_3}z_4^{-i\omega_4}}{(z_3\bm x_{13}^2+z_4\bm x_{14}^2+z_3 z_4 \bm x_{34}^2)^{1-i\omega_1}(\bm x_{12}^2+z_3\bm x_{23}^2+z_4\bm x_{24}^2)^{1-i\omega_2}},\nn\\\label{Lee}
        \eea with 
        \bea 
        \tilde{\lambda}_4'=\tilde{\lambda}_4\frac{\Gamma (1-i \omega_2) \Gamma(1-i\omega_1)}{\Gamma(2-\frac{iw}{2})}.
        \eea In the last line, we have restored the coordinate $\bm x_1$. 
        Now we can change them to two new variables $t,t'$
        \be 
        z_3=\frac{\bm x_{12}^2}{\bm x_{23}^2}t,\quad z_4=\frac{\bm x_{12}^2}{\bm x_{24}^2}t',
        \ee 
        then the four-point Carrollian amplitude becomes 
        \bea 
         && \mathcal{T}(\omega_1,\bm x_1,+;\omega_2,\bm x_2,+;\omega_3,\bm x_3,-;\omega_4,\bm x_4,-)\nn\\&=&\tilde{\lambda}'_4|\bm x_{12}|^{-2+2i\omega_1}|\bm x_{34}|^{-2+2i\omega_1}|\bm x_{23}|^{-2i(\omega_1-\omega_3)}|\bm x_{24}|^{-2i(\omega_1-\omega_4)}\nn\\&&\times\int_0^\infty dt \int_0^\infty dt'\frac{t^{-i\omega_3}t'^{-i\omega_4}}{(1+t+t')^{1-i\omega_2}(tt'+\frac{1}{z\bar z}t+\frac{(1-z)(1-\bar z)}{z\bar z} t')^{1-i\omega_1}}.\label{nonzero}
        \eea 

        The integrand is conformally invariant and then the transformation law of the four-point Carrollian amplitude should satisfy  \eqref{ward}. As a consequence, the tree-level four-point Carrollian amplitude obeys the following Ward identities besides \eqref{Ward}
        \bs\begin{align}
       & \left(\sum_{j=1}^4 x_j^A \frac{\partial}{\partial x_j^A}+\Delta_j\right)\mathcal{T}(\omega_1,\bm x_1,+;\omega_2,\bm x_2,+;\omega_3,\bm x_3,-;\omega_4,\bm x_4,-)=0,\\
      & \left(\sum_{j=1}^4( b^A\bm x_j^2-2\bm b\cdot\bm x_j x_j^A)\frac{\partial}{\partial x_j^A}-2\bm b\cdot\bm x_j \Delta_j\right)\mathcal{T}(\omega_1,\bm x_1,+;\omega_2,\bm x_2,+;\omega_3,\bm x_3,-;\omega_4,\bm x_4,-)=0,\nn\\
        \end{align}\es 
    \end{itemize} where $\bm b$ is a two-dimensional constant vector associates with the special conformal transformations. Stripping off $b^A$, the second Ward identity becomes 
\bea \left(\sum_{j=1}^4 \bm x_j^2 \frac{\partial}{\partial x_{jA}} -2x_{j}^{A} x_j^B\frac{\partial}{\partial x_j^B}-2x_{j}^{A} \Delta_j\right)\mathcal{T}(\omega_1,\bm x_1,+;\omega_2,\bm x_2,+;\omega_3,\bm x_3,-;\omega_4,\bm x_4,-)=0.\eea
    \subsubsection{Non-zero energy Carrollian amplitude (NECA)} We still need to compute the four-point NECA. According to  equation \eqref{Lee}, the relevant integral is 
    \bea 
    &&J(\omega_1,\omega_2,\omega_3,\omega_4;\bm x_{12},\bm x_{13},\bm x_{14},\bm x_{23},\bm x_{24},\bm x_{34})\nn\\&=&\frac{\Gamma \left(1-\frac{i w}{2}\right) \Gamma \left(2-\frac{i w}{2}\right)}{\prod_{j=1}^4\Gamma(1-i\omega_j)}\int_0^\infty \left(\prod_{j\in S} dz_j \right)\frac{\prod_{j\in S} z_j^{-i\omega_j}}{\left(\sum_{j\in S}z_j\bm x_{ 1j}^2+\sum_{j,k\in S,\ j<k}z_jz_k \bm x_{jk}^2\right)^{2-iw/2}}\label{J}
    \eea with the set 
   $ S=\{2,3,4\}.$
    We have inserted back the factors related to the frequency and stripped off the coupling constant.
    Note that this integral is the same as the form of the Lee-Pomeransky representation\cite{Lee:2013hzt} of the usual Feynman integrals. In Lee-Pomeransky representation, any $L$-loop momentum space Feynman integral 
    \bea 
    \int \prod_{k=1}^\ell\frac{d^D\ell_k}{i\pi^{d/2}}\prod_{j=1}^n\frac{1}{(-q_j^2+m_j^2)^{\nu_j}}
    \eea can be represented as \cite{Weinzierl:2022eaz}
    \bea 
    \frac{\Gamma(\frac{D}{2})}{\Gamma(\frac{(L+1)D}{2}-\nu)\prod_{j=1}^n \Gamma(\nu_j)} \int\left(\prod_{j=1}^{n} dz_j\right)\left(\prod_{j=1}^n z_j^{\nu_j-1}\right)\mathcal{G}^{-D/2},\label{Lee2}
    \eea 
    where $\ell_k$ is the loop momentum and $q_j$ is linear superposition of the loop and external momenta. $D$ is the spacetime dimension and  the parameter $\nu$ is the  summation $\nu=\sum_{j=1}^n \nu_j$. The function $\mathcal{G}$ is summation of the first and second Symanzik polynomials which are homogeneous in the Schwinger parameters with degree $L$ and $L+1$, respectively. Comparing \eqref{J} with the above representation, we should identify 
    \bea 
    L=1,\quad \nu_j=1-i\omega_{j+1},\quad D=4-iw,\quad n=3
    \eea and
    \bea 
    \mathcal{G}=\sum_{j\in S}z_j \bm x_{1j}^2+\sum_{j,k\in S,\ j<k}z_j z_k \bm x_{jk}^2.\label{pG}
    \eea Then \eqref{J} is exactly \eqref{Lee2} up to a Gamma function $\Gamma(1-iw/2)$.  However, we still find an amusing fact that the dimension $D$ and $\nu_j$ are complex number since the energies are non-negative $\omega_j\ge 0$ in general. In the dimensional regularization of the standard Feynman integrals, one is often interested in the integer $\nu_j$ and dimension $D$ plus small $\epsilon$ correction. Moreover, the polynomial \eqref{pG} is given in position space which is the dual space of the original Lee-Pomeransky representation. Nevertheless, it has be shown \cite{delaCruz:2019skx,Klausen:2019hrg} that the Feynman integrals can be expressed as GKZ hypergeometric function. In our case, we have six variables $\bm x_{12}^2,\bm x_{13}^2,\bm x_{14}^2,\bm x_{23}^2,\bm x_{24}^2,\bm x_{34}^2$ and the GKZ hypergeometric function is associated with the $4\times 6$ matrix $\mathbb{A}$ and the vector $\bm c$ with four components
    \bea 
    \mathbb{A}=\left(\begin{array}{cccccc}1&1&1&1&1&1\\
    1&0&0&1&1&0\\
    0&1&0&1&0&1\\
    0&0&1&0&1&1\end{array}\right),\quad \bm c=\left(\begin{array}{c}-2+iw/2\\ -1+i\omega_2\\-1+i\omega_3\\-1+i\omega_4\end{array}\right).
    \eea 
    The solution is the superposition of the Appell function of the fourth kind
    \bea 
 \int_0^\infty \left(\prod_{j\in S} dz_j \right)\frac{\prod_{j\in S} z_j^{-i\omega_j}}{\left(\sum_{j\in S}z_j\bm x_{ 1j}^2+\sum_{j,k\in S,\ j<k}z_jz_k \bm x_{jk}^2\right)^{2-iw/2}}=K_1\phi_1+K_2\phi_2+K_3\phi_3+K_4\phi_4,\nn\\
    \eea where the constants $K_i$ and the  Appell functions $\phi_i$ can be found in \cite{delaCruz:2019skx}. Therefore, the tree-level four-point Carrollian amplitude has been solved analytically in principle.  In the following, we transform the result to another integral representation in which the relation between three-point and four-point Carrollian amplitude becomes transparent. 
   We start from the integral
    \eqref{nonzero} and introduce two new variables 
    \bea 
    s=\frac{{\bm x_{14}^2\bm x_{23}^2}}{\bm x_{12}^2\bm x_{34}^2}\frac{1}{t},\quad s'=t',
    \eea 
    the four-point Carrollian amplitude becomes
    \bea 
     &&\hspace{-1cm}\mathcal{T}(\omega_1,\bm x_1,+;\omega_2,\bm x_2,+;\omega_3,\bm x_3,-;\omega_4,\bm x_4,-)\nn\\\hspace{-0.8cm}&\hspace{-1.5cm}=&\hspace{-1cm}\tilde{\lambda}_4|\bm x_{12}|^{-2+2i\omega_3}|\bm x_{34}|^{-2+2i\omega_3}|\bm x_{24}|^{-2i(\omega_1-\omega_4)}|\bm x_{14}|^{2i(\omega_1-\omega_3)}\int_0^\infty ds \int_0^\infty ds'\frac{s^{-i\omega_4}s'^{-i\omega_4}}{(ss'+s+{\tt V})^{1-i\omega_2}(ss'+s'+{\tt U})^{1-i\omega_1}}\nn\\
     \hspace{-0.8cm}&\hspace{-1.5cm}=&\hspace{-1cm}\tilde{\lambda}_4|\bm x_{12}|^{-2+2i\omega_3}|\bm x_{34}|^{-2+2i\omega_3}|\bm x_{24}|^{-2i(\omega_1-\omega_4)}|\bm x_{14}|^{2i(\omega_1-\omega_3)}\mathcal{I}(-i\omega_4,-i\omega_4,1-i\omega_1,1-i\omega_2;{\tt U},{\tt V}),\label{treeT}
    \eea where we have used the integral representation of the function $\mathcal{I}$ defined in \eqref{funI}. This integral has been computed analytically in Appendix \ref{int} using Mellin-Barnes type integrals. The result is also expressed as the superposition of Appell function of the fourth kind. The variables ${\tt U}$ and ${\tt V}$ are related to the cross ratio
    \be 
    {\tt U}=\frac{1}{z\bar z},\quad {\tt V}=\frac{(1-z)(1-\bar z)}{z\bar z}.
    \ee 
    Interestingly, the four-point Carrollian amplitude in $\Phi^4$ theory and the three-point Carrollian amplitude in $\Phi^3$ theory still have the same form even when the frequencies are non-vanishing. Moreover, in the limit of $\bm x_2\to \infty$, the variables ${\tt U}$ and ${\tt V}$ are mapped to ${\tt u}$ and ${\tt v}$ ,respectively \footnote{Here the three-point Carrollian amplitude in $\Phi^3$ theory is built from the three operators inserted at $\bm x_1,\bm x_3$ and $\bm x_4$. Therefore, one should replace $\bm x_2\to \bm x_3,\ \bm x_3\to \bm x_4$ in the definition of ${\tt u}$ and ${\tt v}$.}
    \be 
    \lim_{\bm x_2\to\infty}{\tt U}=\frac{\bm x_{13}^2}{\bm x_{34}^2}={\tt u},\quad \lim_{\bm x_2\to\infty}{\tt V}=\frac{\bm x_{14}^2}{\bm x_{34}^2}={\tt v}.
    \ee When $\omega_2=0$, the conservation of the energy leads to the equation 
    \be 
    \omega_1=\omega_3+\omega_4.
    \ee Then we reproduce the  three-point Carrollian amplitude of $\Phi^3$ theory from the four-point Carrollian amplitude of $\Phi^4$ theory by setting the energy of an external leg to be zero
   \bea 
    \lim_{\bm x_2\to\infty,\ \omega_2\to 0}|\bm x_2|^2\ \mathcal{T}(\omega_2,\bm x_2,+;\omega_1,\bm x_1,+;\omega_4,\bm x_4,-;\omega_3,\bm x_3,-)=\mathcal{F}\times \mathcal{T}(\omega_1,\bm x_1,+;\omega_3,\bm x_3,-;\omega_4,\bm x_4,-).\nn\\\label{soft}
    \eea
    
    The factor $\mathcal{F}$ is independent of the position and is completely determined by the quotient of $\tilde{\lambda}_4'$ and $\tilde{\lambda}_3$. We will call $\mathcal{F}$ the soft form factor due to its striking resemblance to the Weinberg's soft theorem \cite{Weinberg:1965nx} in momentum space. Physically, when a particle becomes soft, we may move it far away and deduce its effect on the scattering amplitude up to a proportional form factor. Then the four-point Carrollian amplitude becomes an effective three-point Carrollian amplitude in the soft limit. In fact, the above equation{{ is a relation between the soft limit of four-point Carrollian amplitude of $\Phi^4$ theory and the three-point Carrollian amplitude of $\Phi^3$ theory. A real soft theorem should connect the soft limit of $n$-point Carrollian amplitude and $(n-1)$-point one in the same theory. We will leave the soft theorem on higher-point Carrollian amplitude  for further study.

\section{Conclusion}\label{conc}
In this work, we have studied the propagators and Carrollian amplitudes in Rindler spacetime. The boundary-to-boundary propagator and the bulk-to-boundary propagator have not been presented in the literature. We can use the bulk-to-boundary propagator to reconstruct the Feynman propagator in the bulk using the split representation. We have also computed the tree-level three-point Carrollian amplitude in $\Phi^3$ and four-point Carrollian amplitude in $\Phi^4$ massless scalar theory. Curiously, their forms are the same up to some kinematic factors in the zero energy limit. In general, the Carrollian amplitude in RRW  breaks the Poincar\'e group to $\text{SO}(1,1)\times\text{ISO}(2)$ in order to fix the position of the Rindler horizon. However, the tree-level four-point Carrollian amplitude in $\Phi^4$ theory preserves a larger symmetry group $\text{SO}(1,1)\times\text{SO}(1,3)$ in which the $\text{SO}(1,3)\simeq SL(2,\mathbb{C})$ is the M$\ddot{o}$bius transformation of the two-dimensional transverse plane. This may not be a surprise since the $\Phi^4$ theory has no dimensional parameters classically. It would be natural to conjecture that the tree-level $\Phi^4$ theory in the RRW is dual to an Euclidean CFT in the transverse plane which is  the Fourier transform of the Carrollian field theory. It is interesting to study the higher-point Carrollian amplitude to check this conjecture.   
\begin{figure}
  \centering
  \usetikzlibrary{decorations.text}
\begin{tikzpicture}

\coordinate (o) at (-3,-2);
\coordinate (a) at (0,3);
\coordinate (b) at (3,-2);
\draw (a)node[above=9pt]{\footnotesize Momentum Space}node[above]{\footnotesize $(\omega,\bm {k})$}--(b)node[below]{\footnotesize ``Celestial'' Space}node[below=9pt]{\footnotesize $(\omega,\bm x)$}--node[above]{\footnotesize Fourier transform }(o)node[below]{\footnotesize Carrollian Space}node[below=9pt]{\footnotesize $(u,\Vec{x})$}--(a);

 \path [decorate,decoration={text along path,text align=center,text={},raise=4pt}](-3,-2)--(0,3);
 \path [decorate,decoration={text along path,text align=center,text={},raise=4pt}](0,3)--(3,-2);
\end{tikzpicture}
  \caption{The triangle among momentum space amplitude, Carrollian amplitude and ``celestial amplitude''. In momentum space, the states are labeled by $(\omega,\bm k)$. This is transformed to Carrollian space which is labeled by $(u,\bm x)$. A Fourier transform on the state in Carrollian representation leads to the state $(\omega,\bm x)$. The integral transforms between momentum space amplitude and Carrollian/``celestial'' amplitude have been omitted.  }\label{amplitude}
\end{figure}
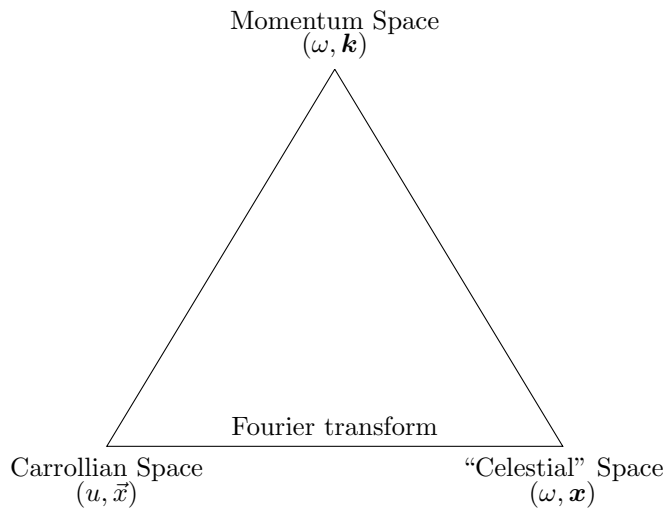

\begin{enumerate}
    \item It would be rather interesting to further study the Carrollian amplitude with non-zero frequency. The tree-level four-point Carrollian amplitude  \eqref{treeT} is already   nontrivial and has the loop integral structure of momentum space Feynman diagrams. This is also crucial for us to explore the conformal invariance of the dual Euclidean field theory where the field operator $\Sigma(\omega,\bm x)$ has the conformal dimension $\Delta=1-i\omega$ with $\omega\ge 0$ as a consequence of the two-point Carrollian amplitude. We notice that the conformal dimension is lying on the  principal continuous series \cite{unitary,Bargmann:1946me} of the M$\ddot{o}$bius group $SL(2,\mathbb{C})$, which is also the one in celestial holography \cite{Pasterski:2017kqt}. Therefore, one may borrow the method of celestial holography to study the operator $\Sigma(\omega,\bm x)$ and its correlators. For example, we may study the OPE expansion of the operator using the inversion formula \cite{Atanasov:2021cje}. However, we should emphasize that  $\Sigma(\omega,\bm x)$ is the Fourier transform of the Carrollian field, while the primary operator in celestial holography comes from the Mellin transform. Nevertheless, regarding the Fourier transform of the Carrollian amplitude as a ``celestial'' amplitude, then the momentum space amplitude, Carrollian amplitude and ``celestial'' amplitude form a triangle, as shown in Figure \ref{amplitude}. The triangle is the analog of the one in \cite{Donnay:2022wvx}. Note that the Rindler coordinates $(u,\bm x)$ are related to the null coordinates $(U,\bm X)$ in Minkowski spacetime through the relation 
    \be 
    U=-e^{-u},\quad \bm X=(X,Y)=\bm x,
    \ee we may rewrite the field $\Sigma(u,\bm x)$ as 
    \be 
    \widetilde{\Sigma}(-U,\bm X)=\Sigma(u,\bm x).
    \ee Then the Fourier transform \eqref{Sf} may switch to the Mellin transform
    \bea 
    \widetilde{\Sigma}(\omega,\bm X)=\Sigma(\omega,\bm x)=\int_{-\infty}^0 d(-U)(-U)^{-1+i\omega}\widetilde{\Sigma}(-U,\bm X)=\int_0^\infty d\widetilde{U} \widetilde{U}^{-1+i\omega}\widetilde{\Sigma}(\widetilde{U},\bm X)
    \eea where $\widetilde{U}=-U$. The formula matches with the one in celestial holography \cite{Pasterski:2017kqt} by the following one-to-one correspondence  \be i\omega\quad\Leftrightarrow\quad \text{conformal dimension in celestial holography}
    \ee and 
    \be 
    \widetilde{U}\quad\Leftrightarrow\quad \text{frequency in celestial holography}.
    \ee 
    Note that the $\Delta'=i\omega$ may also be identified as a conformal dimension from Mellin transform and should be distinguished from $\Delta=1-i\omega$.
    It is rather interesting that the state associated with $\widetilde{\Sigma}(\widetilde{U},\bm X)$ is in the  position space and it maps to the state in the momentum space of celestial holography.
    After the transformation, the field $\Sigma(\omega,\bm x)$ becomes a local operator in the transverse plane. 
    At the tree-level of $\Phi^4$ theory, the symmetry group is enhanced from $SO(1,1)\times ISO(2)$ to $SO(1,3)$. The latter is isomorphic to a two dimensional conformal group and the Carrollian null vector $\partial_u$ becomes the dilatation of this enhanced group. The conformal dimension of $\Sigma(\omega,\bm x)$ transmutes to  $1-i\omega$.

    \item There is an hint \eqref{soft} on the ``soft theorem'' in the amplitude of Rindler spacetime. It would be nice to check the  ``soft theorem'' in the Rindler wedge by considering general $n$-point Carrollian amplitude in the Fourier space. 
    \item Loop correction and symmetry breaking. The loop correction of Carrollian amplitude has been explored in \cite{Liu:2024nfc} in Minkowski spacetime and the method can be applied directly to Rindler spacetime. It would be interesting to understand the  symmetry breaking of the conformal group $\text{SO}(1,3)$ at the loop-level for non-conformal field theories.
    \item In the literature, the Green's functions in Rindler wedge have also been studied for the theories with other spins \cite{Candelas:1978gg,Soffel:1980kx,Hacian:1985gqu}. We can extend the computation of Rindler Carrollian amplitude to these theories. This may open a new window on the Rindler perturbation theory in Minkowski vacuum \cite{Fulling:1972md,Davies:1974th,Candelas:1976jv,Dowker:1977zj,Troost:1978yk, Hill:1986ec, Burgess:2018sou} and even the field theory in curved spacetime \cite{Birrell:1982ix,DeWitt:1975ys}.
    \item  Massive Carrollian field theory. The Carrollian holography works well for massless scattering. However, there could be serious problems when incoming and outgoing particles are massive. Classically, a massless light ray reaches null infinity while a massive particle  arrives at timelike infinity. This is the obstacle to define the massive Carrollian amplitude in asymptotically flat spacetime. One can find recent efforts on this topic in \cite{Borthwick:2024skd, Have:2024dff,  Duary:2024kxl,Figueroa-OFarrill:2021sxz,Campiglia:2024uqq,Kulp:2024scx}. However, when we consider a portion of Minkowski spacetime, the null boundaries are not located at infinity and the Carrollian holography is still valid for massive particles. In our paper, we have shown that the massive propagators and amplitudes are well defined in Rindler spacetime and it is expected to be valid for more general spacetime where future/past null infinity are not important. We believe that a complete Carrollian holography should include the massive fields, even for the case that the future/past null infinity plays an important role.   
    \item Thermal Carrollian field theory. The propagator in Minkowski vacuum is a thermal propagator from the perspective of an accelerating observer. This fact strongly supports the idea that one can also define the dual thermal Carrollian field theory using thermal Carrollian amplitude. We can construct the thermal correlators 
    \be 
    \langle \prod_{j=1}^n \Sigma(u_j,\bm x_{j},\sigma_j)\rangle_{\beta}=\frac{\text{Tr}\left(e^{-\beta H}\prod_{j=1}^n \Sigma(u_j,\bm x_{j},\sigma_j)\right)}{\text{Tr}e^{-\beta H}}
    \ee at the boundary from Feynman rules in the bulk thermal field theory.
    The thermal propagators should still satisfy the KMS condition. It is interesting to extend the 
 imaginary time \cite{Matsubara:1955ws}, real-time  \cite{1957JPSJ...12..570K,Martin:1959jp} and the thermo field dynamics  \cite{Umezawa:1982nv} formulations to Carrollian field theories. 
\end{enumerate}




\vspace{10pt}
{\noindent \bf Acknowledgments.}
The work of J.L. was supported by NSFC Grant No. 12005069. The work of J.-L. Yang is supported by ``the Fundamental Research Funds for the Central Universities'' with
No. YCJJ20242112.

\appendix
\section{The null infinity of RRW}\label{rindlerconformal}
We will discuss the $\mathscr{I}_R^+$ for RRW. Using the coordinate $\bar U$ in \eqref{barU}, we can obtain the identity 
\be 
\bar U^2-2\rho \bar U\sinh\tau=\rho^2+X^2+Y^2.
\ee 
The right hand side tends to $\infty$ since at least one of the coordinates $\rho, X,Y$ tends to $\infty$ when approaching $\mathscr{I}^+$. Similarly, at least one of the coordinates $\rho, \tau$ tends to $\infty$ since $T=\rho\sinh\tau$ tends to  $\infty$. Therefore, the first term on the left hand side may be ignored for our discussion since it is finite. Given that $\bar U$ is finite, we may discuss it case by case. 
\begin{enumerate}
    \item $\rho\to\infty$ with \be X^2+Y^2=o(\rho^2).\ee  In this case, $\rho^2$ is dominant and 
    \be 
    \bar U\sim-e^{-\tau+\log\rho}.
    \ee Therefore, the Rindler time $\tau\to\infty$ and we should keep 
    $ u=\tau-\log\rho
    $ finite. 
    \item $\rho\to\infty$ with 
    \be 
    X^2+Y^2=c\rho^2+o(\rho^2),\quad c>0.
    \ee In this case, the term $\rho^2$ and $X^2+Y^2$ are of the same order and 
    \be 
    \bar U\sim -(1+c)e^{-\tau+\log\rho}.
    \ee  Then we still require $\tau\to\infty$ with $u$ finite although the coefficient before the exponential becomes $-(1+c)$.
    \item $X^2+Y^2\to\infty$ with 
    \be 
    \rho^2=o(X^2+Y^2).
    \ee Now the term $X^2+Y^2$ is dominant and we find 
    \be 
    \bar U\sim -\frac{X^2+Y^2}{2\rho\sinh\tau}.
    \ee We should take $X^2+Y^2\to\infty,\ \rho\sinh\tau\to\infty$ and keep their ratio finite.
\end{enumerate}

\section{Killing vectors in Rindler coordinates}\label{kv}
The ten Killing vectors can be written 
\bs\begin{align}
    \bm\xi_T&=e^u\partial_u-\frac{1}{2\rho}(-e^{-u}+\rho^2e^u)\partial_\rho,\\
    \bm\xi_X&=\partial_x,\\
    \bm\xi_Y&=\partial_y,\\
    \bm\xi_Z&=-e^u \partial_u+\frac{1}{2\rho}(e^{-u}+\rho^2 e^u)\partial_\rho,\\
    \bm\xi_{XY}&=x\partial_y-y\partial_x,\\
    \bm\xi_{YZ}&=y[-e^u \partial_u+\frac{1}{2\rho}(e^{-u}+\rho^2 e^u)\partial_\rho]-\frac{1}{2}(-e^{-u}+\rho^2 e^u)\partial_y,\\
    \bm\xi_{XZ}&=x[-e^u \partial_u+\frac{1}{2\rho}(e^{-u}+\rho^2 e^u)\partial_\rho]-\frac{1}{2}(-e^{-u}+\rho^2 e^u)\partial_x,\\
    \bm\xi_{TX}&=\frac{1}{2}(-e^{-u}+\rho^2 e^u)\partial_x+x[e^u\partial_u-\frac{1}{2\rho}(-e^{-u}+\rho^2e^u)\partial_\rho],\\
    \bm\xi_{TY}&=\frac{1}{2}(-e^{-u}+\rho^2 e^u)\partial_y+y[e^u\partial_u-\frac{1}{2\rho}(-e^{-u}+\rho^2e^u)\partial_\rho],\\
    \bm\xi_{TZ}&=\partial_u
\end{align}\es in advanced coordinates and 
\bs\begin{align}
    \bm\xi_T&=e^{-v}\partial_v-\frac{1}{2\rho}(e^v-\rho^2e^{-v})\partial_\rho,\\
    \bm\xi_X&=\partial_x,\\
    \bm\xi_Y&=\partial_y,\\
    \bm\xi_Z&=e^{-v}\partial_v+\frac{1}{2\rho}(e^v+\rho^2e^{-v})\partial_\rho,\\
    \bm\xi_{XY}&=x\partial_y-y\partial_x,\\
    \bm\xi_{YZ}&=y[e^{-v}\partial_v+\frac{1}{2\rho}(e^v+\rho^2e^{-v})\partial_\rho]-\frac{1}{2}(e^v+\rho^2 e^{-v})\partial_y,\\
    \bm\xi_{XZ}&=x[e^{-v}\partial_v+\frac{1}{2\rho}(e^v+\rho^2e^{-v})\partial_\rho]-\frac{1}{2}(e^v+\rho^2 e^{-v})\partial_x,\\
    \bm\xi_{TX}&=\frac{1}{2}(e^v-\rho^2 e^{-v})\partial_x+x[e^{-v}\partial_v-\frac{1}{2\rho}(e^v-\rho^2e^{-v})\partial_\rho],\\
    \bm\xi_{TY}&=\frac{1}{2}(e^v-\rho^2 e^{-v})\partial_y+y[e^{-v}\partial_v-\frac{1}{2\rho}(e^v-\rho^2e^{-v})\partial_\rho],\\
    \bm\xi_{TZ}&=\partial_v
\end{align}\es in retarded coordinates.
\section{Propagators in general dimensions}\label{massived}
This section is a collection of various propagators of massive/massless scalars in general dimensions. In $d$ dimensions, 
the mode expansion of the bulk field in RRW is \eqref{modePhi} with
\bea 
\chi_{\omega,\bm k}(\rho)=\sqrt{\frac{4\sinh\pi \omega }{(2\pi)^d}}K_{i\omega}(\bar{k}_{}\rho).
\eea Therefore, we find the following the mode expansion of the boundary fields
\begin{align}
    \Sigma(u,\bm x_{},\sigma)&= 
\int_0^\infty \frac{d\omega}{\sqrt{4\pi\omega}}\int_{-\infty}^{\infty}\frac{d\bm k_{}}{\sqrt{(2\pi)^{d-2}}}[a_{\omega,\bm k_{},\sigma}^{} e^{-i\omega u+i\bm k_{}\cdot\bm x_{}}+a_{\omega,\bm k_{},\sigma}^{\dagger} e^{i\omega u-i\bm k_{}\cdot\bm x_{}}],
\end{align} where the creation and annihilation operators are given in \eqref{amap} as four dimensions.
\paragraph{Boundary-to-boundary propagator.}
The boundary-to-boundary propagator for a  massive scalar theory is
\bea 
&&\langle \Sigma(u_1,\bm x_{ 1},+)\Sigma(u_2,\bm x_{ 2},-)\rangle\nn\\&=&-i\frac{1}{(2\pi)^{d/2}}\left(\frac{m}{|\bm x_{ 1}-\bm x_{ 2}|}\right)^{d/2-1}\int_0^\infty \frac{d\omega}{\Gamma(1-i\omega)}\left(\frac{2|\bm x_{ 1}-\bm x_{ 2}|}{m}\right)^{i\omega} e^{-i\omega(u_1-u_2)}K_{\frac{d}{2}-1-i\omega}(m|\bm x_{1}-\bm x_{2}|).\nn\\
\eea 
To prove this result, we can substitute the mode expansion
\bea 
&&\text{LHS}\nn\\&=&\int_0^\infty \frac{d\omega}{\sqrt{4\pi\omega}}\int_{-\infty}^{\infty}\frac{d\bm k}{\sqrt{(2\pi)^{d-2}}}e^{-i\omega u_1+i\bm k\cdot\bm x_{1}} \int_0^\infty \frac{d\omega'}{\sqrt{4\pi\omega'}}\frac{d\bm k'}{\sqrt{(2\pi)^{d-2}}}e^{i\omega' u_2-i\bm k'\cdot\bm x_{ 2}}\langle a_{\omega,\bm k,+}a^\dagger_{\omega,\bm k,-}\rangle\nn\\&=&\int_0^\infty d\omega \int_{-\infty}^{\infty}d\bm k \frac{\sinh\pi\omega\Gamma(i\omega)^2}{(2\pi)^d}(\frac{\bar k}{2})^{-2i\omega} e^{-i\omega(u_1-u_2)+i\bm k\cdot(\bm x_{1}-\bm x_{ 2})}\nn\\&=&\int_0^\infty d\omega \int_0^\infty k^{d-3}dk \int_0^\pi \sin^{d-4}\theta d\theta \Omega_{d-4} \frac{\sinh(\pi\omega)\Gamma(i\omega)^2}{(2\pi)^d}(\frac{\bar k}{2})^{-2i\omega} e^{-i\omega(u_1-u_2)+i k|\bm x_{ 1}-\bm x_{2}|\cos\theta}\nn\\
&=&\frac{1}{(2 \pi )^{\frac{d}{2}+1}(|\bm x_{1}-\bm x_{ 2}|)^{\frac{d}{2}-2}}\int_0^\infty d\omega \int_0^\infty k^{\frac{d}{2}-1}dk \sinh (\pi  \omega ) \Gamma (i \omega )^2  2^{2i\omega}\bar k^{-2i\omega}e^{-i\omega(u_1-u_2)}J_{\frac{d-4}{2}}(k|\bm x_{1}-\bm x_{ 2}|)\nn\\&=&\text{RHS}.
\eea In the third step, we have used the spherical coordinates in $(d-2)$-dimensional momentum space. The relative angle between $\bm k$ and $\bm x_{ 1}-\bm x_{2}$ is denoted as $\theta$. The solid angle of the  unit sphere $S^{d-4}$ is 
\bea 
\Omega_{d-4}=\frac{2\pi^{(d-3)/2}}{\Gamma((d-3)/2)}.
\eea In the fourth step, we have used the integral representation of Bessel function of order $\nu$
\bea 
J_\nu(x)=\frac{(x/2)^{\nu}}{\sqrt{\pi}\Gamma(\nu+1/2)}\int_0^\pi d\theta (\sin\theta)^{2\nu}\cos(x\cos\theta).
\eea In the last step, we have used the following definite integral
\bea 
\int_0^\infty dk k^{\nu+1}(k^2+m^2)^{-1-\mu}J_\nu(k r)=\frac{2^{-\mu } r^{\mu } m^{\nu -\mu } K_{\mu -\nu }(m r)}{\Gamma (\mu +1)},\\  r>0,\quad \text{Re}(\nu)>-1,\quad \text{Re}(2 \mu -\nu )>-\frac{3}{2}.\nn
\eea
In our case, we have 
\be  \mu=i\omega -1,\nu=d/2-2,2\mu-\nu = 2i\omega - d/2 , Re(2\mu-\nu) = -d/2>-3/2\quad\Rightarrow\quad 2<d<3.\ee We should compute the integral for $2<d<3$ at first and then extend it to general dimensions by analytic continuation. Note that the Modified Bessel function of the second type is even under the exchange of the order 
\be 
K_\nu(x)=K_{-\nu}(x).
\ee 
The two-point Carrollian amplitude in the Fourier space is 
\bea 
\hspace{-3pt}\mathcal{T}(\omega,\bm x_{1},+;\omega,\bm x_{ 2},-)=-\frac{1}{(2\pi)^{d/2-2}}\frac{2^{i\omega}m^{d/2-1-i\omega}}{\Gamma(1-i\omega)}\frac{1}{|\bm x_{1}-\bm x_{ 2}|^{d/2-1-i\omega}}K_{\frac{d}{2}-1-i\omega}(m|\bm x_{1}-\bm x_{2}|).
\eea 
In the massless limit, the boundary-to-boundary propagator becomes
\bea 
\langle\Sigma(u_1,\bm x_{ 1},+)\Sigma(u_2,\bm x_{ 2},-)\rangle=-\frac{i}{4\pi^{d/2}}\int_0^\infty d\omega \frac{\Gamma(\frac{d}{2}-1-i\omega)}{\Gamma(1-i\omega)}|\bm x_{ 1}-\bm x_{ 2}|^{-d+2+2 i \omega }{ e^{-i\omega(u_{1}-u_{2})}}.\label{masslessd}
\eea In the Fourier space, the boundary-to-boundary propagator is non-vanishing only for $\omega_1=\omega_2=\omega$
\bea 
\mathcal{T}(\omega,\bm x_{ 1},+;\omega,\bm x_{2},-)=-\frac{1}{\pi^{d/2-2}}\frac{\Gamma(\frac{d}{2}-1-i\omega)}{\Gamma(1-i\omega)}\frac{1}{|\bm x_{1}-\bm x_{ 2}|^{d-2-2 i \omega }}.
\eea In the limit $m|\bm x_{1}-\bm x_{2}|\gg 1$, the boundary-to-boundary propagator in Fourier space decays exponentially
\bea 
\mathcal{T}(\omega,\bm x_{ 1},+;\omega,\bm x_{2},-)\sim -\frac{1}{\pi^{d/2-5/2}}\frac{(m/2)^{\frac{d}{2}-\frac{3}{2}-i\omega} }{\Gamma (1-i \omega )} \frac{1}{|\bm x_{1}-\bm x_{2}|^{d/2-1/2-i\omega}}e^{-m|\bm x_{1}-\bm x_{2}|}+\cdots.
\eea 
\paragraph{Bulk-to-boundary propagator.}
The propagator from bulk to $\mathcal{H}^{++}$ is 
\bea
&& D(u,\bm x_{},+;\tau',\rho',\bm x'_{})\nn\\\hspace{-1.2cm}&=&\frac{2}{(2\pi)^d}\int_0^\infty d\omega \int_{-\infty}^{\infty}d\bm k (\frac{\bar k}{2})^{-i\omega}\sinh\pi\omega \Gamma(i\omega)K_{i\omega}(\bar k \rho')e^{-i\omega(u-\tau')+i\bm k\cdot(\bm x_{}-\bm x'_{})}\nn\\&=&\frac{2}{\pi ^{\frac{d}{2}+1} |\bm x_{}-\bm x'_{}|^{d/2-2}} \int_0^\infty d\omega \int_0^\infty dk \bar k^{-i \omega } k^{d/2-1} 2^{-\frac{d}{2}+i \omega -1} \sinh\pi\omega \Gamma(i\omega) e^{-i \omega  (u-\tau')}K_{i\omega}(\bar k\rho')J_{\frac{d-4}{2}}(k |\bm x_{}-\bm x'_{}|)\nn\\&=&-\frac{i}{(2\pi)^{d/2}}\int_0^\infty d\omega \frac{2^{i\omega}e^{-i\omega(u-u')}}{\Gamma(1-i\omega)}(\frac{m}{\sqrt{\rho'^2+|\bm x_{}-\bm x'_{}|^2}})^{\frac{d}{2}-1-i\omega}K_{\frac{d}{2}-1-i\omega}(m\sqrt{\rho'^2+|\bm x_{}-\bm x'_{}|^2}).\nn\\
\eea In the last step, we have used the integral \cite{2007tisp.book.....G}
\bea 
\int_0^\infty{ dx} J_\nu(b x)K_\mu(a\sqrt{z^2+x^2})x^{\nu+1}(x^2+z^2)^{-\frac{\mu}{2}}=\frac{b^\nu}{a^\mu}(\frac{\sqrt{a^2+b^2}}{z})^{\mu-\nu-1}K_{\mu-\nu-1}(z\sqrt{a^2+b^2}),\nn\\
\text{Re}\ \nu>-1,\quad a>0,\quad b>0,\quad |\text{arg}z|<\frac{\pi}{2}.
\eea In the massless limit, the bulk-to-boundary propagator becomes 
\bea 
&&D(u,\bm x_{},+;\tau',\rho',\bm x'_{})\nn\\&=&-\frac{i}{4\pi^{d/2}}\int_0^\infty d\omega \frac{\Gamma(\frac{d}{2}-1-i\omega)}{\Gamma(1-i\omega)} \frac{e^{-i\omega(u-u')}}{(\rho'^2+|\bm x_{}-\bm x'_{}|^2)^{d/2-1-i\omega}}.\label{dbulk}
\eea We also find the massive  propagator  from bulk to $\mathcal{H}^{--}$
\bea 
&&D(u,\bm x,-;\tau',\rho',\bm x')\nn\\&=&-\frac{i}{(2\pi)^{d/2}}\int_0^\infty d\omega \frac{2^{i\omega}e^{-i\omega(v'-u)}}{\Gamma(1-i\omega)}(\frac{m}{\sqrt{\rho'^2+|\bm x-\bm x'|^2}})^{\frac{d}{2}-1-i\omega}K_{\frac{d}{2}-1-i\omega}(m\sqrt{\rho'^2+|\bm x-\bm x'|^2})\nn\\
\eea whose massless limit is 
\bea 
D(u,\bm x,-;\tau',\rho',\bm x')=-\frac{i}{4\pi^{d/2}}\int_0^\infty d\omega \frac{\Gamma(\frac{d}{2}-1-i\omega)}{\Gamma(1-i\omega)}\frac{e^{-i\omega(v'-u)}}{(\rho'^2+|\bm x-\bm x'|^2)^{d/2-1-i\omega}}.\label{dbulk2}
\eea 
\paragraph{Bulk-to-bulk propagator.}
For massless theory, we use the split representation and the bulk-to-boundary propagator \eqref{dbulk}
\bea 
&&W^+(\tau,\rho,\bm x;\tau',\rho',\bm x')\nn\\&=&\frac{1}{4\pi^{d-1}}\int_0^\infty d\omega \omega \Big|\frac{\Gamma(\frac{d}{2}-1-i\omega)}{\Gamma(1-i\omega)}\Big|^2 e^{-i\omega(u-u')}\int d\bm x'' \frac{1}{(\rho^2+|\bm x-\bm x''|^2)^{d/2-1+i\omega}(\rho'^2+|\bm x'-\bm x''|^2)^{d/2-1-i\omega}}.\nn\\
\eea  We may use the Feynman's integral formula \bea 
&&\frac{1}{A_1^{a_1}A_2^{a_2}\cdots A_n^{a_n}}\nn\\
&=&\frac{\Gamma(a_1+a_2+\cdots+a_n)}{\Gamma(a_1)\cdots\Gamma(a_n)}\int_0^1 t_1\cdots\int_0^1 dt_n \frac{\delta(t_1+\cdots+t_n-1)t_1^{a_1-1}\cdots t_n^{a_n-1}}{(t_1A_1+\cdots+t_nA_n)^{a_1+\cdots+a_n}}\label{feynman}
\eea to obtain 
\bea 
&&W^+(\tau,\rho,\bm x;\tau',\rho',\bm x')\nn\\&=&\frac{1}{4\pi^{d-1}}\int_0^\infty d\omega \omega \frac{\Gamma(d-2)e^{-i\omega(u-u')}}{|\Gamma(1-i\omega)|^2}\int d\bm x''\int_0^1 dt \frac{t^{d/2-2+i\omega}(1-t)^{d/2-2-i\omega}}{[t(\rho^2+|\bm x-\bm x''|^2)+(1-t)(\rho'^2+|\bm x'-\bm x''|^2)]^{d-2}}.\nn\\
\eea 
In general $d$ dimensions, we have the following integral 
\bea 
\int d^d\bm x \frac{1}{(x^2+2\bm c\cdot \bm x+b^2)^a}=\frac{\pi^{d/2}\Gamma(a-\frac{d}{2})}{\Gamma(a)}\frac{1}{(b^2-\bm c^2)^{a-d/2}},\quad \text{Re}(a)>\frac{d}{2},\quad b^2>\bm c^2.\label{twodim}
\eea Therefore, the Wightman function becomes 
\bea 
&&W^+(\tau,\rho,\bm x;\tau',\rho',\bm x')\nn\\&=&\frac{1}{4\pi^{d/2}}\int_0^\infty d\omega \omega \frac{\Gamma(\frac{d}{2}-1)e^{-i\omega(u-u')}}{|\Gamma(1-i\omega)|^2}\int_0^1 dt \frac{t^{d/2-2+i\omega}(1-t)^{d/2-2-i\omega}}{(t\rho^2+(1-t)\rho'^2+t(1-t)|\bm x-\bm x'|^2)^{d/2-1}}\nn\\&=&\frac{1}{4\pi^{d/2}(\rho\rho')^{d/2-1}}\int_0^\infty d\omega \frac{\Gamma(\frac{d}{2}-1)\omega e^{-i\omega(\tau-\tau')}}{|\Gamma(1-i\omega)|^2}\int_0^\infty ds \frac{s^{d/2-2+i\omega}}{(s^2+2\eta s+1)^{d/2-1}},
\eea where $\eta$ is defined in \eqref{eta}. In the second step, we have changed the integral variable from $t$ to $s$
\bea 
\frac{t}{1-t}=\frac{\rho'}{\rho}s.
\eea We introduce the integral 
\bea 
J(\eta;\omega;d)=\frac{\omega\Gamma(\frac{d}{2}-1)}{|\Gamma(1-i\omega)|^2}\int_0^\infty ds  \frac{s^{d/2-2+i\omega}}{(s^2+2\eta s+1)^{d/2-1}}.
\eea 
The Wightman function can be expressed as 
\bea 
W^+(\tau,\rho,\bm x;\tau',\rho',\bm x')=\frac{1}{4\pi^{d/2}(\rho\rho')^{d/2-1}}\int_0^\infty d\omega J(\eta;\omega;d)e^{-i\omega(\tau-\tau')}.
\eea 
Note that the $s$ integral leads to a hypergeometric function
\bea 
J(\eta;\omega;d)&=&\frac{\Gamma(\frac{d}{2}-1)2^{\frac{d-3}{2}}}{\Gamma(d-2)}\omega\frac{|\Gamma(\frac{d}{2}-1-i\omega)|^2}{|\Gamma(1-i\omega)|^2}(\eta+1)^{\frac{3-d}{2}}{}_2F_1(\frac{1}{2}+i\omega,\frac{1}{2}-i\omega,\frac{d-1}{2};\frac{1-\eta}{2}).\nn\\\label{Jfun}
\eea where we have used the integral formula \cite{2007tisp.book.....G}
\bea 
\int_0^\infty ds s^{-1-\nu}(1+2\eta s+s^2)^{\mu-1/2}=2^{-\mu}(\eta^2-1)^{\mu/2}\Gamma(1-\mu)B(\nu-2\mu+1,-\nu)P_{\nu-\mu}^\mu(\eta),\nn\\
\text{Re}\ \nu<0,\quad \text{Re}(2\mu-\nu)<1,\quad |\text{arg}(\eta\pm 1)|<\pi,
\eea where the associated Legendre function is related to the hypergeometric function 
\bea 
P_\nu^\mu(\eta)=\frac{1}{\Gamma(1-\mu)}(\frac{\eta+1}{\eta-1})^{\mu/2}{}_2F_1(-\nu,1+\nu,1-\mu;\frac{1-\eta}{2}).
\eea 
However, we may
use the recurrence relation
\bea 
\frac{\partial}{\partial\eta}J(\eta;\omega;d)=-2J(\eta;\omega;d+2)
\eea to simplify the results. Therefore, we only need the following  results in $d=3$ and $d=4$
\bs\begin{align}
J(\eta;\omega;3)&=\sqrt{\pi}\tanh \pi\omega \ P_{-1/2-i\omega}(\eta),\\
    J(\eta;\omega;4)&=-i \frac{\zeta^{i\omega}-\zeta^{-i\omega}}{\zeta-\zeta^{-1}},\quad \zeta=\eta+\sqrt{\eta^2-1}
\end{align}\es where $P_\nu(\eta)$ is the Legendre function with non-integer order $\nu=-\frac{1}{2}-i\omega$, and it is related to the hypergeometric function 
\bea 
P_{-1/2-i\omega}(\eta)={}_2F_1(\frac{1}{2}+i\omega,\frac{1}{2}-i\omega,1;\frac{1-\eta}{2}).
\eea 
Note that $\eta\ge 1$ and it approaches $\infty$ in the limit $\rho\to0$ or $\rho'\to 0$. Using the asymptotic behaviour of the hypergeometric function, we find 
\bea 
J(\eta;\omega;d)\sim -i(2\eta)^{1-\frac{1}{2}+i\omega}\frac{\Gamma(\frac{d}{2}-1-i\omega)}{\Gamma(1-i\omega)}+i(2\eta)^{1-\frac{d}{2}-i\omega}\frac{\Gamma(\frac{d}{2}-1+i\omega)}{\Gamma(1+i\omega)},\quad \eta\to\infty.
\eea Then one can check that the Wightman function $W^+(\tau,\rho,\bm x;\tau',\rho',\bm x')$ reduces to the bulk-to-boundary propagator \eqref{dbulk} as $\rho\to 0$.

\section{Consistency check of bulk reconstruction}\label{consistency}
The expressions \eqref{Phi1} and \eqref{Phi2} should be equal to each other, \bea 
\int du' d\bm x' D(u',\bm x',+;\tau,\rho,\bm x)|\dot\Sigma(u',\bm x',+)\rangle=\int du' d\bm x' D^*(u',\bm x',-;\tau,\rho,\bm x)|\dot\Sigma(u',\bm x',-)\rangle.\nn\\
\eea Multiplying both sides by the state $\langle \Sigma(u'',\bm x''_{},-)|$, the above equality becomes 
\bea 
&&\int du'd\bm x'_{}D(u',\bm x',+;\tau,\rho,\bm x)\langle \Sigma(u'',\bm x''_{},-)|\dot\Sigma(u',\bm x',+)\rangle\nn\\&=&\int du' d\bm x'_{}D^*(u',\bm x',-;\tau,\rho,\bm x)\langle \Sigma(u'',\bm x''_{},-)| \dot\Sigma(u',\bm x',-)\rangle.\label{checkidentity}
\eea Now we compute the left hand side 
\bea 
\text{LHS}&=&-\frac{i}{4\pi^2}\int du' d\bm x'_{} \int_0^\infty d\omega \frac{e^{-i\omega(u'-u)}}{(\rho^2+|\bm x-\bm x'_{}|^2)^{1-i\omega}}\partial_{u'}\frac{i}{4\pi^2}\int_0^\infty d\omega' \frac{e^{i\omega'(u'-u'')}}{|\bm x'_{}-\bm x''_{}|^{2+2i\omega'}}\nn\\&=&\frac{i}{8\pi^3}\int_0^\infty \omega d\omega \int d\bm x'_{} \frac{e^{i\omega(u-u'')}}{(\rho^2+|\bm x-\bm x'_{}|^2)^{1-i\omega}|\bm x'_{}-\bm x''_{}|^{2+2i\omega}}\nn\\&=&\frac{i}{8\pi^3}\int_0^\infty d\omega \omega e^{i\omega(u-u'')} \int d\bm x'_{}\frac{\Gamma(2)}{\Gamma(1-i\omega)\Gamma(1+i\omega)}\int_0^1 dt  \frac{t^{-i\omega}(1-t)^{i\omega}}{[t(\rho^2+|\bm x-\bm x'_{}|^2)+(1-t)|\bm x'_{}-\bm x''_{}|^2]^2}\nn\\&=&\frac{i}{8\pi^2}\int_0^\infty d\omega \frac{\omega e^{i\omega(u-u'')}}{\Gamma(1-i\omega)\Gamma(1+i\omega)}\int_0^1 dt \frac{t^{-i\omega}(1-t)^{i\omega}}{t\rho^2+t(1-t)|\bm x_{}-\bm x''_{}|^2}\nn\\&=&-\frac{1}{8\pi^2}\int_0^\infty d\omega \frac{e^{i\omega(v-u'')}}{(\rho^2+|\bm x_{}-\bm x''_{}|^2)^{1+i\omega}}.
\eea In the third line, we have used Feynman's integral formula \eqref{feynman}. In the fourth line, we used the two-dimensional integral formula of \eqref{twodim}. The right hand side of \eqref{checkidentity} is 
\bea 
\text{RHS}&=&\frac{i}{4\pi^2}\int du'd\bm x'_{} \int_0^\infty d\omega \frac{e^{i\omega(v-u')}}{(\rho^2+|\bm x-\bm x'_{}|^2)^{1+i\omega}}\partial_{u'}\frac{1}{4\pi}\int_0^\infty \frac{d\omega'}{\omega'} e^{-i\omega'(u''-u')}\delta^{(2)}(\bm x'_{}-\bm x''_{})\nn\\&=&-\frac{1}{8\pi^2}\int_0^\infty d\omega \frac{e^{i\omega(v-u'')}}{(\rho^2+|\bm x-\bm x''_{}|^2)^{1+i\omega}}\nn\\&=&\text{LHS}.
\eea Therefore, we have checked the consistency of equations \eqref{Phi1} and \eqref{Phi2}.
\section{Hurwitz zeta and Polylogarithm functions}\label{special}
In this work, we need several interesting functions such as Hurwitz zeta  and Polylogarithm functions whose properties are collected in this appendix.
\paragraph{Hurwitz zeta function}
The Hurwitz zeta function can be introduced by the series expansion\cite{2007tisp.book.....G}
\bea 
\zeta(s,x)=\sum_{n=0}^\infty (x+n)^{-s}.
\eea The series is convergent in the region \be 
\text{Re}(s)>1
\ee and
\be 
x\not=0,-1,-2,\cdots.
\ee As a function of $s$, Hurwitz zeta function can be extended to the  complex plane except $s=1$ where the residue is 1, which is independent of $x$. The expansion around $s=1$ is 
\be 
\zeta(s,x)=\frac{1}{s-1}-\psi(x)+\mathcal{O}(s-1),
\ee where $\psi(x)$ is the Digamma function which is defined as 
\be 
\psi(x)=\frac{d}{dx}\log\Gamma(x).
\ee The Digamma function obeys the identity \be 
\psi(1+x)-\psi(-x)=-\pi\cot(\pi x).
\ee Therefore, 
\be 
\zeta(s,1+x)-\zeta(s,-x)=\pi\cot(\pi x)+\mathcal{O}(s-1).
\ee 
{Besides,we can expand the combination of $\zeta(s,x)$ near s =1
\bea
\left(\frac{i}{2 \pi }\right)^s \zeta \left(s,\frac{i u}{2 \pi }+1\right)+\left(-\frac{i}{2 \pi }\right)^s \zeta \left(s,-\frac{i u}{2 \pi }\right)=-\frac{1}{2}+\frac{i}{2\pi}[\zeta(s,1+x)-\zeta(s,-x)]+\mathcal{O}(s-1)
\eea}
This identity is useful to obtain the expansion \eqref{hur}.

\paragraph{Polylogarithms}
The classical  polylogarithms can be defined as an iterated integral\cite{Lewin:100000}
\bea 
\text{Li}_n(x)=\int_0^x dt \frac{\text{Li}_{n-1}(t)}{t},\quad\text{Li}_1(x)=-\log(1-x).\label{def}
\eea From the series expansion of the natural logarithm 
\be 
-\log(1-x)=\sum_{m=1}^\infty \frac{x^m}{m},
\ee we find the series expansion of the polylogarithms
\bea 
\text{Li}_n(x)=\sum_{m=1}^\infty \frac{x^m}{m^n}.
\eea 
For $n=2$, it becomes the dilogarithm 
\be 
\text{Li}_2(x)=\sum_{m=1}^\infty \frac{x^m}{m^2}
\ee whose integral representation is
\be 
\text{Li}_2(x)=-\int_0^x dt \frac{\log(1-t)}{t}=-\int_0^1 dt \frac{\log(1-xt)}{t}.
\ee Polylogarithm at the special value $x=1$ is related to Riemann zeta function 
\be 
\text{Li}_n(1)=\sum_{m=1}^\infty \frac{1}{m^n}=\zeta(n).
\ee For $n=2$, we find the special value 
\be 
\text{Li}_2(1)=\frac{\pi^2}{6}.
\ee 
From the integral \eqref{def}, we find the following relation
\be 
x\frac{d}{dx}\text{Li}_n(x)=\text{Li}_{n-1}(x).
\ee Using the above relation, it is easy to prove the following identities for dilogarithm
\bs\begin{align}
   \text{inversion formula},&\quad \text{Li}_2(x)+\text{Li}_2(x^{-1})=-\frac{\pi^2}{6}-\frac{\ln^2(-x)}{2},\\
   \text{reflection formula},&\quad \text{Li}_2(x)+\text{Li}_2(1-x)=\frac{\pi^2}{6}-\ln x\ \ln (1-x),\\
    \text{duplication formula},&\quad \text{Li}_2(x)+\text{Li}_2(-x)=\frac{1}{2}\text{Li}_2(x^2).
\end{align}\es 

\paragraph{Bloch-Wigner polylogarithm}
Bloch-Wigner dilogarithm is defined as \cite{Zagier:1990,Bloch2011HigherRA,Cartier:2007zz}
\bea 
P_2(z)=\text{Im}\left(\text{Li}_2(z)+\log |z|\ \log(1-z)\right),
\eea which is actually the imaginary part of the
classical dilogarithm. 
The dilogarithm has a branch cut at $z>1$ part of the real axis. It has a discontinuity as $z$ crosses the cut. The discontinuity is canceled by the second term in the definition such that the Bloch-Wigner dilogarithm is continuous when $z$ crosses the cut. Bloch-Wigner dilogarithm is analytic on the complex plane except at $z=0,1$ where it is continuous but not differentiable. It is also a real function on the complex plane, obeys the following identity 
\bea 
P_2(z)=\frac{1}{2}[P_2(\frac{z}{\bar z})+P_2(\frac{1-z^{-1}}{1-\bar z^{-1}})+P_2(\frac{1-\bar z}{1-z})], \label{sixfold}
\eea 
and has the 6-fold symmetry
\bea 
P_2(z)=-P_2(1-z)=-P_2(z^{-1})=P_2(\frac{1}{1-z})=-P_2(\frac{z}{z-1})=P_2(\frac{z-1}{z}).\label{six}
\eea By exchanging the role of $z$ and $\bar z$, it follows that 
 \bea P_2(z)=-P_{2}(\bar z)=-P_{2}(\frac{1}{z}). \eea
More general Bloch-Wigner polylogarithm is defined as 
\bea 
P_n(z)=\mathcal{R}_n\left( \sum_{j=0}^{n-1}\frac{2^j B_j}{j!}\log^j|z| \text{Li}_{n-j}(z)\right),
\eea where $\mathcal{R}_n$ denotes the real part if n is odd and the imaginary part if $n$ is even. The Bernoulli numbers $B_j$ are generated by the function 
\be 
\frac{x}{e^x-1}=\sum_{j=0}^\infty B_j \frac{x^j}{j!}.
\ee 

\paragraph{Clausen function}
The Clausen function can be given in terms of the series sum
\bea 
\text{Cl}_n(\theta)=\left\{\begin{array}{cc}\sum_{m=1}^\infty \frac{\sin m\theta}{m^n},& n\ \text{even},\\
\sum_{m=1}^\infty \frac{\cos m\theta}{m^n},&n \ \text{odd}.\end{array}\right.
\eea It can be written formally as the polylogarithms 
\bea 
\text{Cl}_n(\theta)=\mathcal{R}_n\left(\text{Li}_n(e^{i\theta})\right).
\eea 

\section{Cyclic quadrilateral}\label{coll}
Consider three points on the plane whose coordinates are $\bm x_1,\bm x_2,\bm x_3$, they form a triangular in general.  The sufficient and necessary condition for them to be collinear is that the area of the triangular is zero 
\bea 
\bm x_{12}^4+\bm x_{13}^4+\bm x_{23}^4-2\bm x_{12}^2\bm x_{23}^2-2\bm x_{13}^2\bm x_{23}^2-2\bm x_{13}^2\bm x_{12}^2=0
\eea which can be formulated in complex coordinates 
\bea 
z_1 \bar z_{23}+z_2\bar{z}_{13}+z_3 \bar{z}_{12}=0.
\eea Now we consider four points on the plane whose coordinates are $\bm x_j,\ j=1,2,3,4$. In general, they should form a quadrilateral. When the four points are collinear,  we can set $z_1=\bar{z}_1=0$ and rotate the collinear line to the real axis without loss of generality. Then the other three points can be parameterized as 
\bea 
z_j=\lambda_j=\bar z_j,\quad j=2,3,4.
\eea Then it is straightforward to find $Q=0$. 
Inversely, the condition $Q=0$ can lead to more interesting configurations. In complex coordinates, the condition is 
\bea 
z_{12}z_{34}\bar{z}_{13}\bar{z}_{24}-\bar{z}_{12}\bar{z}_{34}z_{13}z_{24}=0.
\eea We can also switch to the Cartesian coordinates, 
\bea 
(\bm x_{12}^2\bm x_{34}^2+\bm x_{13}^2\bm x_{24}^2-\bm x_{23}^2\bm x_{14}^2)^2=4\bm x_{12}^2\bm x_{34}^2\bm x_{23}^2\bm x_{24}^2.
\eea This is equivalent to 
\bea 
\bm x_{12}^2\bm x_{34}^2+\bm x_{13}^2\bm x_{24}^2-\bm x_{23}^2\bm x_{14}^2=\pm 2|\bm x_{12}||\bm x_{34}||\bm x_{13}||\bm x_{24}|
\eea which could be simplified further 
\bs\begin{align}
|\bm x_{12}||\bm x_{34}|+|\bm x_{13}||\bm x_{24}|&=|\bm x_{23}||\bm x_{14}|,\label{40}\\
\text{or}\quad |\bm x_{12}||\bm x_{34}|+|\bm x_{13}||\bm x_{24}|&=-|\bm x_{23}||\bm x_{14}|,\\
\text{or}\quad |\bm x_{12}||\bm x_{34}|-|\bm x_{13}||\bm x_{24}|&=|\bm x_{23}||\bm x_{14}|,\label{41}\\
\text{or}\quad |\bm x_{12}||\bm x_{34}|-|\bm x_{13}||\bm x_{24}|&=-|\bm x_{23}||\bm x_{14}|\label{42}.
\end{align}\es 
Note that this is just the content of the  Ptolemy's theorem \cite{geometry} which states that the sum of the product of the two pairs of opposite sides equals the product of the diagonals in a quadrilateral. For four points $\bm x_1,\bm x_2,\bm x_3,\bm x_4$, there are three distinct ways to form the cyclic quadrilateral, depending on the order of the points, as shown in Figure \ref{Rindlerlimit}. On the left of Figure \ref{Rindlerlimit}, the two pairs of opposite sides are 
\bea 
\bm x_{12},\bm x_{34}\quad\text{and}\quad \bm x_{14},\bm x_{23},
\eea
and the diagonal vectors are 
\be 
\bm x_{13}\quad\text{and}\quad \bm x_{24}.
\ee Therefore, they should satisfy the eqn. \eqref{42}. Similarly, in the middle of Figure \ref{Rindlerlimit}, the quadrilateral satisfies the eqn. \eqref{41}. On the right of Figure \ref{Rindlerlimit}, the quadrilateral satisfies the eqn. \eqref{40}. Note that the second equation should be ruled out since the right hand side is always non-positive.
Conversely, the Ptolemy's inequality states that the product of diagonals is no larger than the sum of the product of its opposite sides for any quadrilateral. The equality is satisfied only if the four points are collinear or on a circle.

\begin{figure}
  \centering
  \subfloat{\scalebox{0.5}{
 \begin{tikzpicture} 
\draw(0,0) circle (4);
\draw (-3,2.646)node[left=1pt]{$\bm x_1$}--(2,3.464)node[above=1pt]{$\bm x_4$}--(3.7,-1.5198)node[right=1pt]{$\bm x_3$}--(-1.5,-3.7)node[below=1pt]{$\bm x_2$}--(-3,2.646);
\draw [dashed](-3,2.646)--(3.7,-1.5198);
\draw [dashed](2,3.464)--(-1.5,-3.7);
\end{tikzpicture}}
  }\hspace{1cm}
  \subfloat{\scalebox{0.5}{
    \begin{tikzpicture} 
\draw(0,0) circle (4);
\draw (-3,2.646)node[left=1pt]{$\bm x_1$}--(2,3.464)node[above=1pt]{$\bm x_4$}--(3.7,-1.5198)node[right=1pt]{$\bm x_2$}--(-1.5,-3.7)node[below=1pt]{$\bm x_3$}--(-3,2.646);
\draw [dashed](-3,2.646)--(3.7,-1.5198);
\draw [dashed](2,3.464)--(-1.5,-3.7);
\end{tikzpicture}}
  }\hspace{1cm}
  \subfloat{\scalebox{0.5}{
    \begin{tikzpicture} 
\draw(0,0) circle (4);
\draw (-3,2.646)node[left=1pt]{$\bm x_1$}--(2,3.464)node[above=1pt]{$\bm x_3$}--(3.7,-1.5198)node[right=1pt]{$\bm x_4$}--(-1.5,-3.7)node[below=1pt]{$\bm x_2$}--(-3,2.646);
\draw [dashed](-3,2.646)--(3.7,-1.5198);
\draw [dashed](2,3.464)--(-1.5,-3.7);
\end{tikzpicture}}
  }
  \caption{\centering{
 Cyclic quadrilaterals.}}\label{Rindlerlimit}
\end{figure}
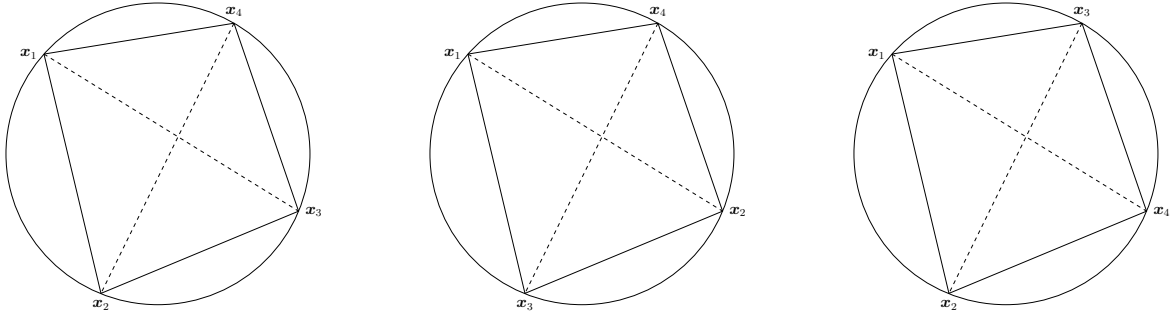
\section{Hypergeometric functions}
\paragraph{Appell function of the fourth kind.} The sum representation of the Appell function of the fourth kind is \cite{1953hft1.book...59E}
\bea 
F_4(a_1,a_2;a_3,a_4;\xi,\eta)=\sum_{m,n=0}^\infty \frac{(a_1)_{m+n}(a_2)_{m+n}}{(a_3)_m(a_4)_n m! n!}\xi^m\eta^n,
\eea where the Pochhammer symbol is defined as 
\be 
(a)_n=a(a+1)\cdots(a+n-1)=\frac{\Gamma(a+n)}{\Gamma(a)},
\ee and the series is convergent in the domain 
\be 
|\xi|^{1/2}+|\eta|^{1/2}<1.
\ee 
Obviously, the Appell function of the fourth kind is invariant  under exchange of positions of parameters and arguments as follows
\bs\begin{align}
    F_4(a_1,a_2;a_3,a_4;\xi,\eta)&=F_4(a_2,a_1;a_3,a_4;\xi,\eta),\\
    F_4(a_1,a_2;a_3,a_4;\xi,\eta)&=F_4(a_1,a_2;a_4,a_3;\eta,\xi).
\end{align}\es 
One may use the integral representation to c continue it to the other region in the complex plane. A useful integral representation is the Mellin-Barnes type double integral representation 
\bea 
&&F_4(a_1,a_2;a_3,a_4;\xi,\eta)\nn\\&=&\frac{\Gamma(a_3)\Gamma(a_4)}{\Gamma(a_1)\Gamma(a_2)}\frac{1}{(2\pi i)^2}\int_{-i\infty}^{i\infty}ds\int_{-i\infty}^{i\infty} dt \frac{\Gamma(a_1+s+t)\Gamma(a_2+s+t)\Gamma(-s)\Gamma(-t)}{\Gamma(a_3+s)\Gamma(a_4+t)}(-\xi)^s(-\eta)^t.\nn\\
\eea The Appell hypergeometric function also satisfies the partial differential equations 
\bs\begin{align}
    &\xi(1-\xi)\frac{\partial^2F_4}{\partial\xi^2}-2\xi\eta\frac{\partial^2F_4}{\partial\xi\partial\eta}-\eta^2\frac{\partial^2F_4}{\partial\eta^2}+(a_3-(a_1+a_2+1)\xi)\frac{\partial F_4}{\partial \xi}-(a_1+a_2+1)\eta\frac{\partial F_4}{\partial \eta}-a_1a_2F_4=0,\\
    &\eta(1-\eta)\frac{\partial^2F_4}{\partial\eta^2}-2\xi\eta\frac{\partial^2F_4}{\partial\xi\partial\eta}-\xi^2\frac{\partial^2F_4}{\partial\xi^2}+(a_4-(a_1+a_2+1)\eta)\frac{\partial F_4}{\partial \eta}-(a_1+a_2+1)\xi\frac{\partial F_4}{\partial \xi}-a_1a_2F_4=0.
\end{align}\es 
\paragraph{GKZ hypergeometric functions.}
A GKZ hypergeometric  system \cite{Cattani2006THREELO,Gelfand:1990bua, delaCruz:2019skx,Klausen:2019hrg} is defined by a $(M+1)\times N$ integer matrix $\mathbb{A}$ 
\bea 
\mathbb{A}=(\bm A_1,\bm A_2,\cdots,\bm A_N),
\eea and a vector $\bm c$ with $M+1$ components 
\bea 
\bm c=\left(\begin{array}{c}c_1\\c_2\\\cdots\\ c_{M+1}\end{array}\right).
\eea It is a system of differential equations for function  $G$ with $N$ variables $x_1,x_2,\cdots,x_{N}$ which satisfies the following two conditions.
\begin{enumerate}
    \item For any vectors $\bm u$ and $\bm v$ with $N$ non-negative integer components
    \bea 
    \bm u=\left(\begin{array}{c}u_1\\ u_2\\\cdots\\u_N\end{array}\right),\quad \bm v=\left(\begin{array}{c}v_1\\ v_2\\\cdots\\v_N\end{array}\right)
    \eea that satisfy the condition 
    \be 
    \mathbb{A}(\bm u-\bm v)=0,
    \ee the GKZ hypergeometric function obeys the following toric differential equations 
    \bea 
   \left( \prod_{j=1}^N (\frac{\partial}{\partial x_j})^{u_j}-\prod_{j=1}^N (\frac{\partial}{\partial x_j})^{v_j}\right)G(x_1,\cdots,x_N)=0\label{GKZ1}.
    \eea 
    \item GKZ function should also satisfies the following $M+1$ homogeneity differential equations 
    \bea 
   \left( \sum_{j=1}^N\bm A_j x_j\frac{\partial}{\partial x_j}-\bm c\right)G(x_1,\cdots,x_N)=0.\label{GKZ2}
    \eea 
\end{enumerate} Note that the first row of the matrix $\mathbb{A}$ is $(1,1,\cdots,1)$.
In this work, we only need the GKZ hypergeometric function as an Euler-Mellin type integral. Given a matrix $\mathbb{A}$ in the following form 
\bea 
\mathbb{A}=\left(\begin{array}{cccc}1&1&\cdots&1\\ a_{11}&a_{12}&\cdots&a_{1N}\\
\cdots&&&\cdots\\
a_{M1}&a_{M2}&\cdots&a_{MN}\end{array}\right),
\eea 
we can always construct a  Laurent polynomial 
\bea 
\mathcal{P}(z_1,z_2,\cdots,z_M;x_1,x_2,\cdots,x_N)=\sum_{j=1}^N x_j \left(\prod_{k=1}^M z_{k}^{a_{kj}}\right), \quad{ a_{kj} \in \mathbb{Z}}.
\eea Now we assume the vector $\bm c$ is 
\bea 
\bm c=\left(\begin{array}{c}-\beta_0\\-\beta_1\\\cdots\\-\beta_M\end{array}\right),
\eea then the Euler-Mellin integral  
\bea 
G(x_1,x_2,\cdots,x_N)=\int \left(\prod_{k=1}^M dz_k z_k^{\beta_k-1}\right)\left(\mathcal{P}(z_1,z_2,\cdots,z_M;x_1,x_2,\cdots,x_N)\right)^{-\beta_0}\label{eM}
\eea is a GKZ hypergeometric function associated with the above matrix $\mathbb{A}$ and vector $\bm c$. It is straightforward to check that the integral indeed obeys the conditions \eqref{GKZ1} and \eqref{GKZ2}. Note that the integral should be evaluated in a certain domain. In our case, the non-vanishing components of $a_{ij}$ are
\bea 
a_{11}=a_{14}=a_{15}=a_{22}=a_{24}=a_{26}=a_{33}=a_{35}=a_{36}=1.
\eea The parameters $\beta_i$ are 
\bea 
\beta_0=2-iw/2,\quad \beta_1=1-i\omega_2,\quad \beta_2=1-i\omega_3,\quad \beta_3=1-i\omega_4.
\eea The variables $x_j$ are
\be 
x_1=\bm x_{12}^2,\quad x_2=\bm x_{13}^2,\quad x_3=\bm x_{14}^2,\quad x_4=\bm x_{23}^2,\quad x_5=\bm x_{24}^2,\quad x_6=\bm x_{34}^2.
\ee The polynomial $\mathcal{P}$ is exactly the polynomial $\mathcal{G}$ in the context and the Euler-Mellin integral \eqref{eM} becomes the  equation \eqref{pG}.

\section{Integrals}\label{int}
\paragraph{Three-point Carrollian amplitude.} We
 should compute the integral \bea 
&&\int_0^\infty dt \frac{\log(t+a)+\log(t+a^{-1})-\log t+\log c}{(t+b)(t+b^{-1})}\nn\\&=&\frac{1}{b^{-1}-b}\int_0^\infty dt[ \frac{\log(t+a)+\log(t+a^{-1})-\log t+\log c}{t+b}-(b\to b^{-1})].\label{3ptp}
\eea  At first,
we will need the following integrals
\bea 
I(a,b)&=&\int_0^\infty dt[ \frac{\log(t+a)}{t+b}-\frac{\log(t+a)}{t+b^{-1}}]\nn\\&=&\int_0^\infty d\log(t+b)\log(t+a)-d\log(t+b^{-1})\log(t+a)\nn\\&=&[\log(t+b)\log(t+a)-\log(t+b^{-1})\log(t+a)] \Big|_0^\infty-\int_0^\infty dt\frac{\log(t+b)-\log(t+b^{-1})}{t+a}\nn\\&=&-2\log a\log b+\int_0^1 ds \frac{\log \left(a+\frac{s}{b}-as\right)-\log (a+b s-as)}{s}.
\eea In the last step, we have introduced a new integral variable 
\bea 
1-s=\frac{t}{t+a}.
\eea Using the integral representation of the dilogarithm,
 we find 
\bea 
I(a,b)=-2\log a\log b+\text{Li}_2(\frac{a-b}{a})-\text{Li}_2(\frac{a-b^{-1}}{a}).
\eea Second, we will compute the integral 
\bea 
I_1(a,b)&=&\int_0^\infty dt [\frac{\log t}{t+b}-\frac{\log t}{t+b^{-1}}]\nn\\&=&\int_{\infty}^0(-\frac{dt}{t^2})\log(t^{-1})[\frac{1}{t^{-1}+b}-\frac{1}{t^{-1}+b^{-1}}]\nn\\&=&-\int_0^\infty dt \log t[\frac{1}{t+b}-\frac{1}{t+b^{-1}}]\nn\\&=&0.
\eea Note that in the second line, we changed the variable from $t$ to $t^{-1}$. 

Finally, the integral 
\bea 
I_2(b)&=&\int_0^\infty dt [\frac{1}{t+b}-\frac{1}{t+b^{-1}}]\nn\\&=&\log\frac{t+b}{t+b^{-1}}\Big|_0^\infty\nn\\&=&-2\log b.
\eea Therefore, we obtain the following result 
\bea 
&&\int_0^\infty dt \frac{\log(t+a)+\log(t+a^{-1})-\log t+\log c}{(t+b)(t+b^{-1})}\nn\\&=&\frac{1}{b^{-1}-b}[I(a,b)+I(a^{-1},b)+2\log c\log b]\nn\\&=&\frac{1}{b^{-1}-b}[\text{Li}_2(1-\frac{b}{a})-\text{Li}_2(1-\frac{1}{ab})+\text{Li}_2(1-ab)-\text{Li}_2(1-\frac{a}{b})-2\log c\log b]
\eea Note that the parameters $a,b,c$ in \eqref{T} are related to $z,\bar z$ 
\bea 
a=\sqrt{z\bar z},\quad b=\sqrt{\frac{z}{\bar z}},\quad c=\frac{\sqrt{z\bar z}}{(1-z)(1-\bar z)}.
\eea 
Utilizing the identity
\bea 
-4iP_2(z)=\text{Li}_2(1-\bar z^{-1})-\text{Li}_2(1-z^{-1})+\text{Li}_2(1-z)-\text{Li}_2(1-\bar z)-\log\frac{\sqrt{z\bar z}}{(1-z)(1-\bar z)}\log\frac{z}{\bar z},\nn\\
\eea  we find the equivalence between \eqref{T} and \eqref{block}.

\paragraph{Four-point Carrollian amplitude.} In the zero energy limit {and $\bm x_{1}=0$}, we should compute 
\bea 
&&\int_0^\infty dz_2\int_0^\infty dz_3\int_0^\infty dz_4 \frac{1}{\tilde{S}_4^2}\nn\\&=&\int_0^\infty dz_4\frac{-\log\bm x_{23}^2-\log\bm x_4^2-\log z_4+\log(\bm x_2^2+z_4\bm x_{24}^2)+\log(\bm x_3^2+z_4\bm x_{34}^2)}{\bm x_{24}^2\bm x_{34}^2 z_4^2+2(\bm x_2^2\bm x_3\cdot\bm x_{34}-\bm x_3^2\bm x_2\cdot\bm x_4+\bm x_4^2\bm x_2\cdot\bm x_3)z_4+\bm x_2^2\bm x_3^2}\nn\\&=&\frac{1}{|\bm x_2||\bm x_3||\bm x_{24}||\bm x_{34}|}\int_0^\infty dt \frac{\log(t+\overline{a})+\log(t+\overline{a}^{-1})-\log t+\log \overline{c}}{t^2+(\overline{b}+\overline{b}^{-1})t+1}
\eea where 
$\overline{a},\overline{b},\overline{c}$ are given in the context. In the last step, we have changed the variable $z_4$ to $t$  \be t = \frac{|\bm x_{24}||\bm x_{34}|z_{4}}{|\bm x_{2}||\bm x_{3}|}.\ee  
The form of the integral is exactly the same as \eqref{3ptp}. Therefore, we can obtain the result in the context immediately.

Now we turn to the non-zero energy Carrollian amplitude where the key integral is 
\bea 
&&\mathcal{I}(a_1,a_2,a_3,a_4;\xi,\eta)\nn\\&=&\int_0^\infty dt \int_0^\infty dt' \frac{t^{a_1}t'^{a_2}}{(t t'+t'+\xi)^{a_3}(t t'+t+\eta)^{a_4}}\nn\\&=&\frac{1}{(2\pi i)^2\Gamma(a_3)\Gamma(a_4)}\int_{-i\infty}^{i\infty}dz \int_{-i\infty}^{i\infty}dz'\int_0^\infty dt \int_0^\infty dt'\frac{\Gamma(a_3+z)\Gamma(-z)\Gamma(a_4+z')\Gamma(-z')t^{a_1}t'^{a_2}{\xi}^z{\eta }^{z'}}{(tt'+t')^{a_3+z}(tt'+t)^{a_4+z'}}\nn\\&=&\frac{1}{(2\pi i)^2\Gamma(a_3)\Gamma(a_4)}\int_{-i\infty}^{i\infty} dz\int_{-i\infty}^{i\infty}dz' \Gamma(-z)\Gamma(-z')\Gamma(a_2-a_3-z+1)\Gamma(a_1-a_4-z'+1)\nn\\&&\times\Gamma(-a_1+a_3+a_4+z+z'-1)\Gamma(-a_2+a_3+a_4+z+z'-1){\xi}^{z}{\eta}^{z'}.\label{intmellin}
\eea We have transform it to the Mellin-Barnes type integrals using the basic formula 
\bea 
\frac{1}{(A+B)^\lambda}=\frac{1}{\Gamma(\lambda)}\frac{1}{2\pi i\ }\int_{-i\infty}^{i\infty} dz\  \Gamma(-z)\Gamma(\lambda+z)\frac{B^z}{A^{\lambda+z}}.
\eea In the last step, we have used the formula \bea 
&&\int_0^\infty dt \int_0^\infty dt' \frac{t^{a_1}t'^{a_2}}{(t t'+t')^{a_{3}}(t t'+t)^{a_4}}\nn\\&=&\frac{\Gamma \left(a_2-a_3+1\right) \Gamma \left(a_1-a_4+1\right) \Gamma \left(-a_1+a_3+a_4-1\right) \Gamma \left(-a_2+a_3+a_4-1\right)}{\Gamma \left(a_3\right) \Gamma \left(a_4\right)},\nn\\ &&\text{Re}(a_3+a_4)>1+\text{Re}(a_1)>\text{Re}(a_4),\quad \text{Re}(a_3+a_4)>1+\text{Re}(a_2)>\text{Re}(a_3).\nn\\
\eea In our case, the previous conditions are not always satisfied. Then, we should evaluate the integral under the conditions and then  analytically continue the result. 
We can use residue theorem to obtain the result
\bea 
&&\mathcal{I}(a_1,a_2,a_3,a_4;{\xi},{\eta })\nn\\&=&C_1F_4(-a_1+a_3+a_4-1,-a_2+a_3+a_4-1;a_4-a_1,a_3-a_2;{\eta},{\xi})\nn\\&+&C_2{\eta}^{1+a_1-a_4} F_4(a_1 -a_2 +a_3,a_3;2+a_1-a_4,a_3-a_2;{\eta},{\xi})\nn\\&+&C_3{\xi }^{1+a_2-a_3}F_4(a_4,-a_1+a_2+a_4;a_4-a_1,2+a_2-a_3;{\eta},{\xi})\nn\\&+&C_4{\xi}^{1+a_2-a_3}{\eta}^{1+a_1-a_4}F_4( 1+a_2, 1+a_1;2+a_1-a_4,2+a_2-a_3;{\eta},{\xi})
\eea where 
\bs\begin{align}
    C_1&=\frac{\Gamma(a_2-a_3+1)\Gamma(a_1-a_4+1)\Gamma(-a_1+a_3+a_4-1)\Gamma(-a_2+a_3+a_4-1)}{\Gamma(a_3)\Gamma(a_4)},\\
    C_2&=\frac{\Gamma(a_3)\Gamma(a_1-a_2+a_3)\Gamma(-1-a_1+a_4)\Gamma(a_2-a_3+1)}{\Gamma(a_3)\Gamma(a_4)},\\
    C_3&=\frac{\Gamma(a_3-a_2-1)\Gamma(a_1-a_4+1)\Gamma(a_4)\Gamma(-a_1+a_2+a_4)}{\Gamma(a_3)\Gamma(a_4)},\\
C_4&=\frac{\Gamma(1+a_1)\Gamma(1+a_2)\Gamma(a_3-a_2-1)\Gamma(a_4-a_1-1)}{\Gamma(a_3)\Gamma(a_4)}.
\end{align}\es 
We have checked that our result matches with the formula in \cite{Boos:1990rg} after identifying the parameters 
\bea 
\mu\leftrightarrow -1-a_1+a_3+a_4,\quad \nu\leftrightarrow a_4,\quad \rho\leftrightarrow a_1-a_2+a_3,\quad n\leftrightarrow 2(a_3+a_4).
\eea We define two matrices
\bea 
\mathbb{C}=\left(
\begin{array}{cccc}
 -a_1+a_3+a_4-1 & -a_2+a_3+a_4-1 & a_1-a_4+1 & a_2-a_3+1 \\
 a_1-a_2+a_3 & a_3 & -a_1+a_4-1 & a_2-a_3+1 \\
 a_4 & -a_1+a_2+a_4 & a_1-a_4+1 & -a_2+a_3-1 \\
 a_2+1 & a_1+1 & -a_1+a_4-1 & -a_2+a_3-1 \\
\end{array}
\right)
\eea and 
\bea 
\mathbb{F}=\left(
\begin{array}{cccc}
 -a_1+a_3+a_4-1 & -a_2+a_3+a_4-1 & a_4-a_1 & a_3-a_2 \\
 a_1-a_2+a_3 & a_3 & a_1-a_4+2 & a_3-a_2 \\
 a_4 & -a_1+a_2+a_4 & a_4-a_1 & a_2-a_3+2 \\
 a_2+1 & a_1+1 & a_1-a_4+2 & a_2-a_3+2 \\
\end{array}
\right),
\eea and then the coefficients $C_i$ are 
\bea 
C_i=\frac{1}{\Gamma(a_3)\Gamma(a_4)}\prod_{j=1}^4\Gamma(\mathbb{C}_{ij}),\quad i=1,2,3,4.
\eea Similarly, the Appell functions associated with $C_i$ are 
\bea 
F_4(\mathbb{F}_{i1},\mathbb{F}_{i2};\mathbb{F}_{i3},\mathbb{F}_{i4};{\tt v},{\tt u}).
\eea 
In our case, we have 
\bea 
a_1=a_2=-i\omega_4,\quad a_3=1-i\omega_1,\quad a_4=1-i\omega_2,\quad \xi={\tt U},\quad \eta={\tt V}.
\eea Therefore, the $\mathbb{C}$ and $\mathbb{F}$ matrices are 
\bs\begin{align} 
\mathbb{C}&=\left(
\begin{array}{cccc}
 1-i\omega_3 &  1-i\omega_3& i \omega _2-i \omega _4 & i \omega _1-i \omega _4 \\
 1-i \omega _1 & 1-i \omega _1 & i \omega _4-i \omega _2 & i \omega _1-i \omega _4 \\
 1-i \omega _2 & 1-i \omega _2 & i \omega _2-i \omega _4 & i \omega _4-i \omega _1 \\
 1-i \omega _4 & 1-i \omega _4 & i \omega _4-i \omega _2 & i \omega _4-i \omega _1 \\
\end{array}
\right),\\ \mathbb{F}&=\left(
\begin{array}{cccc}
 1-i\omega_3 & 1-i\omega_3 & -i \omega _2+i \omega _4+1 & -i \omega _1+i \omega _4+1 \\
 1-i \omega _1 & 1-i \omega _1 & i \omega _2-i \omega _4+1 & -i \omega _1+i \omega _4+2 \\
 1-i \omega _2 & 1-i \omega _2 & -i \omega _2+i \omega _4+1 & i \omega _1-i \omega _4+1 \\
 1-i \omega _4 & 1-i \omega _4 & i \omega _2-i \omega _4+1 & i \omega _1-i \omega _4+1 \\
\end{array}
\right).
\end{align}\es

\bibliography{refs}

\providecommand{\href}[2]{#2}\begingroup\raggedright\begin{thebibliography}{10}

\bibitem{1993gr.qc....10026T}
G.~{'t Hooft}, ``{Dimensional Reduction in Quantum Gravity},'' {\em arXiv
  e-prints} (Oct., 1993) gr--qc/9310026,
  \href{http://www.arXiv.org/abs/gr-qc/9310026}{{\tt gr-qc/9310026}}.

\bibitem{Susskind:1994vu}
L.~Susskind, ``{The World as a hologram},'' {\em J. Math. Phys.} {\bf 36}
  (1995) 6377--6396, \href{http://www.arXiv.org/abs/hep-th/9409089}{{\tt
  hep-th/9409089}}.

\bibitem{Polchinski:1999ry}
J.~Polchinski, ``{S matrices from AdS space-time},''
  \href{http://www.arXiv.org/abs/hep-th/9901076}{{\tt hep-th/9901076}}.

\bibitem{Susskind:1998vk}
L.~Susskind, ``{Holography in the flat space limit},'' {\em AIP Conf. Proc.}
  {\bf 493} (1999), no.~1, 98--112,
  \href{http://www.arXiv.org/abs/hep-th/9901079}{{\tt hep-th/9901079}}.

\bibitem{Giddings:1999jq}
S.~B. Giddings, ``{Flat space scattering and bulk locality in the AdS / CFT
  correspondence},'' {\em Phys. Rev. D} {\bf 61} (2000) 106008,
  \href{http://www.arXiv.org/abs/hep-th/9907129}{{\tt hep-th/9907129}}.

\bibitem{deBoer:2003vf}
J.~de~Boer and S.~N. Solodukhin, ``{A Holographic reduction of Minkowski
  space-time},'' {\em Nucl. Phys. B} {\bf 665} (2003) 545--593,
  \href{http://www.arXiv.org/abs/hep-th/0303006}{{\tt hep-th/0303006}}.

\bibitem{Arcioni:2003xx}
G.~Arcioni and C.~Dappiaggi, ``{Exploring the holographic principle in
  asymptotically flat space-times via the BMS group},'' {\em Nucl. Phys. B}
  {\bf 674} (2003) 553--592,
  \href{http://www.arXiv.org/abs/hep-th/0306142}{{\tt hep-th/0306142}}.

\bibitem{Arcioni:2003td}
G.~Arcioni and C.~Dappiaggi, ``{Holography in asymptotically flat space-times
  and the BMS group},'' {\em Class. Quant. Grav.} {\bf 21} (2004) 5655,
  \href{http://www.arXiv.org/abs/hep-th/0312186}{{\tt hep-th/0312186}}.

\bibitem{Mann:2005yr}
R.~B. Mann and D.~Marolf, ``{Holographic renormalization of asymptotically flat
  spacetimes},'' {\em Class. Quant. Grav.} {\bf 23} (2006) 2927--2950,
  \href{http://www.arXiv.org/abs/hep-th/0511096}{{\tt hep-th/0511096}}.

\bibitem{Pasterski:2016qvg}
S.~Pasterski, S.-H. Shao, and A.~Strominger, ``{Flat Space Amplitudes and
  Conformal Symmetry of the Celestial Sphere},'' {\em Phys. Rev. D} {\bf 96}
  (2017), no.~6, 065026, \href{http://www.arXiv.org/abs/1701.00049}{{\tt
  1701.00049}}.

\bibitem{Pasterski:2017kqt}
S.~Pasterski and S.-H. Shao, ``{Conformal basis for flat space amplitudes},''
  {\em Phys. Rev. D} {\bf 96} (2017), no.~6, 065022,
  \href{http://www.arXiv.org/abs/1705.01027}{{\tt 1705.01027}}.

\bibitem{Pasterski:2017ylz}
S.~Pasterski, S.-H. Shao, and A.~Strominger, ``{Gluon Amplitudes as 2d
  Conformal Correlators},'' {\em Phys. Rev. D} {\bf 96} (2017), no.~8, 085006,
  \href{http://www.arXiv.org/abs/1706.03917}{{\tt 1706.03917}}.

\bibitem{Donnay:2022aba}
L.~Donnay, A.~Fiorucci, Y.~Herfray, and R.~Ruzziconi, ``{Carrollian Perspective
  on Celestial Holography},'' {\em Phys. Rev. Lett.} {\bf 129} (2022), no.~7,
  071602, \href{http://www.arXiv.org/abs/2202.04702}{{\tt 2202.04702}}.

\bibitem{Bagchi:2022emh}
A.~Bagchi, S.~Banerjee, R.~Basu, and S.~Dutta, ``{Scattering Amplitudes:
  Celestial and Carrollian},'' {\em Phys. Rev. Lett.} {\bf 128} (2022), no.~24,
  241601, \href{http://www.arXiv.org/abs/2202.08438}{{\tt 2202.08438}}.

\bibitem{Une}
J.~M. L\'evy-Leblond, ``{Une nouvelle limite non-relativiste du groupe de
  Poincar\'e},'' {\em Ann. Inst. H Poincar\'e} {\bf 3} (1965), no.~1, 1--12.

\bibitem{Gupta1966OnAA}
N.~Gupta, ``On an analogue of the galilei group,'' {\em Nuovo Cimento Della
  Societa Italiana Di Fisica A-nuclei Particles and Fields} {\bf 44} (1966)
  512--517.

\bibitem{Duval:2014uva}
C.~Duval, G.~W. Gibbons, and P.~A. Horvathy, ``{Conformal Carroll groups and
  BMS symmetry},'' {\em Class. Quant. Grav.} {\bf 31} (2014) 092001,
  \href{http://www.arXiv.org/abs/1402.5894}{{\tt 1402.5894}}.

\bibitem{Duval:2014lpa}
C.~Duval, G.~W. Gibbons, and P.~A. Horvathy, ``{Conformal Carroll groups},''
  {\em J. Phys. A} {\bf 47} (2014), no.~33, 335204,
  \href{http://www.arXiv.org/abs/1403.4213}{{\tt 1403.4213}}.

\bibitem{Liu:2022mne}
W.-B. Liu and J.~Long, ``{Symmetry group at future null infinity: Scalar
  theory},'' {\em Phys. Rev. D} {\bf 107} (2023), no.~12, 126002,
  \href{http://www.arXiv.org/abs/2210.00516}{{\tt 2210.00516}}.

\bibitem{Donnay:2022wvx}
L.~Donnay, A.~Fiorucci, Y.~Herfray, and R.~Ruzziconi, ``{Bridging Carrollian
  and celestial holography},'' {\em Phys. Rev. D} {\bf 107} (2023), no.~12,
  126027, \href{http://www.arXiv.org/abs/2212.12553}{{\tt 2212.12553}}.

\bibitem{Salzer:2023jqv}
J.~Salzer, ``{An embedding space approach to Carrollian CFT correlators for
  flat space holography},'' {\em JHEP} {\bf 10} (2023) 084,
  \href{http://www.arXiv.org/abs/2304.08292}{{\tt 2304.08292}}.

\bibitem{Nguyen:2023miw}
K.~Nguyen, ``{Carrollian conformal correlators and massless scattering
  amplitudes},'' {\em JHEP} {\bf 01} (2024) 076,
  \href{http://www.arXiv.org/abs/2311.09869}{{\tt 2311.09869}}.

\bibitem{Mason:2023mti}
L.~Mason, R.~Ruzziconi, and A.~Yelleshpur~Srikant, ``{Carrollian Amplitudes and
  Celestial Symmetries},'' \href{http://www.arXiv.org/abs/2312.10138}{{\tt
  2312.10138}}.

\bibitem{Liu:2024nfc}
W.-B. Liu, J.~Long, and X.-Q. Ye, ``{Feynman rules and loop structure of
  Carrollian amplitudes},'' {\em JHEP} {\bf 05} (2024) 213,
  \href{http://www.arXiv.org/abs/2402.04120}{{\tt 2402.04120}}.

\bibitem{Stieberger:2024shv}
S.~Stieberger, T.~R. Taylor, and B.~Zhu, ``{Carrollian Amplitudes from
  Strings},'' \href{http://www.arXiv.org/abs/2402.14062}{{\tt 2402.14062}}.

\bibitem{Adamo:2024mqn}
T.~Adamo, W.~Bu, P.~Tourkine, and B.~Zhu, ``{Eikonal amplitudes on the
  celestial sphere},'' \href{http://www.arXiv.org/abs/2405.15594}{{\tt
  2405.15594}}.

\bibitem{Alday:2024yyj}
L.~F. Alday, M.~Nocchi, R.~Ruzziconi, and A.~Yelleshpur~Srikant, ``{Carrollian
  Amplitudes from Holographic Correlators},''
  \href{http://www.arXiv.org/abs/2406.19343}{{\tt 2406.19343}}.

\bibitem{Ruzziconi:2024zkr}
R.~Ruzziconi, S.~Stieberger, T.~R. Taylor, and B.~Zhu, ``{Differential
  Equations for Carrollian Amplitudes},''
  \href{http://www.arXiv.org/abs/2407.04789}{{\tt 2407.04789}}.

\bibitem{Liu:2024llk}
W.-B. Liu, J.~Long, H.-Y. Xiao, and J.-L. Yang, ``{On the definition of
  Carrollian amplitudes in general dimensions},''
  \href{http://www.arXiv.org/abs/2407.20816}{{\tt 2407.20816}}.

\bibitem{Barnich:2010eb}
G.~Barnich and C.~Troessaert, ``{Aspects of the BMS/CFT correspondence},'' {\em
  JHEP} {\bf 05} (2010) 062, \href{http://www.arXiv.org/abs/1001.1541}{{\tt
  1001.1541}}.

\bibitem{Sachs:1962wk}
R.~K. Sachs, ``{Gravitational waves in general relativity. 8. Waves in
  asymptotically flat space-times},'' {\em Proc. Roy. Soc. Lond. A} {\bf 270}
  (1962) 103--126.

\bibitem{Campiglia:2014yka}
M.~Campiglia and A.~Laddha, ``{Asymptotic symmetries and subleading soft
  graviton theorem},'' {\em Phys. Rev. D} {\bf 90} (2014), no.~12, 124028,
  \href{http://www.arXiv.org/abs/1408.2228}{{\tt 1408.2228}}.

\bibitem{Campiglia:2015yka}
M.~Campiglia and A.~Laddha, ``{New symmetries for the Gravitational
  S-matrix},'' {\em JHEP} {\bf 04} (2015) 076,
  \href{http://www.arXiv.org/abs/1502.02318}{{\tt 1502.02318}}.

\bibitem{Liu:2023qtr}
W.-B. Liu and J.~Long, ``{Symmetry group at future null infinity II: Vector
  theory},'' {\em JHEP} {\bf 07} (2023) 152,
  \href{http://www.arXiv.org/abs/2304.08347}{{\tt 2304.08347}}.

\bibitem{Liu:2023gwa}
W.-B. Liu and J.~Long, ``{Symmetry group at future null infinity III:
  Gravitational theory},'' {\em JHEP} {\bf 10} (2023) 117,
  \href{http://www.arXiv.org/abs/2307.01068}{{\tt 2307.01068}}.

\bibitem{Li:2023xrr}
A.~Li, W.-B. Liu, J.~Long, and R.-Z. Yu, ``{Quantum flux operators for
  Carrollian diffeomorphism in general dimensions},'' {\em JHEP} {\bf 11}
  (2023) 140, \href{http://www.arXiv.org/abs/2309.16572}{{\tt 2309.16572}}.

\bibitem{Liu:2023jnc}
W.-B. Liu, J.~Long, and X.-H. Zhou, ``{Quantum flux operators in higher spin
  theories},'' {\em Phys. Rev. D} {\bf 109} (2024), no.~8, 086012,
  \href{http://www.arXiv.org/abs/2311.11361}{{\tt 2311.11361}}.

\bibitem{Liu:2024nkc}
W.-B. Liu and J.~Long, ``{Holographic dictionary from bulk reduction},'' {\em
  Phys. Rev. D} {\bf 109} (2024), no.~6, L061901,
  \href{http://www.arXiv.org/abs/2401.11223}{{\tt 2401.11223}}.

\bibitem{Liu:2024rvz}
W.-B. Liu, J.~Long, and X.-H. Zhou, ``{Electromagnetic helicity flux operators
  in higher dimensions},'' \href{http://www.arXiv.org/abs/2407.20077}{{\tt
  2407.20077}}.

\bibitem{Dowker:1978aza}
J.~S. Dowker, ``{Thermal properties of Green's functions in Rindler, de Sitter,
  and Schwarzschild spaces},'' {\em Phys. Rev. D} {\bf 18} (1978), no.~6, 1856.

\bibitem{Birrell:1982ix}
N.~D. Birrell and P.~C.~W. Davies, {\em {Quantum Fields in Curved Space}}.
\newblock Cambridge Monographs on Mathematical Physics. Cambridge Univ. Press,
  Cambridge, UK, 2, 1984.

\bibitem{Bacry:1974af}
H.~Bacry, P.~Combe, and P.~Sorba, ``{Connected subgroups of the poincare group.
  1.},'' {\em Rept. Math. Phys.} {\bf 5} (1974) 145--186.

\bibitem{1966AmJPh..34.1174R}
W.~{Rindler}, ``{Kruskal Space and the Uniformly Accelerated Frame},'' {\em
  American Journal of Physics} {\bf 34} (Dec., 1966) 1174--1178.

\bibitem{Unruh:1976db}
W.~G. Unruh, ``{Notes on black hole evaporation},'' {\em Phys. Rev. D} {\bf 14}
  (1976) 870.

\bibitem{Sewell:1982zz}
G.~L. Sewell, ``{Quantum fields on manifolds: PCT and gravitationally induced
  thermal states},'' {\em Annals Phys.} {\bf 141} (1982) 201--224.

\bibitem{Unruh:1983ac}
W.~G. Unruh and N.~Weiss, ``{Acceleration Radiation in Interacting Field
  Theories},'' {\em Phys. Rev. D} {\bf 29} (1984) 1656.

\bibitem{Bisognano:1975ih}
J.~J. Bisognano and E.~H. Wichmann, ``{On the Duality Condition for a Hermitian
  Scalar Field},'' {\em J. Math. Phys.} {\bf 16} (1975) 985--1007.

\bibitem{Bisognano:1976za}
J.~J. Bisognano and E.~H. Wichmann, ``{On the Duality Condition for Quantum
  Fields},'' {\em J. Math. Phys.} {\bf 17} (1976) 303--321.

\bibitem{Wightman:1956zz}
A.~S. Wightman, ``{Quantum Field Theory in Terms of Vacuum Expectation
  Values},'' {\em Phys. Rev.} {\bf 101} (1956) 860--866.

\bibitem{Takagi:1986kn}
S.~Takagi, ``{Vacuum Noise and Stress Induced by Uniform Acceleration:
  Hawking-Unruh Effect in Rindler Manifold of Arbitrary Dimension},'' {\em
  Prog. Theor. Phys. Suppl.} {\bf 88} (1986) 1--142.

\bibitem{2008RvMP...80..787C}
L.~C.~B. {Crispino}, A.~{Higuchi}, and G.~E.~A. {Matsas}, ``{The Unruh effect
  and its applications},'' {\em Reviews of Modern Physics} {\bf 80} (July,
  2008) 787--838, \href{http://www.arXiv.org/abs/0710.5373}{{\tt 0710.5373}}.

\bibitem{Israel:1976ur}
W.~Israel, ``{Thermo field dynamics of black holes},'' {\em Phys. Lett. A} {\bf
  57} (1976) 107--110.

\bibitem{Gibbons:1976pt}
G.~W. Gibbons and M.~J. Perry, ``{Black Holes and Thermal Green's Functions},''
  {\em Proc. Roy. Soc. Lond. A} {\bf 358} (1978) 467--494.

\bibitem{Christensen:1978tw}
S.~M. Christensen and M.~J. Duff, ``{FLAT SPACE AS A GRAVITATIONAL
  INSTANTON},'' {\em Nucl. Phys. B} {\bf 146} (1978) 11--19.

\bibitem{Troost:1978yk}
W.~Troost and H.~van Dam, ``{Thermal Propagators and Accelerated Frames of
  Reference},'' {\em Nucl. Phys. B} {\bf 152} (1979) 442--460.

\bibitem{Linet:1995mq}
B.~Linet, ``{Euclidean scalar and spinor Green's functions in Rindler space},''
  \href{http://www.arXiv.org/abs/gr-qc/9505033}{{\tt gr-qc/9505033}}.

\bibitem{Usyukina:1992jd}
N.~I. Usyukina and A.~I. Davydychev, ``{An Approach to the evaluation of three
  and four point ladder diagrams},'' {\em Phys. Lett. B} {\bf 298} (1993)
  363--370.

\bibitem{Usyukina:1994iw}
N.~I. Usyukina and A.~I. Davydychev, ``{New results for two loop off-shell
  three point diagrams},'' {\em Phys. Lett. B} {\bf 332} (1994) 159--167,
  \href{http://www.arXiv.org/abs/hep-ph/9402223}{{\tt hep-ph/9402223}}.

\bibitem{Chavez:2012kn}
F.~Chavez and C.~Duhr, ``{Three-mass triangle integrals and single-valued
  polylogarithms},'' {\em JHEP} {\bf 11} (2012) 114,
  \href{http://www.arXiv.org/abs/1209.2722}{{\tt 1209.2722}}.

\bibitem{Lee:2013hzt}
R.~N. Lee and A.~A. Pomeransky, ``{Critical points and number of master
  integrals},'' {\em JHEP} {\bf 11} (2013) 165,
  \href{http://www.arXiv.org/abs/1308.6676}{{\tt 1308.6676}}.

\bibitem{Weinzierl:2022eaz}
S.~Weinzierl, {\em {Feynman Integrals. A Comprehensive Treatment for Students
  and Researchers}}.
\newblock UNITEXT for Physics. Springer, 2022.

\bibitem{delaCruz:2019skx}
L.~de~la Cruz, ``{Feynman integrals as A-hypergeometric functions},'' {\em
  JHEP} {\bf 12} (2019) 123, \href{http://www.arXiv.org/abs/1907.00507}{{\tt
  1907.00507}}.

\bibitem{Klausen:2019hrg}
R.~P. Klausen, ``{Hypergeometric Series Representations of Feynman Integrals by
  GKZ Hypergeometric Systems},'' {\em JHEP} {\bf 04} (2020) 121,
  \href{http://www.arXiv.org/abs/1910.08651}{{\tt 1910.08651}}.

\bibitem{Weinberg:1965nx}
S.~Weinberg, ``{Infrared photons and gravitons},'' {\em Phys. Rev.} {\bf 140}
  (1965) B516--B524.

\bibitem{unitary}
I.~Gelfand and M.~Neumark, ``{Unitary representations of the lorentz group},''
  {\em Acad. Sci. USSR. J. Phys} {\bf 10} (1946) 93--94.

\bibitem{Bargmann:1946me}
V.~Bargmann, ``{Irreducible unitary representations of the Lorentz group},''
  {\em Annals Math.} {\bf 48} (1947) 568--640.

\bibitem{Atanasov:2021cje}
A.~Atanasov, W.~Melton, A.-M. Raclariu, and A.~Strominger, ``{Conformal block
  expansion in celestial CFT},'' {\em Phys. Rev. D} {\bf 104} (2021), no.~12,
  126033, \href{http://www.arXiv.org/abs/2104.13432}{{\tt 2104.13432}}.

\bibitem{Candelas:1978gg}
P.~Candelas and D.~Deutsch, ``{Fermion Fields in Accelerated States},'' {\em
  Proc. Roy. Soc. Lond. A} {\bf 362} (1978) 251--262.

\bibitem{Soffel:1980kx}
M.~Soffel, B.~Muller, and W.~Greiner, ``{DIRAC PARTICLES IN RINDLER SPACE},''
  {\em Phys. Rev. D} {\bf 22} (1980) 1935--1937.

\bibitem{Hacian:1985gqu}
S.~Hacian, ``{Massless Fields of Arbitrary Spin, Uniformly Accelerated Frames,
  and the Zero Point Energy},'' {\em Phys. Rev. D} {\bf 32} (1985) 3216--3220.

\bibitem{Fulling:1972md}
S.~A. Fulling, ``{Nonuniqueness of canonical field quantization in Riemannian
  space-time},'' {\em Phys. Rev. D} {\bf 7} (1973) 2850--2862.

\bibitem{Davies:1974th}
P.~C.~W. Davies, ``{Scalar particle production in Schwarzschild and Rindler
  metrics},'' {\em J. Phys. A} {\bf 8} (1975) 609--616.

\bibitem{Candelas:1976jv}
P.~Candelas and D.~J. Raine, ``{Quantum Field Theory in Incomplete
  Manifolds},'' {\em J. Math. Phys.} {\bf 17} (1976) 2101--2112.

\bibitem{Dowker:1977zj}
J.~S. Dowker, ``{Quantum Field Theory on a Cone},'' {\em J. Phys. A} {\bf 10}
  (1977) 115--124.

\bibitem{Hill:1986ec}
C.~T. Hill, ``{One Loop Operator Matrix Elements in the Unruh Vacuum},'' {\em
  Nucl. Phys. B} {\bf 277} (1986) 547--574.

\bibitem{Burgess:2018sou}
C.~P. Burgess, J.~Hainge, G.~Kaplanek, and M.~Rummel, ``{Failure of
  Perturbation Theory Near Horizons: the Rindler Example},'' {\em JHEP} {\bf
  10} (2018) 122, \href{http://www.arXiv.org/abs/1806.11415}{{\tt 1806.11415}}.

\bibitem{DeWitt:1975ys}
B.~S. DeWitt, ``{Quantum Field Theory in Curved Space-Time},'' {\em Phys.
  Rept.} {\bf 19} (1975) 295--357.

\bibitem{Borthwick:2024skd}
J.~Borthwick, M.~Chantreau, and Y.~Herfray, ``{Ti and Spi, Carrollian extended
  boundaries at timelike and spatial infinity},''
  \href{http://www.arXiv.org/abs/2412.15996}{{\tt 2412.15996}}.

\bibitem{Have:2024dff}
E.~Have, K.~Nguyen, S.~Prohazka, and J.~Salzer, ``{Massive carrollian fields at
  timelike infinity},'' \href{http://www.arXiv.org/abs/2402.05190}{{\tt
  2402.05190}}.

\bibitem{Duary:2024kxl}
S.~Duary and S.~Upadhyay, ``{Flat limit of AdS/CFT from AdS geodesics:
  scattering amplitudes and antipodal matching of Li\'enard-Wiechert fields},''
  \href{http://www.arXiv.org/abs/2411.08540}{{\tt 2411.08540}}.

\bibitem{Figueroa-OFarrill:2021sxz}
J.~Figueroa-O'Farrill, E.~Have, S.~Prohazka, and J.~Salzer, ``{Carrollian and
  celestial spaces at infinity},'' {\em JHEP} {\bf 09} (2022) 007,
  \href{http://www.arXiv.org/abs/2112.03319}{{\tt 2112.03319}}.

\bibitem{Campiglia:2024uqq}
M.~Campiglia and A.~Sudhakar, ``{Gravitational Poisson brackets at null
  infinity compatible with smooth superrotations},'' {\em JHEP} {\bf 12} (2024)
  170, \href{http://www.arXiv.org/abs/2408.13067}{{\tt 2408.13067}}.

\bibitem{Kulp:2024scx}
J.~Kulp and S.~Pasterski, ``{Multiparticle States for the Flat Hologram},''
  \href{http://www.arXiv.org/abs/2501.00462}{{\tt 2501.00462}}.

\bibitem{Matsubara:1955ws}
T.~Matsubara, ``{A New approach to quantum statistical mechanics},'' {\em Prog.
  Theor. Phys.} {\bf 14} (1955) 351--378.

\bibitem{1957JPSJ...12..570K}
R.~{Kubo}, ``{Statistical-Mechanical Theory of Irreversible Processes. I},''
  {\em Journal of the Physical Society of Japan} {\bf 12} (June, 1957)
  570--586.

\bibitem{Martin:1959jp}
P.~C. Martin and J.~S. Schwinger, ``{Theory of many particle systems. 1.},''
  {\em Phys. Rev.} {\bf 115} (1959) 1342--1373.

\bibitem{Umezawa:1982nv}
H.~Umezawa, H.~Matsumoto, and M.~Tachiki, {\em {THERMO FIELD DYNAMICS AND
  CONDENSED STATES}}.
\newblock 1982.

\bibitem{2007tisp.book.....G}
I.~S. {Gradshteyn}, I.~M. {Ryzhik}, A.~{Jeffrey}, and D.~{Zwillinger}, {\em
  {Table of Integrals, Series, and Products}}.
\newblock 2007.

\bibitem{Lewin:100000}
L.~Lewin, {\em {Polylogarithms and associated functions}}.
\newblock North-Holland, New York, NY, 1981.

\bibitem{Zagier:1990}
D.~Zagier, ``{The Bloch-Wigner-Ramakrishnan polylogarithm function},'' {\em
  Math. Ann.} {\bf 286} (1990) 613--624.

\bibitem{Bloch2011HigherRA}
S.~Bloch, {\em Higher Regulators, Algebraic K-Theory, and Zeta Functions of
  Elliptic Curves}.
\newblock Amer Mathematical Society, 2000.

\bibitem{Cartier:2007zz}
P.~Cartier, B.~Julia, P.~Moussa, and P.~Vanhove, eds., {\em {Proceedings, Les
  Houches School of Physics: Frontiers in Number Theory, Physics and Geometry
  II: On Conformal Field Theories, Discrete Groups and Renormalization}: {Les
  Houches, France, March 9-21, 2003}}.
\newblock Springer, Berlin, Germany, 2007.

\bibitem{geometry}
H.~S.~M. Coxeter and S.~Greitzer, {\em {Geometry Revisited}}.
\newblock American Mathematical Society, 2021.

\bibitem{1953hft1.book...59E}
A.~{Erdelyi}, ``{Higher Transcendental Functions},'' in {\em Higher
  Transcendental Functions}, p.~59.
\newblock 1953.

\bibitem{Cattani2006THREELO}
E.~Cattani, ``Three lectures on hypergeometric functions,''
\newblock 2006.

\bibitem{Gelfand:1990bua}
I.~M. Gelfand, M.~M. Kapranov, and A.~V. Zelevinsky, ``{Generalized Euler
  integrals and A-hypergeometric functions },'' {\em Adv. Math.} {\bf 84}
  (1990), no.~2, 255--271.

\bibitem{Boos:1990rg}
E.~E. Boos and A.~I. Davydychev, ``{A Method of evaluating massive Feynman
  integrals},'' {\em Theor. Math. Phys.} {\bf 89} (1991) 1052--1063.

\end{thebibliography}\endgroup

\end{document}